\documentclass[12pt,a4paper]{article}
\pdfoutput=1
\usepackage{graphicx}

\usepackage{amsmath,amsfonts,amssymb}
\usepackage{cite}
\usepackage{colortbl}
\usepackage{pifont}

\providecommand{\openone}{\leavevmode\hbox{\small1\kern-3.8pt\normalsize1}}

\newcommand{\RE}{\operatorname{Re}}
\newcommand{\IM}{\operatorname{Im}}

\begin{document}

\vspace*{-2.5cm}
\begin{flushright}
IFT-UAM/CSIC-19-71 \\
CFTP/19-017
\end{flushright}
\vspace{0.cm}

\begin{center}
\begin{Large}
{\bf The minimal stealth boson: models and benchmarks}
\end{Large}

\vspace{0.5cm}
J.~A.~Aguilar--Saavedra$^{a,b,}$\footnote{jaas@ugr.es}, F.~R. Joaquim$^{c,}$\footnote{filipe.joaquim@tecnico.ulisboa.pt} \\[1mm]
\begin{small}
{$^a$ Instituto de F\'isica Te\'orica UAM-CSIC, Campus de Cantoblanco, E-28049 Madrid, Spain} \\ 
{$^b$ Universidad de Granada, E-18071 Granada, Spain (on leave)} \\ 
{$^c$ Departamento de F\'{\i}sica and CFTP, Instituto Superior T\'ecnico, Universidade de Lisboa, Av. Rovisco Pais 1, 1049-001 Lisboa, Portugal} 
\end{small}
\end{center}

\begin{abstract}
	Stealth bosons are relatively light boosted particles with a cascade decay $S \to A_1 A_2 \to q \bar q q \bar q$, reconstructed as a single fat jet. In this work, we establish minimal extensions of the Standard Model that allow for such processes. Namely, we consider models containing a new (leptophobic) neutral gauge boson $Z'$ and two scalar singlets, plus extra matter required to cancel the $\text{U}(1)'$ anomalies. Our analysis shows that, depending on the model and benchmark scenario, the expected statistical significance of stealth boson signals (yet uncovered by current searches at the Large Hadron Collider) is up to nine times larger than for the most sensitive of the standard leptophobic $Z'$ signals such as dijets, $t \bar t$ pairs or dibosons. These results provide strong motivation for model-independent searches that cover these complex signals.
\end{abstract}

\section{Introduction}
New heavy resonances are easy to spot at the Large Hadron Collider (LHC) when they decay into charged leptons, e.g. $Z' \to e^+ e^- / \mu^+ \mu^-$, $W' \to e \nu / \mu \nu$, but they are quite more difficult to detect in hadronic final states, since the production of quarks and gluons by QCD interactions has a very large cross section. Still, heavy resonances decaying into boosted hadronically-decaying $W$, $Z$ or Higgs bosons, or top quarks, may be separated from the background. In the last decade, great progress has been made in this direction with the development of jet substructure techniques~\cite{Butterworth:2008iy,Thaler:2008ju,Kaplan:2008ie,Almeida:2008yp, Plehn:2009rk,Plehn:2010st,Thaler:2010tr,Hook:2011cq,Jankowiak:2011qa,Thaler:2011gf,Larkoski:2013eya,Moult:2016cvt,Datta:2017rhs} and grooming algorithms~\cite{Butterworth:2008iy,Krohn:2009th,Ellis:2009me,Larkoski:2014wba}. These tools allow to distinguish jets originating from boosted hadronically-decaying bosons and top quarks from the Standard Model (SM) background, composed mainly by quark and gluon jets produced in QCD processes. In this way, searches for diboson~\cite{Khachatryan:2014hpa,Aad:2015owa,Khachatryan:2016cfa,Sirunyan:2017wto,Aaboud:2017ahz,Aaboud:2017eta,Sirunyan:2017acf}, $t \bar t$~\cite{Sirunyan:2017uhk} and $t \bar b$~\cite{Aad:2014xra,Sirunyan:2017ukk} resonances in purely hadronic channels have been performed, with a sensitivity that turns out to be competitive with channels involving leptons in the final state. Nevertheless, these searches are insensitive to heavy resonances decaying into more complex hadronic final states, giving rise to multi-pronged jets. 

One example of a multi-pronged jet signature is given by the `stealth bosons' introduced in ref.~\cite{Aguilar-Saavedra:2017zuc}, which are boosted particles $S$ decaying hadronically into four collimated quarks, via two (equal or different) intermediate particles $A_1$, $A_2$, namely
\begin{equation}
S \to A_1 A_2 \to q \bar q q \bar q \,.
\label{ec:sb}
\end{equation}
The particles $A_{1,2}$ in the above decay chain may be SM weak bosons $W$, $Z$, a Higgs boson, or new relatively light (pseudo-)scalars. When $S$ is produced in the decay of a much heavier parent resonance $R$,
\begin{equation}
R \to S + X \,,
\label{ec:RtoS}
\end{equation}
(with $X$ an additional particle) its experimental signature is a fat jet with four-pronged structure. Jet substructure observables designed to distinguish two-pronged $Z$, $W$ and Higgs decays from the QCD background, for example the so-called $D_2$~\cite{Larkoski:2013eya} and $\tau_{21}$~\cite{Thaler:2010tr,Thaler:2011gf} variables respectively used by the ATLAS and CMS Collaborations, classify four-pronged jets as QCD-like. Therefore, should a new resonance involve one or more decay products of this type, it would be very hard to identify it in current searches. On the other hand, generic searches that use a multivariate tool like an anti-QCD tagger~\cite{Aguilar-Saavedra:2017rzt} to pin down multi-pronged jets from the QCD background are sensitive to this type of signals. Notice that if $S$ weakly couples to SM particles, for example if it is a neutral scalar, its direct production cross section may be too small for this particle to be directly observed. 

The aim of this paper is to investigate the minimal SM extension in which stealth boson signals may appear, and contextualise the relevance of these signatures as a discovery channel for new leptophobic resonances, when compared to the usual decay modes searched for at the LHC, like dijets, dibosons or $t \bar t$ pairs. In section~\ref{sec:2} we find, following a bottom-up approach, that the minimal additional content that allows for the cascade decays in (\ref{ec:sb}) is a $Z'$ boson and two scalars that are singlets under the SM group but charged under the extra $\text{U}(1)'$ (further details of the models are given in appendix~\ref{sec:b}). 
An extension with one $Z'$ boson, a new scalar doublet and a scalar singlet, which is also attractively simple, does not serve our purposes, as briefly discussed in appendix~\ref{sec:a}. Section~\ref{sec:3a} is devoted to the discussion of how benchmark scenarios for the scalar masses and mixings are tested, to ensure that the reconstructed model parameters lead to an absolute minimum of the scalar potential. In section~\ref{sec:4} and appendix~\ref{sec:c} those benchmark scenarios are studied in detail performing fast simulations of the various $Z'$ signals in the decays into stealth bosons, dijets, and $t \bar t$, as well as their SM backgrounds. We discuss our results in section~\ref{sec:5}.

\section{The minimal stealth boson models}
\label{sec:2}

Following a minimalistic approach, we assume that the heavy resonance $R$ in (\ref{ec:RtoS}) is a neutral colour-singlet $Z'$ boson, so that the gauge symmetry of the SM is extended by an extra $\text{U}(1)'$.\footnote{Alternatives in the context of left-right models can easily be worked out from the results in Ref.~\cite{Aguilar-Saavedra:2015iew}. In that work we focused on the `resolved' signatures where three or four well-separated bosons are produced from the cascade decay of a $W'$ boson into new scalars. When the masses of the intermediate particles are lighter, their bosonic decay products are merged giving rise to signatures such as in (\ref{ec:sb}). Cascade decays can also be produced in a variety of other non-minimal scenarios, see for example refs.~\cite{Agashe:2016kfr,Agashe:2018leo}.} We require the $Z'$ boson to be leptophobic, i.e. the left-handed lepton doublets $\ell_L$ and right-handed singlets $e_R$ have zero hypercharge $Y'_{\ell} = Y'_{e} = 0$ under the new $\text{U}(1)'$. Otherwise, the leptonic signals $Z' \to e^+ e^-$ and $Z' \to \mu^+ \mu^-$ would be easy to observe at the LHC. Gauge invariance of the Yukawa couplings with the SM Higgs doublet $\Phi$,
\begin{equation}
\mathcal{L}_Y = - y_u \bar q_L \tilde \Phi u_R - y_d \bar q_L \Phi d_R - y_e \bar \ell_L \Phi e_R + \text{h.c.} \,,
\label{ec:Y}
\end{equation}
then requires that $Y'_{\Phi} = 0$, and that the left-handed quark doublets $q_L$ and right-handed quark singlets $u_R$, $d_R$ have the same hypercharge $Y'_{q} = Y'_{u} = Y'_{d} \equiv z$ (for simplicity we omit generation indices and generically denote the Yukawa couplings by $y_{u}$, $y_d$, $y_e$). The hypercharge assignments are collected in table~\ref{tab:Ysm}, where $z$ is unspecified.

Cancellation of the anomalies associated to $\text{U}(1)'$ requires introducing extra matter, which we assume to be vector-like under the SM gauge group, to preserve SM anomaly cancellation. Two simple choices for these extra degrees of freedom, which we will denote as model 1 and 2, are:
%

\smallskip
$\bullet$ {\bf Model 1:} One set of vector-like quarks, comprising a doublet $(T_1 \; B_1)$ with SM hypercharge $Y=1/6$ plus vector-like singlets $T_2$, $B_2$ of charge $2/3$ and $-1/3$, respectively.

$\bullet$ {\bf Model 2:} One set of vector-like leptons, with a doublet $(N_1 \; E_1)$ with SM hypercharge $Y=-1/2$ plus vector-like singlets $N_2$, $E_2$ with charges 0 and $-1$, respectively.
%
\smallskip

\noindent The hypercharge assignments for these fields is summarised in table~\ref{tab:Ynew}. We note that model 2, with $z=1/3$, has been considered in previous literature~\cite{Caron:2018yzp}, motivated by the search for an anomaly-free $Z'$ dark matter mediator (the dark matter particle corresponds to the singlet $N_2$) with weak constraints from direct detection experiments. A similar model, with a three-fold replication of the new lepton set, $(N_i \; E_i)_{L}$, $(N_i \; E_i)_{R}$, $N_{jL}$, $N_{jR}$, $E_{jL}$, $E_{jL,}$, with $i=1,2,3$ and $j=4,5,6$, also preserves the cancellation of anomalies. In this case, the lepton hypercharges are $1/3$ of the values quoted in table~\ref{tab:Ynew} for model 2. The phenomenology, except for the signals associated to the new fermions which we do not address here, is the same of model 1. Other more baroque possibilities exist for the choice of new fermions by, for example, introducing vector-like quark doublets with $Y= 7/6$ or $Y = -5/6$ and $\mathcal{O}\sim 10$ quark singlets. Notice that kinetic mixing would modify our hypercharge assignments but, since both the $\text{U}(1)'$ coupling $g_{Z^\prime}$ and the hypercharge parameter $z$ are unspecified, it has no effect in our analysis and we do not consider it.

\begin{table}[t]
\begin{center}
\begin{tabular}{ccccccc}
& $Y$ & $Y'$ & & & $Y$ & $Y'$ \\
$(u\;d)_L$ & $1/6$ & $z$ & \quad & $u_R$  & $2/3$  & $z$ \\
                 &           &        &           & $d_R$  & $-1/3$ & $z$ \\ 
$(\nu\;e)_L$ & $-1/2$ & $0$ & \quad & $e_R$  & $-1$  & $0$ \\
\end{tabular}
\end{center}
\caption{Hypercharge assignments for the SM fermions, with $z$ a free parameter. The columns labelled with $Y$ collect the standard hypercharges with the normalisation $Q=T_3+Y$.}
\label{tab:Ysm}
\end{table}

\begin{table}[t]
\begin{center}
\begin{tabular}{ccccccc}
\underline{Model 1} & $Y$ & $Y'$ & & & $Y$ & $Y'$ \\
$(T_1\;B_1)_L$ & $1/6$ & $-3z/2$ & \quad & $(T_1\;B_1)_R$ & $1/6$ &  $3z/2$  \\
$T_{2L}$       & $2/3$     & $3z/2$ & \quad & $T_{2R}$       & $2/3$ &  $-3z/2$  \\
$B_{2L}$       & $-1/3$    & $3z/2$ & \quad & $B_{2R}$       & $-1/3$ &  $-3z/2$
\end{tabular}
\vspace{5mm}

\begin{tabular}{ccccccc}
\underline{Model 2} & $Y$ & $Y'$ & & & $Y$ & $Y'$ \\
$(N_1\;E_1)_L$ & $-1/2$ & $-9 z/2$ & \quad & $(N_1\;E_1)_R$ & $-1/2$ &  $9z/2$  \\
$N_{2L}$       & $0$     & $9 z/2$ & \quad & $N_{2R}$       & $0$ &  $-9z/2$  \\
$E_{2L}$       & $-1$    & $9 z/2$ & \quad & $E_{2R}$       & $-1$ &  $-9z/2$
\end{tabular}

\end{center}
\caption{Two minimal extensions of the fermion sector with $\text{U}(1)'$ anomaly cancellation (top: vector-like quarks, bottom: vector-like leptons). The $Y^\prime$ hypercharges are given in terms of $z$.}
\label{tab:Ynew}
\end{table}

The scalar sector of the SM must be extended in order to break the $\text{U}(1)'$ symmetry and generate the $Z'$ boson mass. The simplest possibility is to consider a neutral complex $\text{SU}(2)_L$ singlet $\chi$ with non-zero hypercharge $Y'_\chi$ under $\text{U}(1)'$. Having the $Z'$ mass generated by a higher $\text{SU}(2)_L$ multiplet is problematic, as its vacuum expectation value (VEV) would also contribute to the weak boson masses. Notice that the heavy fermion masses can also be generated with the same singlet provided $Y'_{\chi} = 3 z$ (for model 1) or $Y'_{\chi} = 9 z$ (for model 2), and that the new fermions do not have Yukawa interactions with the SM ones. Moreover, in model 2, if the lightest new fermion is a neutral singlet, it may possibly be a dark matter particle~\cite{Caron:2018yzp}, while in model 1 the lightest new quark would have exotic signatures~\cite{Langacker:2008yv}, not addressed here. 

Additional scalars, besides this singlet, are required to yield the cascade decays (\ref{ec:sb}) and (\ref{ec:RtoS}) (ref.~\cite{Caron:2018yzp} only considers one scalar singlet). As discussed in appendix~\ref{sec:a}, adding a second scalar doublet is not a viable option. Thus, we instead consider a scalar sector comprising the SM doublet $\Phi$ and two complex singlets $\chi_1$, $\chi_2$ with the same hypercharge. Further extensions of the scalar sector that allow interactions of the new fermions with the SM ones are possible, but they are not required for our purposes. The most general gauge-invariant scalar potential is $V = V_{Z_2} + V_{\not Z_2}$, with
\begin{eqnarray}
V_{Z_2} & = & m_{0}^2 \Phi^\dagger \Phi + m_{11}^2 \chi_1^\dagger \chi_1 + m_{22}^2 \chi_2^\dagger \chi_2 
\notag \\
& &+ \frac{\lambda_0}{2} (\Phi^\dagger \Phi)^2  + \frac{\lambda_1}{2} (\chi_1^\dagger \chi_1)^2
 + \frac{\lambda_2}{2} (\chi_2^\dagger \chi_2)^2
+ \lambda_3 (\chi_1^\dagger \chi_1) (\chi_2^\dagger \chi_2) \notag \\
& & + \frac{1}{2} \left[  \lambda_4 (\chi_1^\dagger \chi_2) (\chi_1^\dagger \chi_2) + \text{h.c.} \right]
+ \frac{\lambda_5}{2} (\Phi^\dagger \Phi) (\chi_1^\dagger \chi_1)
+ \frac{\lambda_6}{2} (\Phi^\dagger \Phi) (\chi_2^\dagger \chi_2)
\,, \notag \\
V_{\not Z_2} & = &  m_{12}^2  \chi_1^\dagger \chi_2 +\frac{1}{2} \left[  \lambda_7 (\chi_1^\dagger \chi_2) (\chi_1^\dagger \chi_1)
+ \lambda_8 (\chi_1^\dagger \chi_2) (\chi_2^\dagger \chi_2) 
+ \lambda_9 (\Phi^\dagger \Phi)  (\chi_1^\dagger \chi_2)\right] + \text{h.c.}  \,,
\label{ec:VS}
\end{eqnarray}
where $V_{Z_2}$ ($V_{\not Z_2}$) contains the terms which are invariant under (break) a $Z_2$ symmetry for which $\chi_2\rightarrow - \chi_2$ and all remaining fields transform trivially. Among all the above parameters,  $m_0^2$, $m_{11}^2$, $m_{22}^2$, $\lambda_{0-3}$ and $\lambda_{5,6}$ are real, while $m_{12}^2$, $\lambda_{4}$ and $\lambda_{7-9}$ can be, in general, complex. We write the neutral scalar in $\Phi = (\phi^+ \; \phi^0)^T$ and  the singlets $\chi_{1,2}$ as
\begin{align}
& \phi^0 = \frac{1}{\sqrt 2} (\rho_0+v + i \eta_0 ) \,, \quad \chi_1 = \frac{1}{\sqrt 2} (\rho_1+u_1 + i \eta_1 ) \,, \quad
\chi_2 = \frac{1}{\sqrt 2} (\rho_2 + i \eta_2+ u_2 e^{i \varphi}) \,, 
\end{align}
such that the VEVs are
\begin{align}
& \langle \phi^0 \rangle = \frac{v}{\sqrt 2} \,, \quad \langle \chi_1 \rangle = \frac{u_1}{\sqrt 2}  \,, \quad
\langle \chi_2 \rangle = \frac{u_2 e^{i \varphi}}{\sqrt 2} \,.
\end{align}
Rephasing $\chi_2$ to $\chi_2' = e^{-i \varphi} \chi_2$ (which has real VEV $\langle \chi_2' \rangle = u_2/\sqrt 2$), the potential $V$ can be written in terms of $\chi_2'$ as in (\ref{ec:VS}) with the replacements
\begin{align}
m_{12}^2 \to m_{12}^{\prime \, 2} = m_{12}^2  e^{i \varphi} \,, \quad
\lambda_4 \to \lambda_4' = \lambda_4 e^{i 2\varphi} \,, \quad
\lambda_{7-9} \to \lambda_{7-9}' = \lambda_{7-9} e^{i \varphi} \,,
\label{ec:primes}
\end{align}
while the remaining parameters stay invariant. Therefore, in the general complex case one can always assume $\varphi = 0$ without loss of generality in order to simplify the expressions, with a possible non-vanishing phase absorbed by the above redefinition. On the other hand, if the (unprimed) parameters in the potential are all real this cannot be done, and a non-zero phase $\varphi$ could break CP spontaneously. 

There are four minimisation conditions corresponding to the four parameters $v$, $u_{1,2}$ and $\varphi$,
\begin{align}
& 0 = m_0^2 + \frac{1}{2} v^2 \lambda_0 + \frac{1}{4} ( u_1^2 \lambda_5 + u_2^2 \lambda_6 ) + \frac{1}{2} u_1 u_2 \RE (\lambda'_9) \,, \notag \\
& 0 = u_1 m_{11}^2 + u_2 \RE (m_{12}^{\prime \, 2})  + \frac{u_1}{2} (u_1^2 \lambda_1 + u_2^2 \lambda_3 )
+ \frac{1}{2} u_1 u_2^2 \RE (\lambda'_4) + \frac{3}{4} u_1^2 u_2 \RE (\lambda'_7)  \notag \\
& \quad + \frac{1}{4} u_2^3 \RE (\lambda'_8)  + \frac{1}{4} v^2 u_1 \lambda_5 + \frac{1}{4} v^2 u_2 \RE (\lambda'_9)  \,, \notag \\
& 0 = u_2 m_{22}^2 + u_1 \RE (m_{12}^{\prime \, 2}) + \frac{u_2}{2} (u_1^2 \lambda_3 + u_2^2 \lambda_2 )
+ \frac{1}{2} u_1^2 u_2 \RE (\lambda'_4)  + \frac{3}{4} u_1 u_2^2 \RE (\lambda'_8) \notag \\
& \quad + \frac{1}{4} u_1^3 \RE (\lambda'_7)  + \frac{1}{4} v^2 u_2 \lambda_6 + \frac{1}{4} v^2 u_1 \RE (\lambda'_9)  \,, \notag \\
& 0 = u_1 u_2 \left\{ \IM (m_{12}^{\prime \, 2}) + \frac{1}{2} u_1 u_2 \IM (\lambda'_4)  + \frac{1}{4} \left[ u_1^2 \IM (\lambda'_7)  + u_2^2 \IM (\lambda'_8) + v^2 \IM (\lambda'_9)\, \right] \right\} \,.
\label{mincond}
\end{align}
Since we will be interested in those vacuum configurations with $u_{1,2}\neq 0$, we adopt the common definitions:
\begin{align}
u=\sqrt{u_1^2 + u_2^2} \;,\;\tan \beta = \frac{u_2}{u_1}\,. 
\end{align}
The minimisation conditions are used to express $m_0^2$, $m_{11}^2$, $m_{22}^2$ and $\IM (m_{12}^{\prime \, 2})$ as functions of the remaining potential parameters and VEVs. The two would-be Goldstone bosons are $G_1^0 = \eta_0$ and $G_2^0 = \cos \beta \, \eta_1 + \sin \beta \, \eta_2$. The orthogonal state $A^0 = - \sin \beta \, \eta_1 + \cos \beta \, \eta_2$ is CP-odd, being a mass eigenstate if the parameters in the scalar potential are real and $\varphi = 0$. In the basis $H'_i = (\rho_0 \; \rho_1 \; \rho_2 \; A^0)$, the squared mass matrix, denoted as $M_{ij}$, has elements (with $M_{ij} = M_{ji}$)
\begin{align}
& M_{11} = v^2 \lambda_0 \,, \notag \\
& M_{12} = \frac{1}{2} u v\, [\,  \RE  (\lambda'_9)\sin \beta+ \lambda_5\,\cos \beta\,]  \,, \notag \\
& M_{13} = \frac{1}{2} u v\, [  \, \RE  (\lambda'_9) \cos \beta + \lambda_6 \sin \beta  \,  ] \,, \notag \\
& M_{14} = - \frac{1}{2} u v \IM (\lambda'_9)  \,, \notag \\
& M_{22} = -  \, \left[ \RE (m_{12}^{\prime \, 2}) + \frac{1}{4} v^2 \RE (\lambda'_9) 
+ \frac{1}{4} u^2 \RE (\lambda'_8)  \sin^2 \beta  \right] \tan \beta \notag \\
& \quad + u^2 \left[\lambda_1 \cos^2 \beta  + \frac{3}{8} \RE (\lambda'_7) \sin (2\beta)   \right] \,, \notag \\
& M_{23} = \RE (m_{12}^{\prime \, 2}) + \frac{1}{4} v^2 \RE (\lambda'_9) 
+\frac{1}{2} u^2 \left\{\, \sin (2\beta) [\, \lambda_3 + \RE (\lambda'_4)\,]  \right. \notag \\
& \quad \left. + \frac{3}{2}\, [\,\RE (\lambda'_7)\cos^2 \beta  + \RE (\lambda'_8)\sin^2 \beta  \,] \right\} \,, \notag \\
\displaybreak
& M_{24} = - \frac{1}{2} u^2 [\,\IM (\lambda'_7)\cos \beta  + \IM (\lambda'_4)\sin \beta\,  ] \,, \notag \\
& M_{33} = -  \left[ \RE (m_{12}^{\prime \, 2}) + \frac{1}{4} v^2 \RE (\lambda'_9) 
+ \frac{1}{4} u^2 \RE (\lambda'_7) \cos^2 \beta  \right]\cot \beta \notag \\
& \quad + u^2 \left[ \lambda_2\sin^2 \beta  +\frac{3}{8} \RE (\lambda'_8) \sin (2\beta)  \right] \,,  \notag \\
& M_{34} = - \frac{1}{2} u^2 [\,\IM (\lambda'_4)\cos \beta  + \IM (\lambda'_8)\sin \beta  \,] \,, \notag \\
& M_{44} = - \frac{2}{\sin( 2\beta)} \left[ \RE (m_{12}^{\prime \, 2}) + \frac{1}{4} v^2 \RE (\lambda'_9)  \right] \notag \\
& \quad - u^2 \left\{\RE (\lambda'_4) - \frac{1}{4} [\,\RE (\lambda'_7)\cot \beta  + \RE (\lambda'_8)\tan \beta \, ] \right\} \,.
\label{ec:M}
\end{align}
We remark again that the minimum conditions and the expressions for $M_{ij}$ are valid both for a general potential with complex parameters, in which case one can just drop the primes and assume $\varphi = 0$, or for a potential with real parameters, in which case the primed parameters are defined by (\ref{ec:primes}). We also note in passing that, should we have chosen $\chi_1$ and $\chi_2$ with different hypercharges,  $m_{12}$ and $\lambda_{4-6}$ would vanish, in which case $M_{i4} = 0$ and $A^0$ would be massless.

The scalar interactions with the $Z'$ boson field originate from the term
\begin{equation}
\mathcal{L} = i g_{Z'} Y'_\chi \left(\chi_1^* \overleftrightarrow{\partial_\mu} \chi_1 + \chi_2^* \overleftrightarrow{\partial_\mu} \chi_2 \right) B^{\prime \mu} \,.
\end{equation}
Since for the scalar doublet $Y'_\Phi = 0$, there is no $Z-Z'$ mixing and $B^\prime_\mu \equiv Z^{\prime}_\mu$ is a mass eigenstate, with mass 
\begin{align}
M_{Z'}^2 = (g_{Z'} Y'_{\chi})^2 \, u^2\,.
\label{eq:MZpdef}
\end{align}
We express the scalar weak eigenstates as $H'_i = O_{ij} H_j$, where $H_i$ are the mass eigenstates with mass $M_{H_i}$, and $O$ is a $4 \times 4$ real orthogonal matrix. The $Z' H_i H_j$ ($i < j$) couplings are then
\begin{equation}
\mathcal{L}_{Z' H_i H_j} = g_{Z'} Y'_{\chi} R_{ij} H_i \overleftrightarrow{\partial_\mu} H_j \, Z^{\prime \mu} \,,
\end{equation}
with mixing factors
\begin{equation}
R_{ij} = \cos \beta \left[ O_{4i} O_{3j} - O_{4j} O_{3i} \right] - \sin \beta  \left[ O_{4i} O_{2j} - O_{4j} O_{2i} \right] \,.
\label{ec:Rij}
\end{equation}
Notice that $R_{ij}$ are anti-symmetric and therefore $R_{ii} = 0$, reflecting the fact that $Z' \to H_i H_i$ is forbidden. Also, it can be shown that $\sum_{i<j} R_{ij}^2 = 1$ due to the orthogonality of the mixing matrix $O$.
The Lagrangian for the interaction of the SM $Z$ boson with two neutral scalars is
\begin{equation}
\mathcal{L} = - i \frac{g_W}{c_W}  \phi^{0\,*} \overleftrightarrow{\partial_\mu} \phi^0  \; Z^{\mu} 
= \frac{g_W}{c_W}  \rho_0 \overleftrightarrow{\partial_\mu} \eta_0  \; Z^{\mu} \,,
\end{equation}
with $g_W$ and $c_W$ being the weak coupling and the cosine of the weak angle, respectively. This term does not yield interactions among the $Z$ and two physical neutral scalars since $G_1^0 = \eta_0$. The three-scalar couplings in the mass eigenstate basis are complicated functions of the potential parameters, the VEVs and the mixing matrix $O$, but can generically be written as
\begin{equation}
\mathcal{L}_{3H} = - u \frac{\lambda_{ijk}}{S_{ijk}} H_i H_j H_k \,.
\end{equation}
The constants $\lambda_{ijk}$ are totally symmetric under interchange of two indices, and their expressions are collected in appendix~\ref{sec:b}. For convenience we introduce symmetry factors $S_{ijk}$, obeying $S_{ijk} = 1$ for three different indices, $S_{ijk} = 2 $ if two of them are equal, and $S_{ijk} = 6$ if $i=j=k$. The coupling of the scalar mass eigenstates to the SM fermions $f$ and weak bosons $V=W,Z$ arise from their $\rho_0$ component,
\begin{align}
& \mathcal{L}_{H_i ff} = - \frac{m_f }{v} O_{1i} \bar f f H_i \,, \notag \\
& \mathcal{L}_{H_i VV} = 2 \frac{M_W^2}{v}  O_{1i} W_\mu^- W^{+ \mu} H_i + \frac{M_Z^2}{v} O_{1i} Z_\mu Z^\mu H_i \,.
\end{align}
Notice that the interaction of $H_i$ to fermions is purely vectorial, even if they have a non-vanishing CP-odd component. This is so because the $\rho_0$ interaction with fermions is vectorial, and the matrix $O$ is real.

By inspection of the mass matrix \eqref{ec:M} one sees that it is rather easy to make our model compatible with experimental data. Taking $\lambda_{5,6,9}$ small, the mixing of SM-like Higgs boson $H \equiv H_1$ with the new singlets (given by $O_{i1}$, $i\neq 1$) can be made as small as desired, in particular fulfilling the current constraints~\cite{ATLAS:2019slw}. The masses and mixing of the additional scalars depend on $m_{12}$, $u$, $v$, $\lambda_{1-4}$, $\lambda_{7-9}$ and $\beta$. If $u$ is at the TeV scale, masses for the new scalars around the electroweak scale can naturally be obtained with small $\lambda_i$ couplings, without the need of fine-tuned cancellations. Mixing between the CP-even states $\rho_1$, $\rho_2$ and the CP-odd one $A^0$ is possible with complex $\lambda_{4-6}$ without affecting the properties of the SM-like Higgs boson. If these parameters are real, $A^0$ is itself a mass eigenstate and $H_{2,3}$ are CP-even and an admixture of $\rho_1$ and $\rho_2$. 

The framework described above naturally accommodates the $Z'$ cascade decays we are seeking for. The decay widths of the $Z'$ boson into SM quarks and scalars are
\begin{align}
& \Gamma(Z' \to q \bar q) = \frac{N_c g_{Z'}^2 z^2}{12 \pi} M_{Z'}
\left[ 1+2 \frac{m_q^2}{M_{Z'}^2} \right] \left[1-4 \frac{m_q^2}{M_{Z'}^2} \right]^{1/2} \,, \notag \\
& \Gamma(Z' \to H_i H_j) = \frac{g_{Z'}^2 Y_{\chi}^{\prime \, 2}}{48 \pi M_{Z'}^5} R_{ij}^2 \lambda^{3/2}(M_{Z'}^2,M_{H_i}^2,M_{H_j}^2) \,,
\end{align}
where $N_c = 3$ is the quark colour factor and $Y'_\chi = 3z \; (9z)$ in model 1 (model 2). We have defined the kinematical function
\begin{equation}
\lambda(x,y,z)=x^2+y^2+z^2-2xy-2xz-2yz \,.
\end{equation}
Neglecting the masses of the decay products, the total width of the $Z'$ boson into scalars is
\begin{equation}
\sum_{i<j}  \Gamma(Z' \to H_i H_j) = \frac{g_{Z'}^2 Y_{\chi}^{\prime \, 2}}{48 \pi } M_{Z'} \,,
\end{equation}
and the total branching ratio into scalars is of order 10\% (50\%) in model 1 (model 2). The mixing factor $R_{ij}$ can be of order unity for some pairs of scalars, in which case the partial width $Z' \to H_i H_j$ saturates the above sum. The decay widths of the scalars are
\begin{align}
& \Gamma(H_i \to f \bar f) = \frac{N_c}{8\pi}  \frac{m_f^2}{v^2} M_{H_i} O_{1i}^2  \left[ 1 - 4 \frac{m_f^2}{M_{H_i}^2} \right]^{3/2} \,, \notag \\
&  \Gamma(H_i \to W^+ W^- ) = \frac{1}{16\pi} \frac{M_{H_i}^3}{v^2} O_{1i}^2 \left[ 1 - 4 \frac{M_W^2}{M_{H_i}^2} \right]^{1/2} \left[ 1 - 4 \frac{M_W^2}{M_{H_i}^2} + 12 \frac{M_W^4}{M_{H_i}^4} \right] \,, \notag \\ 
&  \Gamma(H_i \to Z Z ) = \frac{1}{32\pi} \frac{M_{H_i}^3}{v^2} O_{1i}^2 \left[ 1 - 4 \frac{M_Z^2}{M_{H_i}^2} \right]^{1/2} \left[ 1 - 4 \frac{M_Z^2}{M_{H_i}^2} + 12 \frac{M_Z^4}{M_{H_i}^4} \right] \,, \notag \\ 
& \Gamma(H_i \to H_j H_k) = \frac{u^2 \lambda_{ijk}^2}{16 \pi M_{H_i}^3 (1 + \delta_{jk}) } 
 \lambda^{1/2}(M_{H_i}^2,M_{H_j}^2,M_{H_k}^2) \,.
 \label{ec:Hpw}
\end{align}
In the last equation, the symmetry factor $(1+\delta_{jk})$ accounts for the presence of two identical particles in the final state when $j=k$. Since $O_{1j} \ll 1$, the scalars will dominantly decay to lighter scalars, if kinematically allowed. Otherwise, they will decay into $W^+ W^-$, $ZZ$ or fermion pairs, like a Higgs boson with a mass $M_{H_i}$ (decays such as $H_i \to H_j Z$ are absent). For lighter masses the dominant mode will be $H_i \to b \bar b$.

\section{Methodology for the parameter space scan}
\label{sec:3a}

The number of parameters in the scalar potential (\ref{ec:VS}) is large enough to reproduce any pattern of scalar masses and mixing. Thus, we will focus on setting benchmark scenarios representative of the signals we are interested in. Within this approach, and with the goal of reducing the number of parameters, we will consider a simpler version of the model with $\lambda_{7-9} = 0$ in the scalar potential. This corresponds to having a softly broken $Z_2$ symmetry under which $\chi_2 \to - \chi_2$, i.e. the only term remaining in $V_{\not Z_2} = 0$ is the bilinear $m_{12}^2$ soft-breaking term. 
We are then left with twelve real parameters in the scalar mass matrix: $\RE (m_{12}^2)$, $\lambda_{0-3,5,6}$, $\RE (\lambda_4)$, $\IM (\lambda_4)$,  $\beta$, $v$ and $u$ (determined by the $Z^\prime$ mass through eq.~\eqref{eq:MZpdef}), of which only ten are independent due to the relations
\begin{align}
& M_{14} = 0 \,, \quad
\tan \beta = \frac{M_{24}}{M_{34}} \,.
\label{ec:rel}
\end{align}
These ten parameters match the four scalar masses and six independent parameters of the (real) $4 \times 4$ orthogonal scalar mixing matrix. The first equation in (\ref{ec:rel}) determines one of the masses $M_{H_i}$ through
\begin{align}
M_{14} =  \sum_{i=1}^4 O_{1i} O_{4i} M_{H_i}^2  = 0 \,,
\label{ec:re2}
\end{align}
while the second one determines $\tan \beta$. Taking $O_{ij}$, $v$, $M_{Z^\prime}$, $\lambda_2$ and three of the masses $M_{H_i}$ as inputs, the remaining parameters in the potential can be determined as
\begin{align}
& \RE (m_{12}^2) = ( u^2 \lambda_2 \sin \beta - M_{33} )\tan \beta  \,, \notag \\
& \lambda_0 = \frac{M_{11}}{v^2} \,,  \lambda_1 = \frac{\RE (m_{12}^2) \tan \beta + M_{22}}{u^2} \,, \quad 
 \lambda_3 = - \RE (\lambda_4) - 2\,\frac{M_{23} - \RE (m_{12}^2)}{u^2} \,, \notag \\
& \RE (\lambda_4) =  \frac{M_{44}}{u^2} - \frac{\RE (m_{12}^2)}{u^2 \sin (2 \beta)} \,, \quad
 \IM (\lambda_4) = - \frac{M_{24}}{u^2 \sin \beta} \,,
  \quad \lambda_5 = \frac{2 M_{12}}{v \, u \cos \beta} \,, \quad
 \lambda_6 = \frac{2 M_{13}}{v \, u \sin \beta} \,,
 \label{rec:lamb}
 \end{align}
where $M_{ij}$ are expressed in terms of $M_{H_i}$ and $O_{ij}$ using
\begin{equation}
M_{ij} = \sum_{k=1}^4 O_{ik} O_{jk} M_{H_k}^2 \,.
\label{ec:Mij}
\end{equation}
The $4 \times 4$ mixing matrix is paremeterised by the product of $2 \times 2$ rotations as
\begin{equation}
O = \widehat{O}_{34} \widehat{O}_{24} \widehat{O}_{14} \widehat{O}_{23} \widehat{O}_{13} \widehat{O}_{12} \,,
\end{equation}
being $\widehat{O}_{kl}$ a rotation in the $(k,l)$ plane by an angle $\theta_{kl}$, whose $(i,j)$ matrix element can be written as
\begin{equation}
(\widehat{O}_{kl})_{ij} = \delta_{ij} +  (\delta_{ik} \delta_{jk} + \delta_{il}\delta_{jl} ) (\cos \theta_{kl} -1) + 
 (\delta_{ik} \delta_{jl} - \delta_{il}\delta_{jk} ) \sin \theta_{kl} \,.
\end{equation}
The constraints on the couplings of the SM 125 GeV Higgs boson ($H \equiv H_1$ in our models) obtained by the ATLAS Collaboration~\cite{ATLAS:2019slw} imply $O_{21}^2 + O_{31}^2 + O_{41}^2 \leq 0.05$ at 95\% confidence level (CL),\footnote{The strength parameter $\mu = 1.11^{+0.09}_{-0.08}$ corresponds to $O_{11}^2$ in our notation, from which limits on the other matrix elements can be obtained using unitarity and approximating the Gaussian distribution by a symmetric one. Ignoring the fact that $O_{11}^2 \leq 1$ due to unitarity, one arrives at the bound $O_{21}^2 + O_{31}^2 + O_{41}^2 \leq 0.05$ at 95\% CL. By restricting ourselves to the physical region $O_{11}^2 \leq 1$, the 95\% CL limit is relaxed to 0.11. In all our benchmark scenarios the matrix elements $O_{1j}$ and $O_{j1}$ ($j \neq 1$) are more than two orders of magnitude below these bounds. }
 in which case $O_{1j}$ and $O_{j1}$ are small for $j\neq 1$. This, in turn, implies that the mixing angles $\theta_{12}$, $\theta_{13}$ and $\theta_{14}$ are small. 

The minimisation of the scalar potential and the viability analysis of a given vacuum configuration with $v,u \neq 0$ proceeds as follows. Setting the mass of the SM Higgs boson to $M_{H_1}\equiv M_H = 125$~GeV, for a given set of input parameters $O_{ij}$, $v$, $M_{Z^\prime}$, $\lambda_2$ and $M_{H_{3,4}}$ we first determine $M_{H_2}$ and $\tan\beta$ using eqs.~\eqref{ec:rel}. Afterwards, the parameters in \eqref{rec:lamb} are computed using also eq.~\eqref{ec:Mij}. At this point, for the chosen set of inputs, the potential is completely defined. It now remains to check whether stability and perturbative unitarity criteria are fulfilled, and ensure that our VEV corresponds to the global minimum of the potential. For the stability analysis we follow the method of refs.~\cite{Kannike:2012pe} and \cite{Kannike:2016fmd} based on requiring copositivity of the quartic coupling matrix $\Lambda$. Parameterising the field bilinears as $|\Phi|^2\equiv h^2$,  $|\chi_{1,2}|^2\equiv h_{1,2}^2$ and $\chi_1^\ast \chi_2\equiv \rho h_1 h_2 e^{i\varphi}$ (with $|\rho|\in [0,1]$), and taking $\varphi=0$, we have
\begin{equation}
\Lambda = \left( \! \begin{array}{cccc}
2\lambda_0 & \lambda_2 & \lambda_6  \\
\lambda_5 &2 \lambda_1 &  2\rho^2 \RE(\lambda_4)+2\lambda_3 \\
\lambda_6 &  2\rho^2 \RE(\lambda_4)+2\lambda_3 & 2\lambda_2 
\end{array} \! \right) 
\,,
\label{ec:Lambda}
\end{equation}
defined in the basis $(h^2,h_1^2,h_2^2)$. The stability of the potential is ensured if the above matrix is copositive, i.e., if the following conditions hold~\cite{Kannike:2012pe}:
\begin{align}
&\Lambda_{ii} \geq 0\,, \notag \\
&\Lambda^\prime_{ij}=\Lambda_{ij} + \sqrt{\Lambda_{ii} \Lambda_{jj}}\geq 0 \quad (i<j=1,2,3)\,, \notag \\
&\sqrt{\Lambda_{11}\Lambda_{22}\Lambda_{33}}+\Lambda_{12}\sqrt{\Lambda_{33}}+\Lambda_{13}\sqrt{\Lambda_{22}}+
+\Lambda_{23}\sqrt{\Lambda_{11}}+\sqrt{2\Lambda^\prime_{12}\Lambda^\prime_{23}\Lambda^\prime_{13}}\geq 0\,.
\label{ec:copos}
\end{align}
In the above relations $\rho=0,1$ depending on whether $\RE(\lambda_4)>0$ or $\RE(\lambda_4)<0$, respectively. Since in the cases we are interested in the quartic couplings $\lambda_i$ are typically very small (due to the fact that $M_{H_i}/u \ll 1$), perturbative unitarity is automatically ensured. Thus, we do not perform the complete analysis of $S$-matrix unitarity for elastic scattering of two scalar boson states. 

It now remains to check whether our minimum is the global minimum of the potential. For that, we must compare the value of the potential at our minimum,
%
%
%
%
\begin{align}
V_0=-\frac{1}{8} \left[\lambda_1u_1^4+\lambda_2u_2^4+2\lambda_3u_1^2u_2^2+2u_1^2 u_2^2\RE(\lambda_4)+v^2(\lambda_0 v^2 + \lambda_5 u_1^2 + \lambda_6 u_2^2)\,\right]\,,
\label{ec:copos1}
\end{align}
with the values at any other minima obeying the minimisation conditions \eqref{mincond}. The most straightforward alternative solutions correspond to vacua with vanishing VEVs. Namely, we have
\begin{align}
&u_1=u_2=v=0\;\rightarrow \;V_1=0\nonumber\\
&u_1=u_2=0\,,\,v^2=-\frac{2m_{0}^2}{\lambda_0}\;\rightarrow \; V_2=-\frac{m_{0}^4}{2\lambda_0}\,,
\label{ec:copos2}
\end{align}
where $V_i$ corresponds to the value of the potential at the corresponding set of VEVs. Notice that these two solutions must be discarded as being the global minimum of the potential since they imply no spontaneous symmetry breaking of the SM and/or the U(1)$^\prime$ gauge symmetry. A class of nontrivial minima with $\varphi \neq 0$ and $v,u_{1,2}\neq 0$ may exist. Since for these cases the analytical treatment is quite involved, we use a numerical routine to spot those solutions. For all alternative minima we check positivity of the scalar masses and if $V_i < V_0$. At the end, only those sets of input parameters which lead to a stable potential and to a global minimum are considered viable in the parameter space scan performed in the next section.

\section{Stealth boson benchmarks}
\label{sec:3}

We are interested in scenarios with $Z'$ and $H_i$ masses close to those used for the anti-QCD tagger in ref.~\cite{Aguilar-Saavedra:2017rzt}. We remark that this assumption is done only for the sake of simplicity, and with the purpose of using the performance for signals and backgrounds obtained in previous work without the need of training neural networks for new taggers.  Thus, we restrict our study to benchmarks where the $Z'$ boson decays into two stealth bosons, that is for example
\begin{equation}
Z' \to H_3 H_4 \,, \quad H_{3,4} \to H_2 H_2 \,,
\end{equation}
with $H_2$ subsequently decaying into quark pairs. Scenarios with $Z' \to H_{3,4} H_2$, $H_{3,4} \to H_2 H_2$ are also interesting and lead to signals that are quite elusive as well, since a light boosted $H_2 \to q \bar q$ produces two-pronged jets that closely resemble one-pronged QCD jets. However, their analysis requires the development of new taggers, which is out of the scope of this work.

In the following we will identify three representative scenarios. In scenario 1 with relatively light scalars the decay pattern is quite simple, with dominant decays $H_{3,4} \to H_2 H_2$. For heavier scalars $H_{3,4}$, their decays into $WW$, $ZZ$, $t \bar t$ and $H_1 H_1$ are possible, besides $H_2 H_2$, if the latter is kinematically open. For illustration, we set scenario 2 where decays $H_{3,4} \to H_2 H_2$ dominate, and scenario 3 where $H_{3,4} \to WW, ZZ$ dominate. Notice that these are extreme cases and, in general, for $H_{3,4}$ one could have similar branching ratios for $WW/ZZ$ and $H_2 H_2$ final states. One of the virtues of a generic tagger is that it is sensitive to all of them at once. For simplicity, the extra fermions (quarks in model 1 and leptons in model 2) are assumed heavy enough not to be produced in the decays of the $Z'$ boson.

\subsection{Scenario 1}

\begin{figure}[t]
\begin{center}
\begin{tabular}{cc}
\includegraphics[height=6cm,clip=]{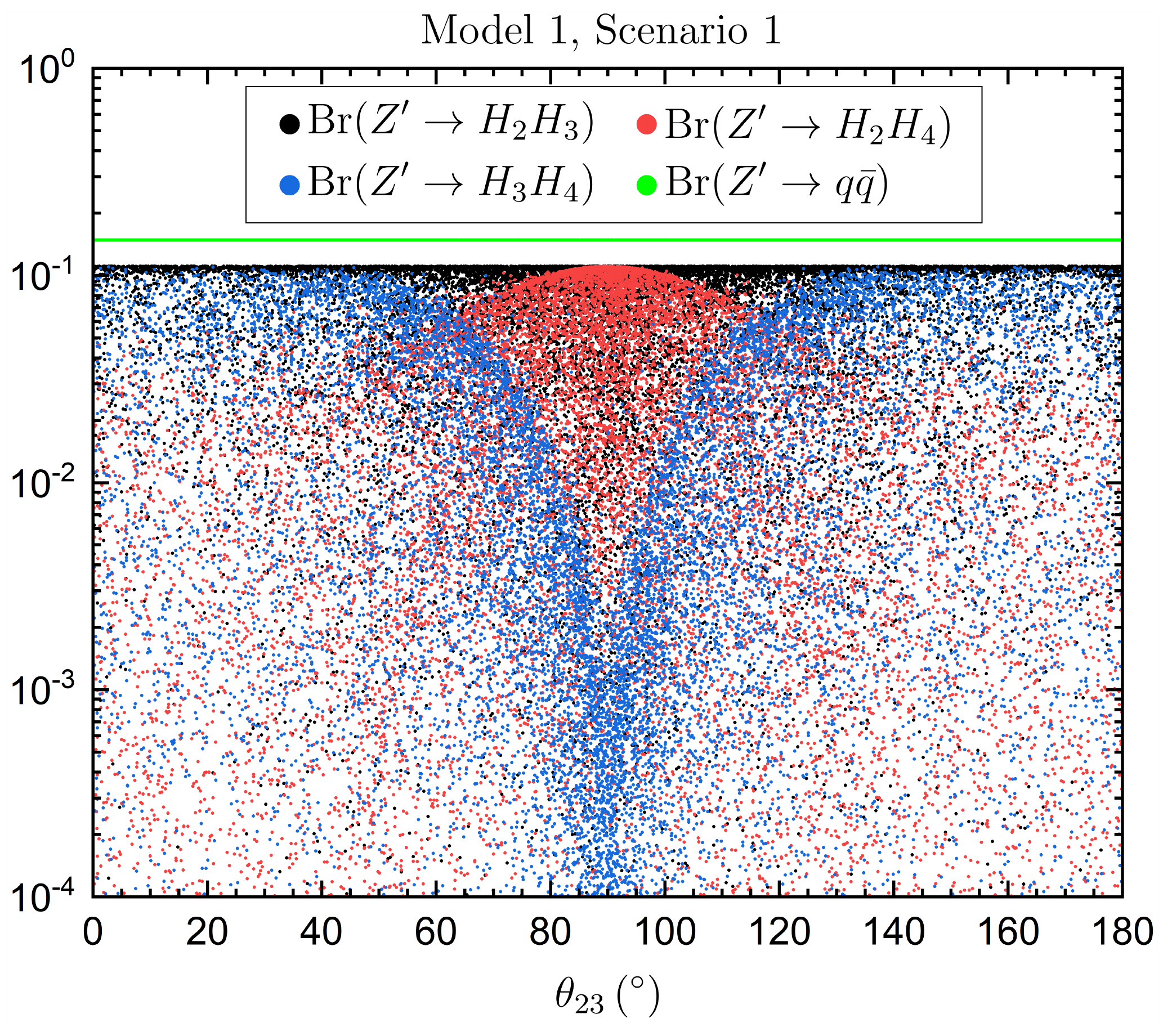} & \includegraphics[height=6cm,clip=]{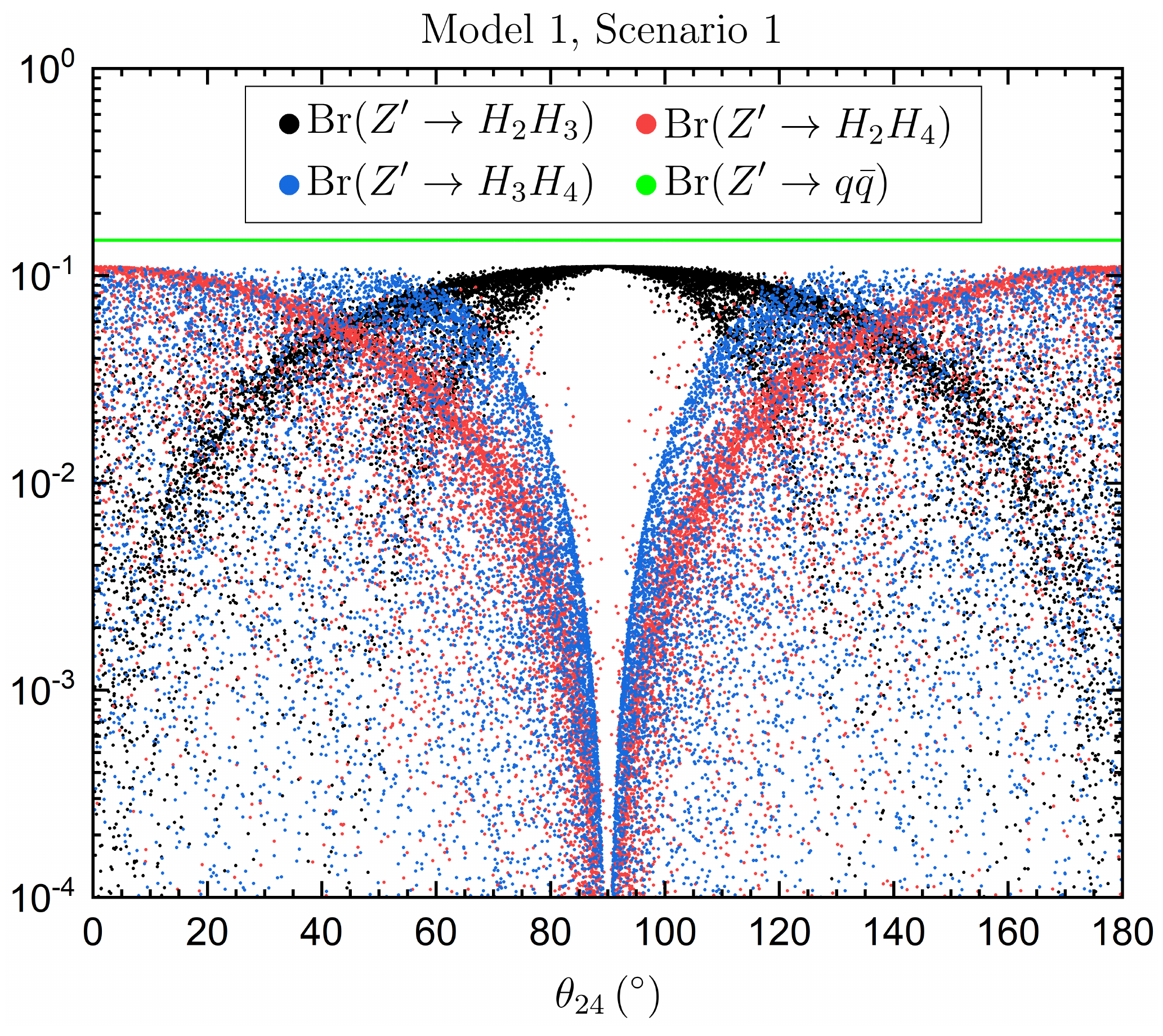} \\ \includegraphics[height=6cm,clip=]{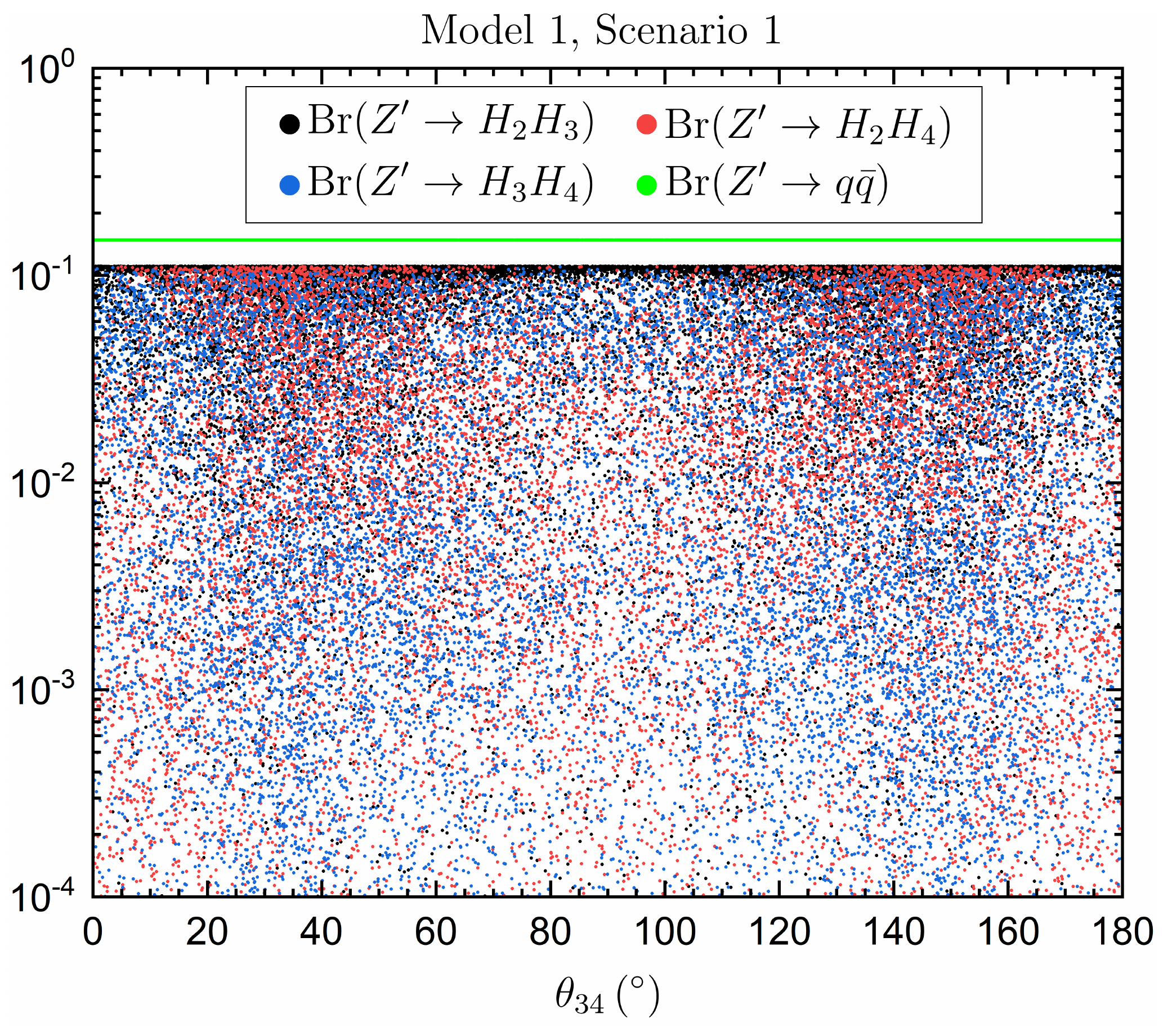} & \includegraphics[height=6cm,clip=]{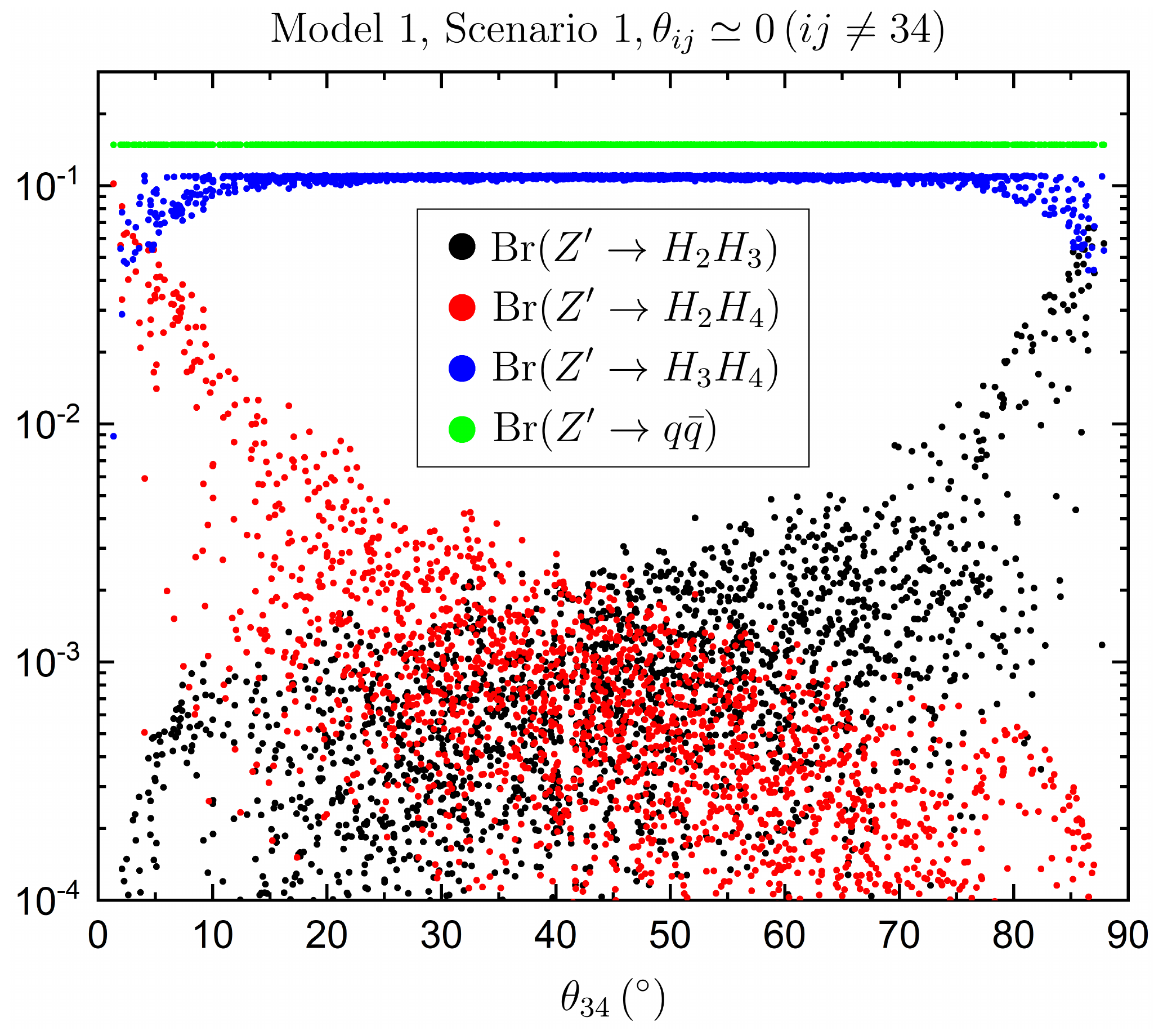} 
\end{tabular}
\caption{$Z'$ branching ratio to different scalar pairs in scenario 1 of model 1. We scan over the input parameters $\theta_{ij}$ and $\lambda_2$ keeping only those points which lead to a viable minimum of the potential. The mixing angles $\theta_{23}$, $\theta_{24}$ and $\theta_{34}$ are unrestricted, except in the bottom right panel where $\theta_{23}$ and $\theta_{24}$ are taken small (see the text). The U(1)$^\prime$ coupling is such that $g_{Z'} z = 0.1$.}
\label{fig:scan1Z}
\end{center}
\end{figure}

We choose $M_{Z'} = 2.2$ TeV, $M_{H_3} \simeq M_{H_4} \simeq 80$ GeV, $M_{H_2} \simeq 30$ GeV, as in one of the benchmark points used in the tagger labelled as {\tt std1000} in ref.~\cite{Aguilar-Saavedra:2017rzt}. The coupling is set to $g_{Z'} z = 0.1$.\footnote{The results for decay branching ratios are quite independent of the actual value used for the coupling, which determines $u$ for fixed $Z'$ mass and sets the scale for the $\lambda_{1-6}$ couplings.} We perform a scan of the allowed parameter space by varying $\theta_{23}$, $\theta_{24}$ and $\theta_{34}$ with a flat distribution (keeping the other mixing angles small as required by constraints on the couplings of the SM Higgs) and compute the $Z'$ branching ratio to scalars. The results for model 1 are presented in figure~\ref{fig:scan1Z}. The branching ratio for quark pairs (not summed over flavours) is included for comparison. For model 2 the branching ratios for scalars trivially scale by a factor of $4.8$, and the branching ratios to quark pairs by $0.53$. 

The results show that the decay $Z' \to H_3 H_4$ can be dominant in wide regions of the parameter space, as long as $\theta_{23}$ and $\theta_{24}$ are not close to $\pi/2$. In particular, if these two angles are small, the mixing matrix is approximately
\begin{equation}
O \simeq \left( \! \begin{array}{cccc}
1 & \varepsilon_{12} & \varepsilon_{13} & \varepsilon_{14} \\
\varepsilon_{21} & 1 &  \varepsilon_{23} &  \varepsilon_{24} \\
\varepsilon_{31} &  \varepsilon_{32} & \cos \theta_{34} & \sin \theta_{34} \\
\varepsilon_{41} &  \varepsilon_{42} & -\sin \theta_{34} & \cos \theta_{34}
\end{array} \! \right) \,,
\end{equation}
with $\varepsilon_{ij} \lesssim 0.01$. Neglecting these small parameters, the $Z'$ couplings to the mass eigenstates are
\begin{equation}
\mathcal{L} = g_{Z'} Y'_{\chi} \left[ - \sin \beta \sin \theta_{34} H_2 \overleftrightarrow{\partial_\mu} H_3 
+  \sin \beta \cos \theta_{34} H_2 \overleftrightarrow{\partial_\mu} H_4
 - \cos \beta H_3 \overleftrightarrow{\partial_\mu} H_4 \right]
 B^{\prime \mu} \,,
\end{equation}
with $\sin \beta \ll \cos \beta$, from which it can be clearly seen that the dominant $Z'$ decay to scalars is $Z' \to H_3 H_4$. This is seen in the bottom right panel of figure~\ref{fig:scan1Z}, where we restrict $|\theta_{23,24}| \leq 0.01$. 

\begin{figure}[t]
\begin{center}
\begin{tabular}{cc}
\includegraphics[height=6cm,clip=]{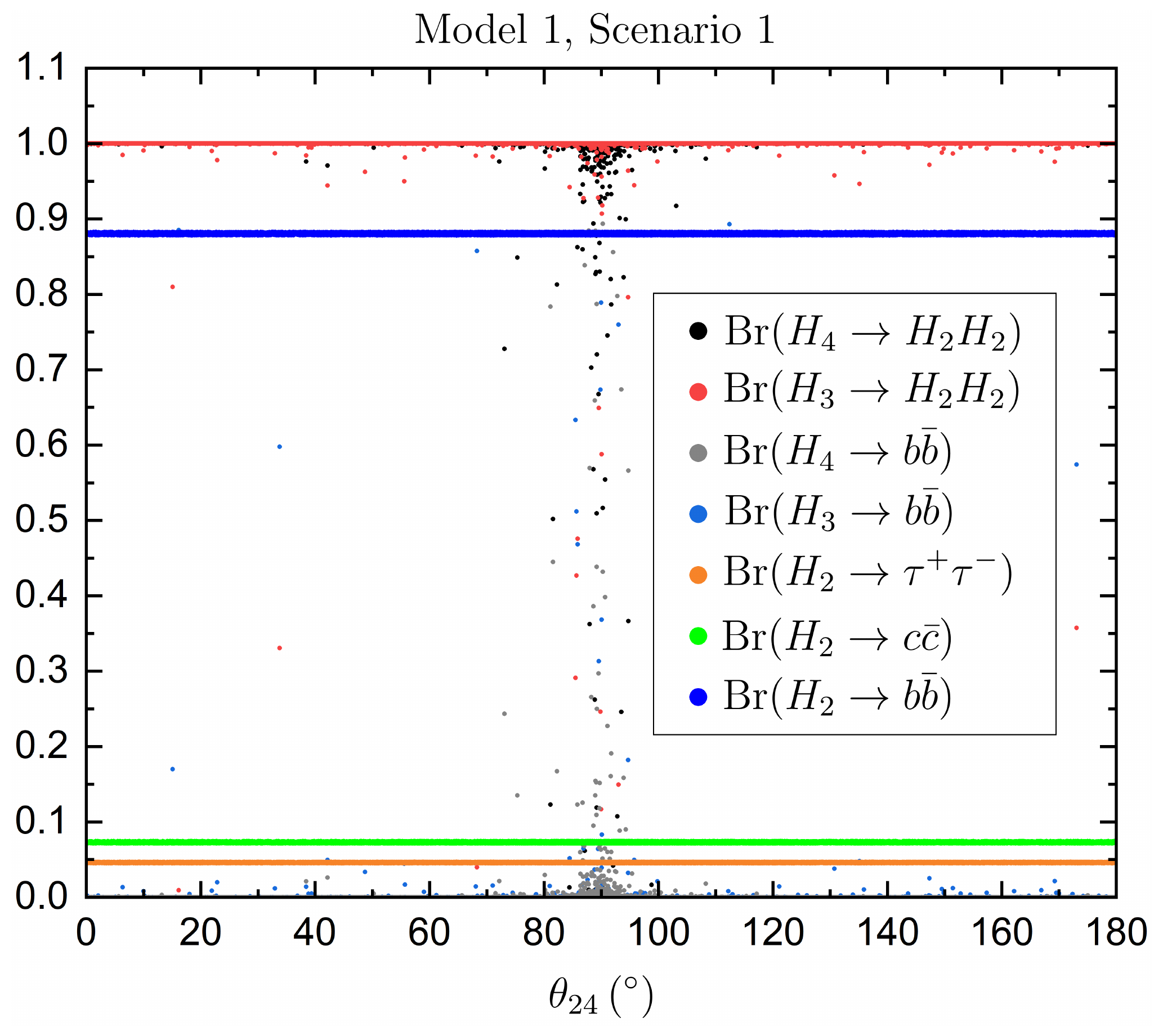} & \includegraphics[height=6cm,clip=]{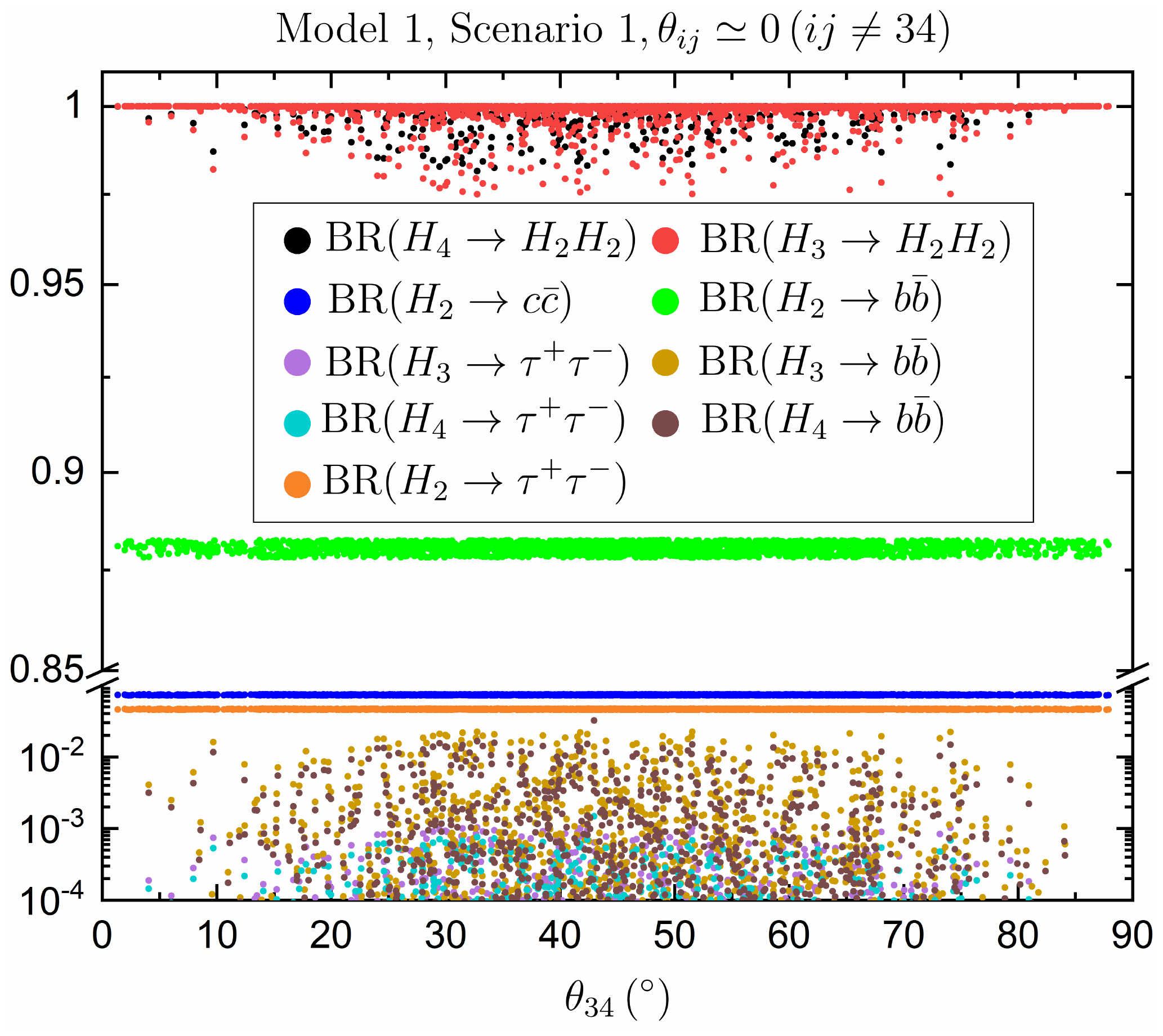} 
\end{tabular}
\caption{Branching ratio for the new scalars into different final states for scenario 1 of model 1, resulting from a scan of allowed points in parameter space. In the left panel $\theta_{23,24}$ vary in the interval $[0,\pi]$ while in the right panel $|\theta_{23,24}| \leq 0.01$. For completeness we include the branching ratios of $H_2$, which do not depend on the mixing angles.}
\label{fig:scan1H}
\end{center}
\end{figure}

For the unrestricted scan (with $\theta_{23}$, $\theta_{24}$ and $\theta_{34}$ free), $\text{Br}(H_{3,4} \to H_2 H_2) \simeq 1$ in most of the parameter space of model 1, as can be seen in the left panel of figure~\ref{fig:scan1H} (for model 2 the results are similar). This is expected in the sense that these decays are controlled by the couplings $\lambda_{223}$ and $\lambda_{224}$, respectively, which are not suppressed. The small mixing with $\rho_0$ allows the decay of $H_2$ into quarks, but has little effect on the decay of the heavier particles.\footnote{The decay of $H_2$ does not produce displaced vertices even with this small mixing. For example, for $M_{H_2} = 30$ GeV and $O_{12} = 0.01$, the decay length is of 1.5 nm.} The same holds when $|\theta_{23,24}| \leq 0.01$, as shown in the right panel of the same figure.

\subsection{Scenario 2}

The masses are set to $M_{Z'} = 3.3$ TeV, $M_{H_3} \simeq M_{H_4} \simeq 400$ GeV and $M_{H_2} \simeq 80$ GeV as in the tagger benchmark labelled as {\tt std1500} in ref.~\cite{Aguilar-Saavedra:2017rzt}. The U(1)$^\prime$ coupling is set to $g_{Z'} z = 0.2$. For the $Z'$ branching ratios the results obtained from the parameter space scan are the same as in scenario 1, and are omitted for brevity. This is so because the scalars are much lighter than the $Z'$ boson, and kinematical effects are unimportant. The decay $Z' \to H_3 H_4$ can dominate the scalar decays of the $Z'$ boson, in particular when $\theta_{23}$ and $\theta_{24}$ are small. 

\begin{figure}[t]
\begin{center}
\begin{tabular}{cc}
\includegraphics[height=6.5cm,clip=]{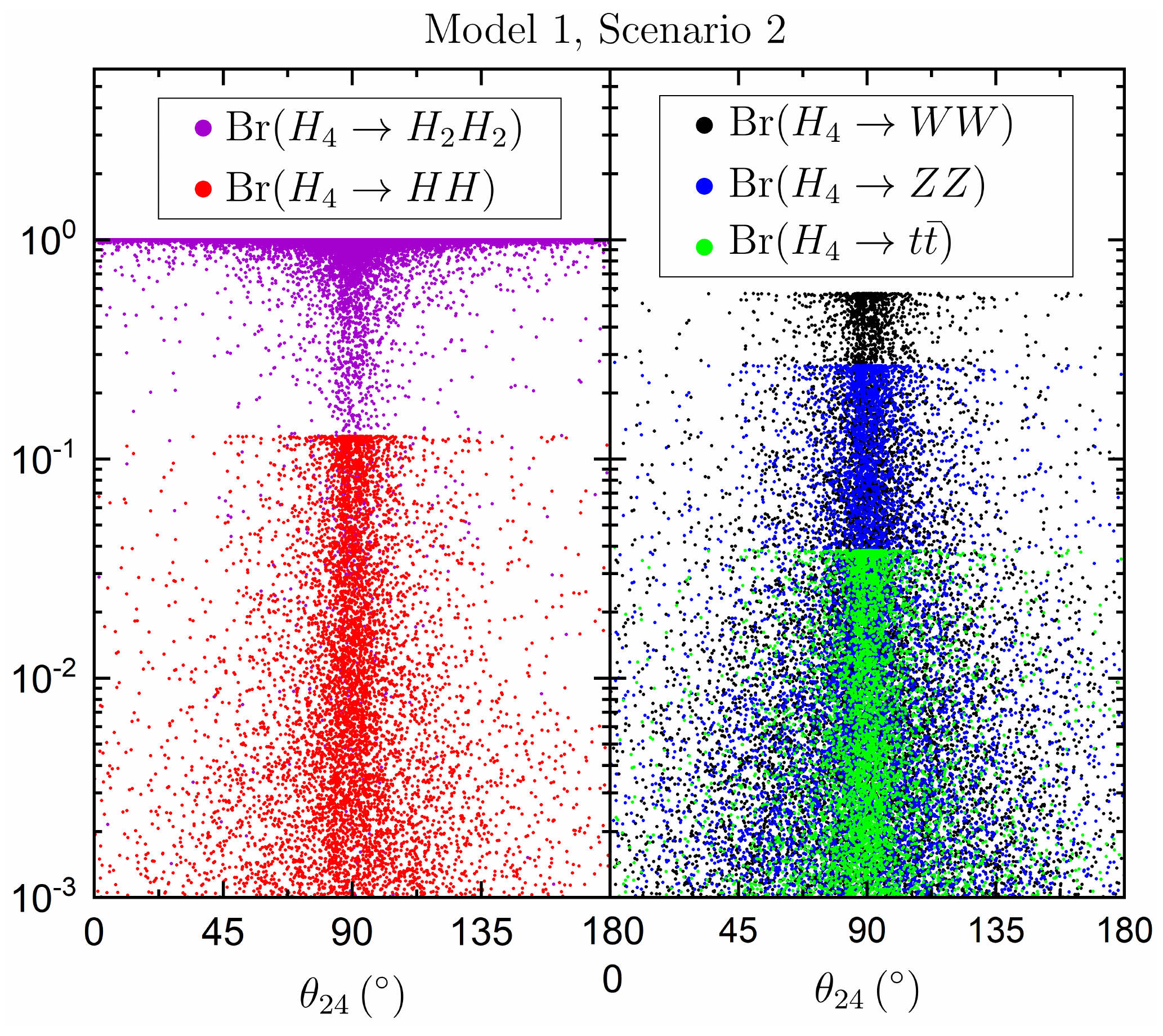} & \includegraphics[height=6.5cm,clip=]{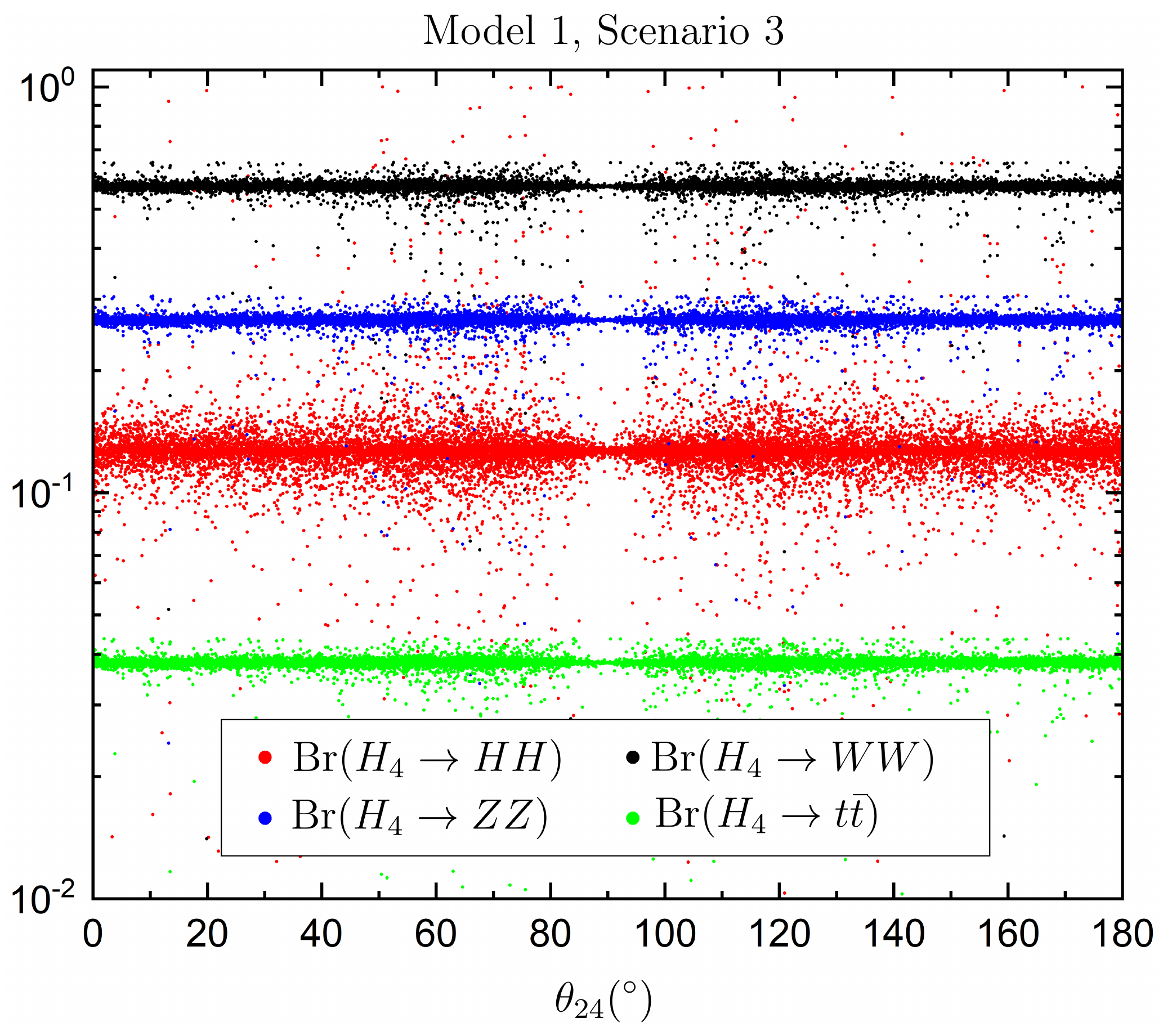} 
\end{tabular}
\caption{Left: Branching ratio for $H_4$ decays into different final states in scenario 2 of model 1, resulting from a scan of allowed points in parameter space. Right: the same as in the left panel but for scenario 3 of model 1.}
\label{fig:scan2H}
\end{center}
\end{figure}

The results for the decay of the heaviest scalar $H_4$ are shown in figure~\ref{fig:scan2H} (left panel). For $H_3$, which has nearly the same mass, the outcome is similar. In most of the parameter space $\text{Br}(H_{3,4} \to H_2 H_2) \simeq 1$, and the same happens for model 2.

\subsection{Scenario 3}

The masses are set to $M_{Z'} = 3.3$ TeV, $M_{H_3} \simeq M_{H_4} \simeq 400$ GeV as in the tagger becnhmark labelled as {\tt std1500} in ref.~\cite{Aguilar-Saavedra:2017rzt} but, in contrast with scenario 2,  $M_{H_2} \simeq 300$ GeV in order to forbid the decay $H_{3,4} \to H_2 H_2$. The coupling is set to $g_{Z'} z = 0.2$. The results of the parameter space scan are the same as in scenarios 1 and 2 for the $Z'$ branching ratios and, thus, they are not presented. Regarding $H_4$ decays, the results are shown in the right panel of figure~\ref{fig:scan2H} (for $H_3$, with nearly the same mass, the outcome is similar). Both scalars decay into pairs of SM particles with branching ratios that are nearly independent of the mixing angles. The partial widths are determined by the matrix element $O_{1i}$ and the small triple couplings $\lambda_{11i}$ ($i=3,4$), as seen from eqs.~(\ref{ec:Hpw}).

\section{Stealth boson signals}
\label{sec:4}

The potential relevance of stealth boson signals as a discovery channel is assessed in this section by a comparative study of the sensitivity of three searches,
\begin{itemize}
\item $Z' \to jj$.
\item $Z' \to t \bar t$.
\item A generic search, using the efficiencies for signals and background previously obtained for the anti-QCD tagger.
\end{itemize}
In addition, for scenario 1 we investigate whether the decay $Z' \to H_3 H_4$ is visible in a diboson resonance search. The various processes in our analysis are generated using {\scshape MadGraph5}~\cite{Alwall:2014hca}, followed by hadronisation and parton showering with {\scshape Pythia~8}~\cite{Sjostrand:2007gs} and detector simulation using {\scshape Delphes 3.4}~\cite{deFavereau:2013fsa} using the configuration for the CMS detector. The reconstruction of jets and their substructure analysis is done using {\scshape FastJet}~\cite{Cacciari:2011ma}. For the signal processes the relevant Lagrangian is implemented in {\scshape Feynrules}~\cite{Alloul:2013bka} and interfaced to {\scshape MadGraph5} using the universal Feynrules output~\cite{Degrande:2011ua}. As background processes we consider QCD dijet production $pp \to jj$, with $j$ being a light jet, $pp \to b \bar b$, and $t \bar t$ production. In order to populate with sufficient Monte Carlo statistics the entire mass and transverse momentum range under consideration, we split the samples in 100 GeV slices in the transverse momentum of the leading jet (or top quark), from 300 GeV to 2.2 TeV and above, generating $2 \times 10^5$ events for $jj$, $10^5$ events for $b \bar b$ and $10^5$ events for $t \bar t$ in each slice. The different samples are then recombined with weights proportional to the cross sections. Signal samples for $Z' \to jj$ (including $b \bar b$), $Z' \to t \bar t$ and $Z' \to H_3 H_4$ have $10^5$ events each, except for $Z' \to t \bar t$ in scenarios 2 and 3 and $Z' \to H_3 H_4$ in scenario 3, which have  $2 \times 10^5$ events.

The dijet and $t \bar t$ analyses are common to the two signal benchmark scenarios studied. We do not recast any specific experimental search but we choose event selections similar to the ones commonly adopted by the ATLAS and CMS Collaborations:
\begin{itemize}
\item Dijet resonance analysis: jets are reconstructed with the anti-$k_T$ algorithm~\cite{Cacciari:2008gp} using a radius  $R=0.8$, and groomed using Soft Drop~\cite{Larkoski:2014wba} with the parameters $z_\text{cut} = 0.05$ and $\beta = 0$. The use of large-radius jets is motivated by the possible presence of hard radiation accompanying the energetic decay products of a heavy resonance, and the grooming is implemented in order to clean the jets from pile-up and initial state radiation (see for example ref.~\cite{CMS:2018wxx}). The leading and subleading jets are required to have pseudo-rapidity $|\eta| \leq 2.5$ and transverse momentum $p_T \geq 500$ GeV, while their pseudo-rapidity difference must satisfy $|\Delta \eta| \leq 1.1$. 
\item $t \bar t$ resonance analysis: we use large-radius jets with $R=0.8$ reconstructed and groomed as in the dijet analysis. The leading and subleading jets are required to have pseudo-rapidity $|\eta| \leq 2.5$, $|\Delta \eta| \leq 1.1$ and transverse momentum $p_T \geq 500$ GeV. For $b$ tagging, a collection of `track jets' of radius $R=0.2$, reconstructed with tracks only, is used. A large-$R$ jet is considered as $b$-tagged if a $b$-tagged track jet (using the 70\% efficiency working point) within $\Delta R = 0.2$ of its centre is found. The large dijet background is reduced by requiring that both leading and sub-leading jets are $b$-tagged, have a (groomed) mass $100 \leq m_J \leq 220$ GeV, and a value of the (ungroomed) subjettiness variable~\cite{Thaler:2010tr} $\tau_{32} \leq 0.7$.

\end{itemize}

\subsection{Scenario 1}

In this scenario we have $M_{Z'} = 2.2$ TeV and we focus on the decay $Z' \to H_3 H_4$, with $M_{H_3} \simeq M_{H_4} \simeq 80$ GeV. For the signal coupling we choose $g_{Z'} z = 0.15$, yielding a total $Z'$ production cross section of 142.6 fb.\footnote{Larger couplings are compatible with current bounds from dijet, diboson and $t \bar t$ resonance searches. We prefer to select for our benchmarks in this section small values of the couplings, still yielding a significance around $5\sigma$ for the generic searches.}  The total $Z'$ width is $\Gamma = 26.5$ GeV (model 1) and $\Gamma = 50.2$ GeV (model 2). The small $Z'$ width, compared to the experimental resolution, justifies using the same signal samples for both models. We set the mixing factor $R_{34}$ in eq.~(\ref{ec:Rij}) to unity for simplicity --- as seen in the previous section for the numerical examples provided, $R_{34}$ is often very close to one. Therefore, we obtain for the branching ratios
\begin{align}
& \text{Br}(Z' \to H_3 H_4) = 0.11 \quad \quad \text{(model 1)} \,, \notag \\
& \text{Br}(Z' \to H_3 H_4) = 0.53 \quad \quad \text{(model 2)} \,,
\end{align}
and $\text{Br}(H_{3,4} \to H_2 H_2)\simeq 1$. ($Z'$ decays to other scalar pairs are suppressed.) We select $M_{H_2} = 30$ GeV as one of the benchmark points studied in ref.~\cite{Aguilar-Saavedra:2017rzt}. With $M_{H_2} = 15$ GeV, the tagger efficiency is quite close but the acceptance of the stealth boson signal in usual diboson resonance searches is slightly larger. That scenario is examined in detail in appendix~\ref{sec:c}. The branching ratios for the decay of the lighter scalar into quark pairs are
\begin{align}
& \text{Br}(H_2 \to b \bar b) = 0.88 \,, \notag \\
& \text{Br}(H_2 \to c \bar c) = 0.07 \,.
\end{align}
It is expected that the tagger performance for $b \bar b b \bar b$, $b \bar b c \bar c$ and $c \bar c c \bar c$ multi-pronged jets is similar, so we include both channels. With four scalars $H_2$ from the $Z'$ cascade decay, the branching ratio factor is $\text{Br}(H_2 \to b \bar b,c\bar c)^4 = 0.824$.  

\begin{figure}[p]
\begin{center}
\begin{tabular}{cc}
\includegraphics[height=5.9cm,clip=]{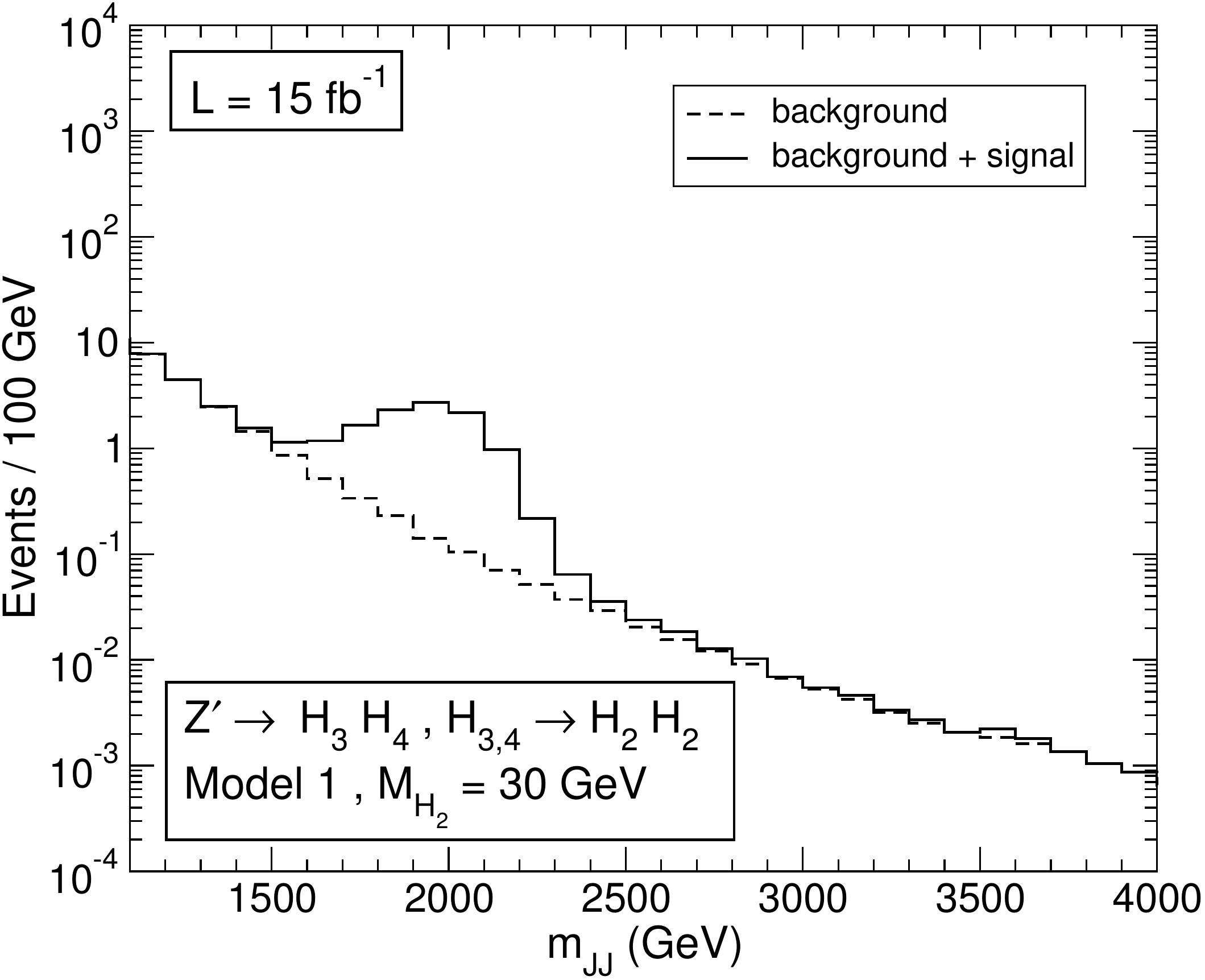} & 
\includegraphics[height=5.9cm,clip=]{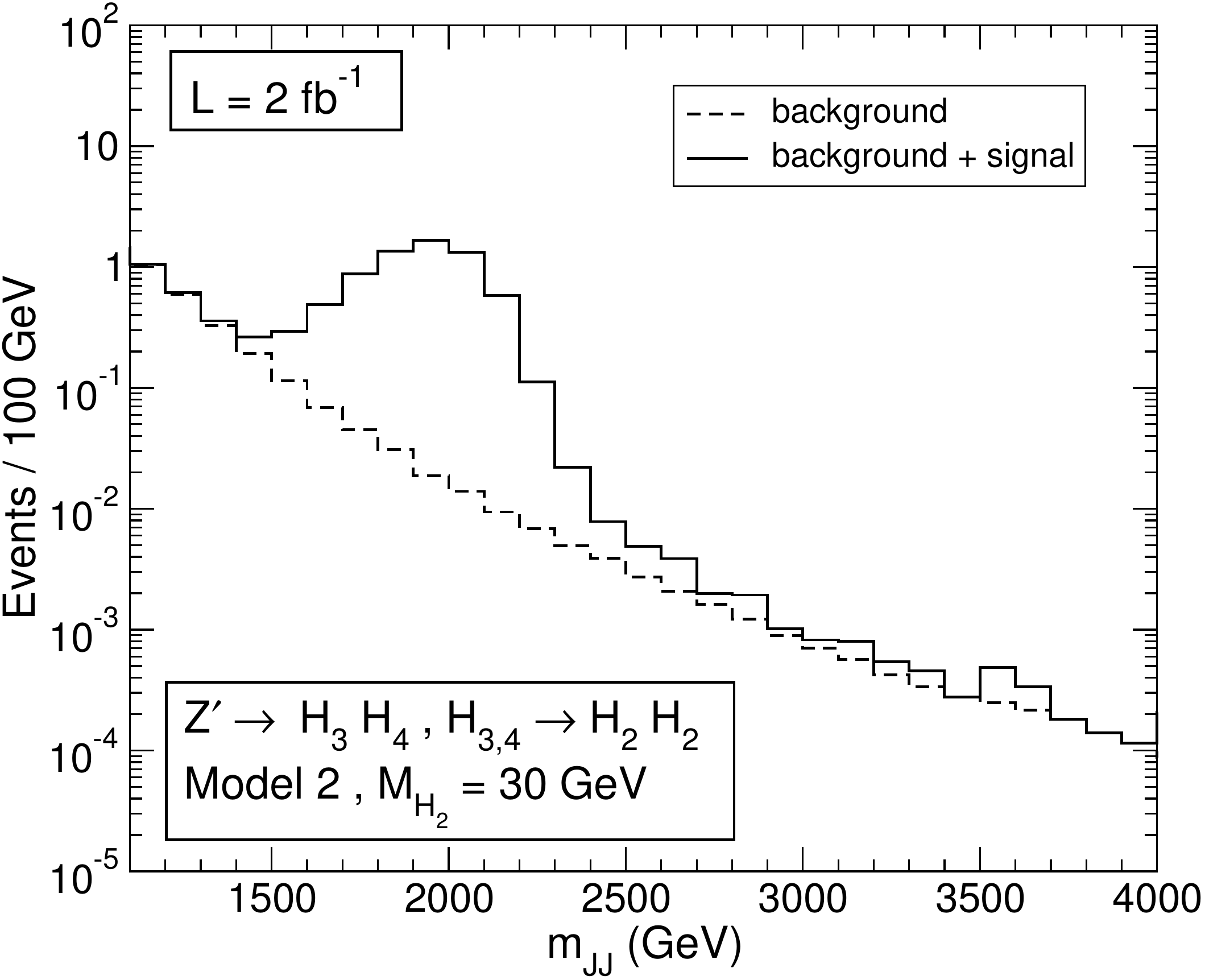} \\
\includegraphics[height=5.9cm,clip=]{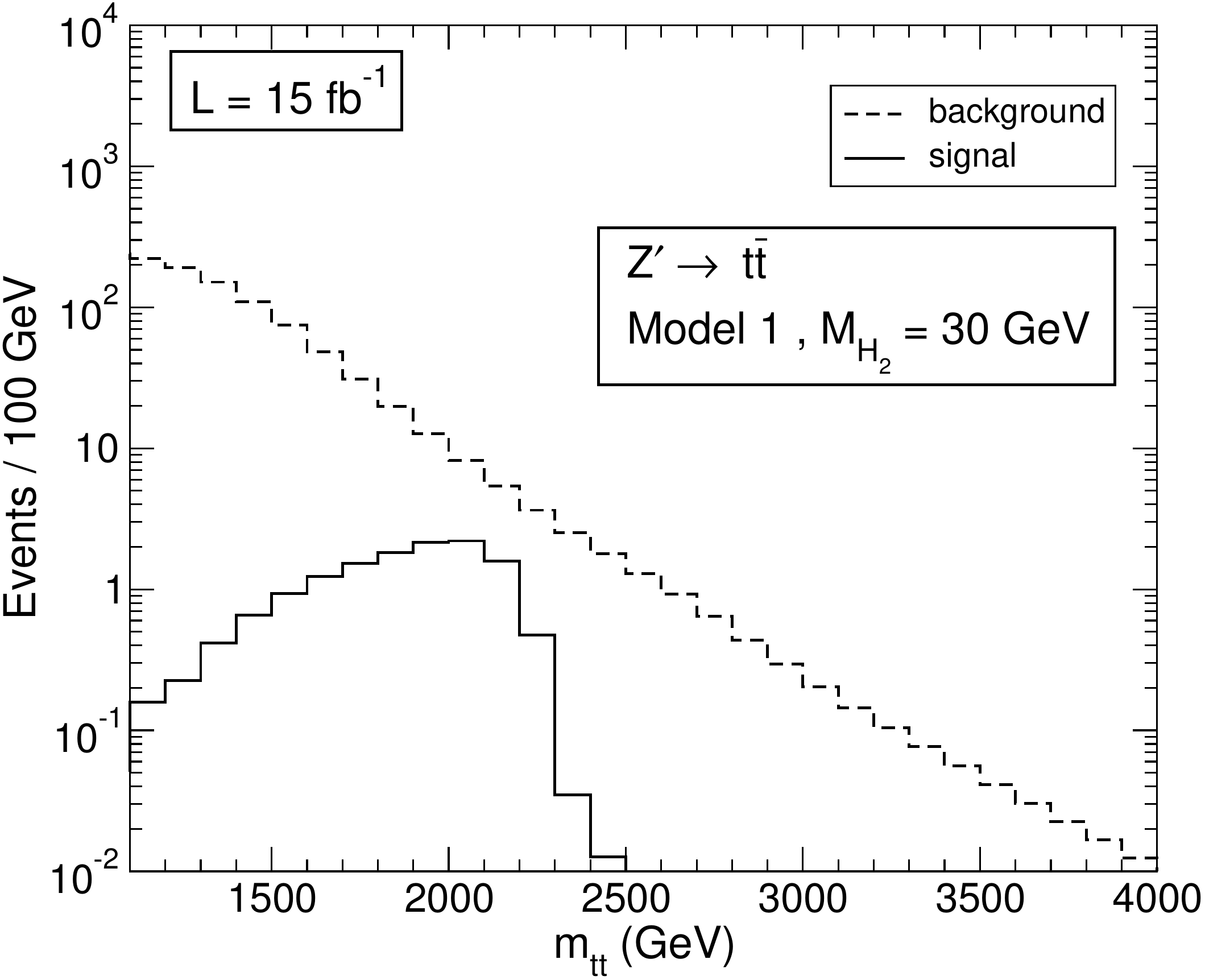}  &
\includegraphics[height=5.9cm,clip=]{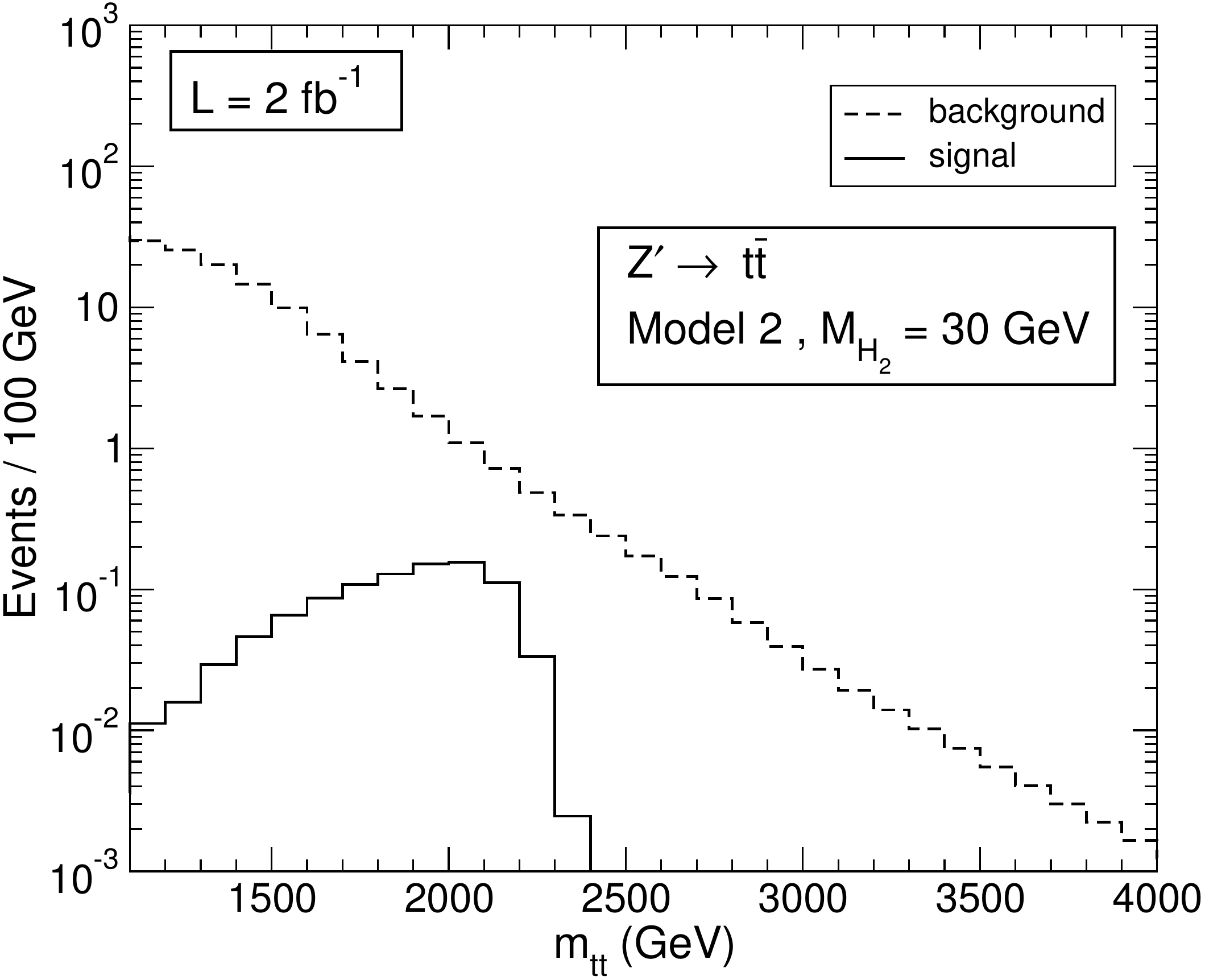} \\
\includegraphics[height=5.9cm,clip=]{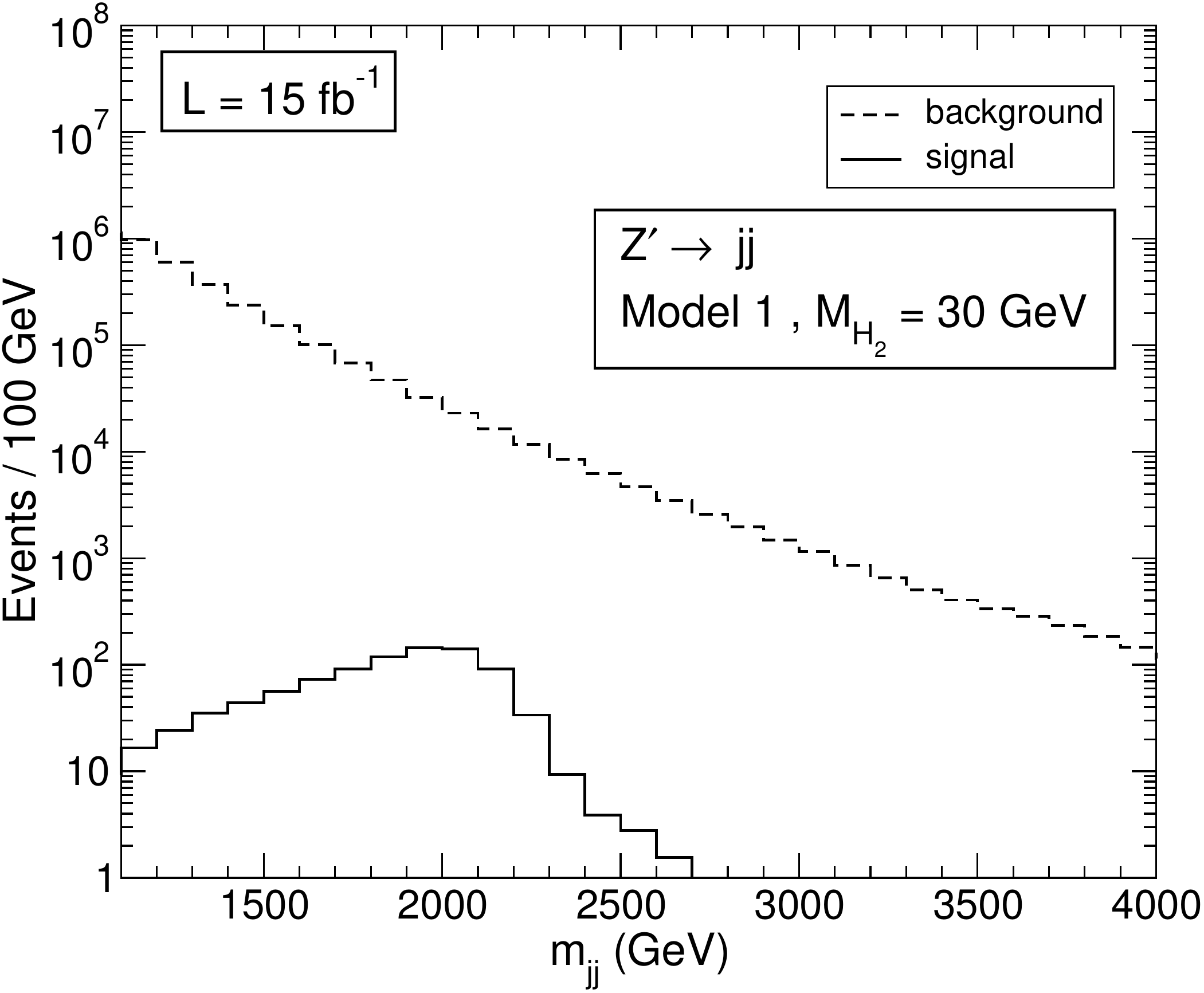}  &
\includegraphics[height=5.9cm,clip=]{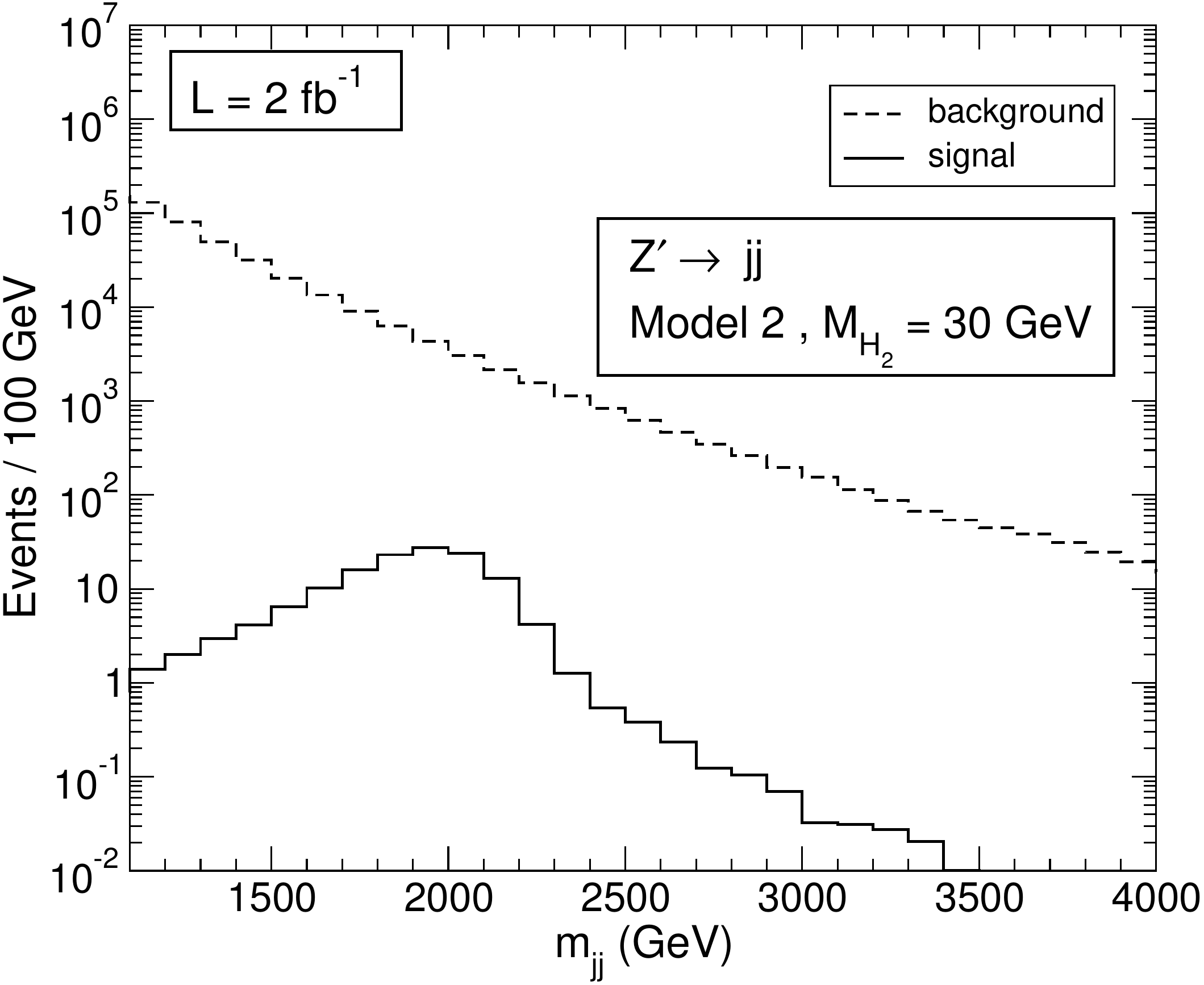}
\end{tabular}
\caption{Invariant mass distribution for the $Z'$ signals in scenario 1 and their backgrounds, in the stealth boson (top), $t \bar t$ (middle) and dijet (bottom) analyses, for model 1 (left) and model 2 (right). }
\label{fig:inv1}
\end{center}
\end{figure}

The event selection for the generic and diboson analyes is the same as for the dijet search, but requiring groomed jet masses $40 \leq m_J \leq 100$ GeV, and
\begin{itemize}
\item For the generic search we apply the tagger performance efficiency factors obtained in ref.~\cite{Aguilar-Saavedra:2017rzt} of $0.01$ for QCD jets and 0.31 for the signal jets ($M_{H_{3,4}} = 80$ GeV, $M_{H_2} = 30$ GeV).
\item For the diboson search we require $\tau_{21} \leq 0.4$ for both jets. The mass window is wider than the usual ones for diboson resonance searches (for example, $65 \leq m_J \leq 105$ GeV is commonly used for $W$ and $Z$ jets by the CMS Collaboration) but we prefer to keep the same window as in the generic search for better comparison. 
\end{itemize}
The presence of the heavy $Z'$ resonance can be detected as a bump in the dijet or $t \bar t$ invariant mass distribution. We present in figure~\ref{fig:inv1} these distributions for the generic (top panels), $t \bar t$ (middle panels) and dijet (bottom panels) analyses. For model 1 we use an integrated luminosity of $L = 15$ fb$^{-1}$, while for model 2, for which the signal is much larger, we take $L = 2$ fb$^{-1}$. The signal and background cross sections for the different event selections are collected in table~\ref{tab:sig1}.

\begin{table}[t]
\begin{center}
\begin{tabular}{lcccc}
                        & Generic      & dijet& $t \bar t$ & diboson \\
$Z'$ (model 1) & 0.71 fb & 27 fb & 0.40 fb & 0.073 fb \\
$Z'$ (model 2) & 3.4 fb   & 31 fb & 0.21 fb & 0.057 fb \\
$jj$                   & 2.3 fb & 280 pb & 21.5 fb & 830 fb \\
$b \bar b$        & 5 ab     & 0.89 pb & 1.9 fb  & 2.5 fb  \\
$t \bar t$          & ---        & ---          & 78 fb & --- 
\end{tabular}
\caption{Signal and background cross sections for the $Z'$ signal in scenario 1 and main SM backgrounds (in rows) under the four different event selections for generic, dijet, $t \bar t$ and diboson resonance searches, in columns.}
\label{tab:sig1} 
\end{center}
\end{table}

Although the resonance is relatively narrow, the detector resolution effects yield a wider distribution, especially for the decays into scalars which produce four-pronged jets. As it has previously been shown~\cite{Aguilar-Saavedra:2017zuc,Aguilar-Saavedra:2018xpl}, standard grooming algorithms are not adequate for multi-pronged jets, shifting jet masses and momenta from their original values. The effect can clearly be seen in the signal profile for the dijet analysis, which is much wider in model 2, where more than half of the dijet events are actually $Z' \to H_3 H_4$.

\begin{figure}[t]
\begin{center}
\begin{tabular}{cc}
\includegraphics[height=5cm,clip=]{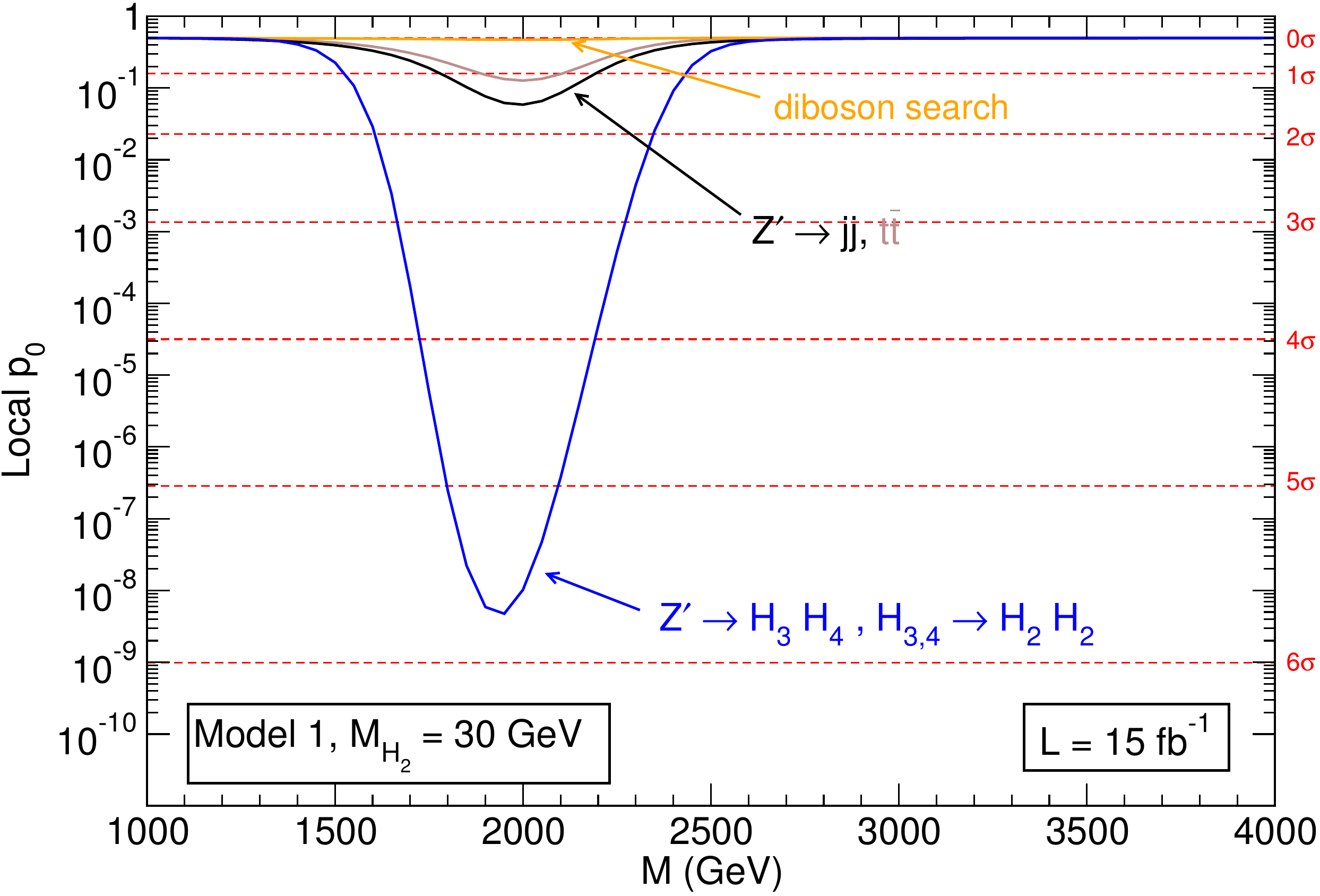} & 
\includegraphics[height=5cm,clip=]{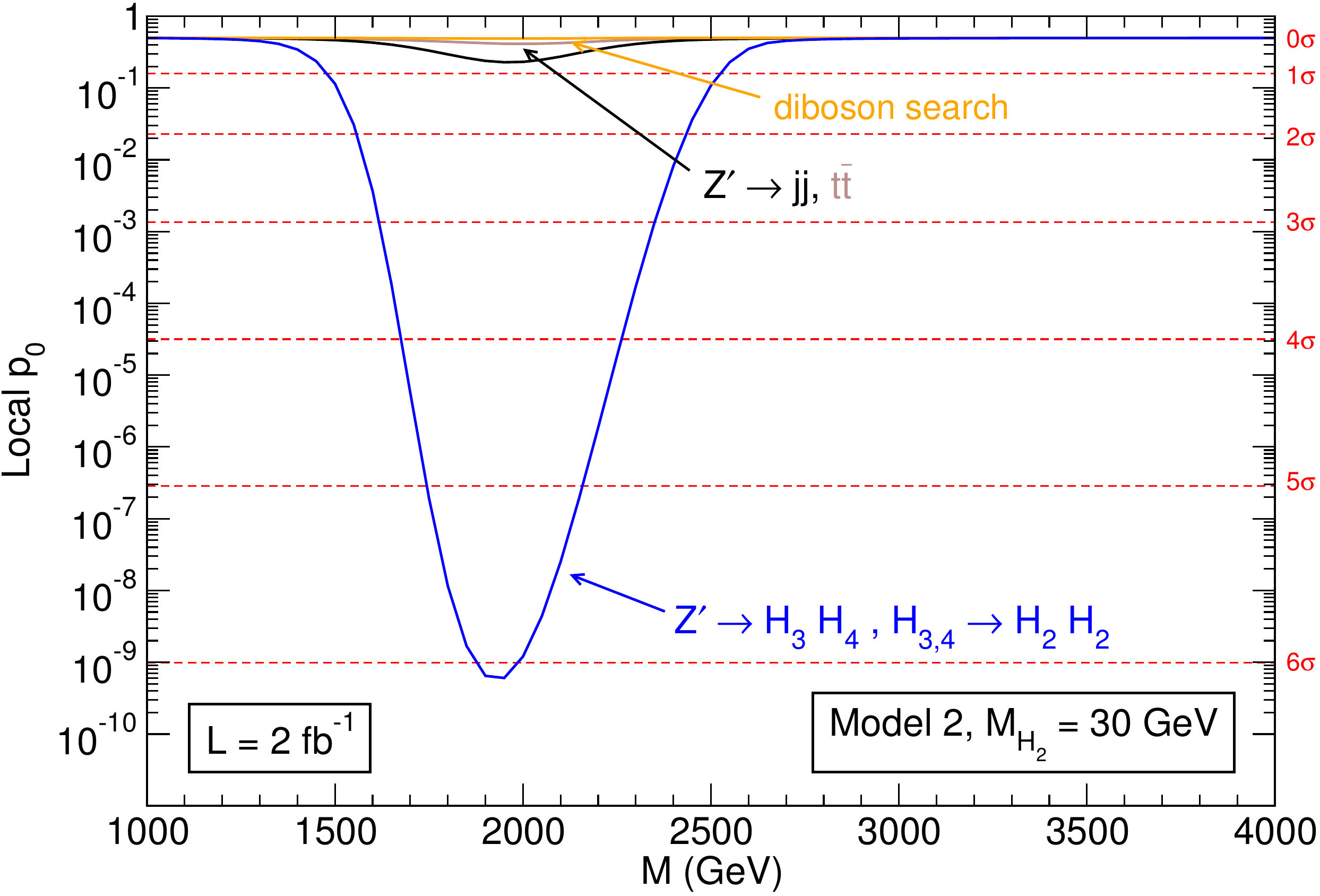}
\end{tabular}
\caption{Expected local $p$-value for the $Z'$ signal in the various searches, for scenario 1 of model 1 (left) and model 2 (right).}
\label{fig:Pval1}
\end{center}
\end{figure}

The expected significance of the $Z'$ signal in the different searches is computed by using the Monte Carlo predictions for signal plus background as pseudo-data,
performing likelihood tests for the presence of narrow resonances over the expected background, using the $\text{CL}_\text{s}$ method~\cite{Read:2002hq} with the asymptotic approximation of ref.~\cite{Cowan:2010js}, and computing the $p$-value corresponding to each hypothesis for the resonance mass. The probability density functions of the potential narrow resonance signals are Gaussians with centre $M$ (i.e. the resonance mass probed) and standard deviation of $0.065 M$. 
The likelihood function is
\begin{equation}
L(\mu) = \prod_{i} \frac{e^{-(b_i+ \mu s_i)} (b_i + \mu s_i) ^{n_i}}{n_i!} \,,
\label{ec:Lik}
\end{equation}
where $i$ runs over the different bins with numbers of events $n_i$, $b_i$ is the predicted number of background events and $s_i$ the predicted number of signal events in each bin, and $\mu$ a scale factor. We do not include any systematic uncertainty in the form of nuisance parameters. For each mass hypothesis the value $\mu_b$ that maximises the likelihood function (\ref{ec:Lik}) is calculated, and local $p$-value is computed as
\begin{equation}
p_0 = 1 - \Phi(\sqrt{2[L(\mu_b)-L(0)]}) \,,
\end{equation}
with
\begin{equation}
\Phi(x) = \frac{1}{2} \left[ 1 + \text{erf}\left(\frac{x}{\sqrt 2}\right) \right] \,.
\end{equation}
The results, assuming luminosities of 15 fb$^{-1}$ (model 1) and 2 fb$^{-1}$ (model 2) are presented in figure~\ref{fig:Pval1}. As it can be seen, stealth boson signals in the generic search are by far more significant than standard dijet signals. In terms of standard deviations, the significance in generic searches is 4 (8) times larger for model 1 (model 2). As we have noted at the beginning of this section, the actual values depend on the product $g_{Z'} z$, which is a free parameter. Several additional comments and clarifications are in order.

\begin{itemize}
\item In our generic search, sensitive to stealth boson signals, we have focused on a jet mass window $40 \leq m_J \leq 100$ GeV, adequate for the benchmark point of the anti-QCD tagger considered. In order to cover all masses for the new scalars, experimental searches should explore bidimensional phase space, also using varying jet mass windows (see for example ref.~\cite{Aaboud:2017ecz}).
\item The use of $b$ tagging in the generic search would significantly improve the significance for stealth boson signals. Requiring one $b$ tag in either jet enhances the ratio $S/\sqrt B$ by a factor of 2, and requiring two $b$ tags by $3.4$, where $S$ stands for signal and $B$ for background cross sections. We have chosen not to make use of $b$ tagging in our analysis because the signals are already quite conspicuous, especially for model 2, and the background is already small. In this regard, our results are quite conservative. The use of $b$ tagging would be very useful for large luminosities, to further reduce the background keeping the same signal efficiency for the anti-QCD jet tagger.
\item We have not considered systematic uncertainties in our estimation of the significance of the different signals. These uncertainties will be more important in the channels where the background is larger, that is, $jj$ and $t \bar t$. In the generic search, where the expected background lies between $0.01 - 0.1$ event per bin near the resonance mass, the impact of systematic uncertainties is expected to be small.
\item As aforementioned, existing grooming algorithms are not designed nor optimised for multi-pronged jets and may shift the mass peaks. (Several other grooming algorithms and parameters were explored in ref.~\cite{Aguilar-Saavedra:2018xpl} with similar results.) This is clearly observed in figure~\ref{fig:Pval1}, where the maximum significance is for the stealth boson signal is near 2 TeV while the $Z'$ mass is 2.2 TeV. We have not attempted any mass recalibration because this small shift does not affect our results and conclusions. 
\item The relative (in)significance of the $Z'$ signal in dijet, $t \bar t$ and diboson searches depends on the model and the mass of the lightest scalar, as the signal for the dijet and diboson event selections receive contributions from various $Z'$ decay modes.
\item For all the final states considered, the global significances of the deviations --- which depend on the mass range studied in each experimental search --- are smaller than the local significances in figure~\ref{fig:Pval1}. In any case, for the point addressed in this section, namely to show that the sensitivity of a generic search is a factor of ten (in terms of significance) or more than for current searches, local significances suffice.
\item For lighter $H_2$, the four-pronged stealth boson jets have a more two-pronged structure, and the acceptance in diboson searches is slightly larger (see appendix~\ref{sec:c}). Actually, for $M_{H_2} = 30$ GeV most of the signal that passes the diboson event selection is $Z' \to jj$ and $Z' \to t \bar t$, not $Z' \to H_3 H_4$. 
\end{itemize}

For completeness, let us comment about the direct production of the new scalars $H_{2-4}$, which can be produced in the same processes (gluon-gluon fusion, associated production with a $W/Z$ boson, etc.) as the SM Higgs $H$. The cross sections are the ones that would correspond to a SM Higgs of the same mass $M_{H_{2-4}}$, multiplied by the small factor $O_{1i}^2$. In the benchmark scenarios considered, with mixing angles $|\theta_{1j}| \leq 0.01$ ($j=2,3,4$), all cross sections for processes mediated by SM particles are suppressed by a factor of $O_{1i}^2 \lesssim 10^{-4}$, leading to unobservable signals.

The most stringent constraints on the lightest scalar $H_2$ result from the Higgs boson searches at LEP experiments~\cite{Barate:2003sz}. For a mass of 30 GeV, the LEP constraints imply $O_{1i}^2 \leq 0.02$, two orders of magnitude above the maximum used in our benchmark.
For $H_{3,4}$ with $M_{H_{3,4}} = 80$ GeV there are no searches targeting the production and decay $pp \to H_{3,4} \to H_2 H_2$, with subsequent decay of $H_2$. Still, one can use similar analyses performed for Higgs decays into light pseudoscalars, $H \to aa$, to obtain an estimate of the sensitivity. In gluon-gluon fusion, a search for $H \to aa \to b \bar b \mu^+ \mu^-$ by the ATLAS Collaboration~\cite{Aaboud:2018esj} obtains an upper limit on cross section times branching ratio of $\sigma(H) \times \text{Br}(b \bar b \mu^+ \mu^-) \leq 0.2$ fb for $m_a = 30$ GeV. Using the cross sections from ref.~\cite{Anastasiou:2016hlm} for a 80 GeV SM-like Higgs, we find that in our benchmark $\sigma (H_{3,4}) \times \text{Br}(b \bar b \mu^+ \mu^-) \lesssim  1.3$ ab, two orders of magnitude below the limit. In $WH/ZH$ associated production, a search for $H \to aa \to b \bar b b \bar b $~\cite{Aaboud:2018iil} sets the upper limit $\sigma(H) \times \text{Br}(b \bar b b \bar b) \leq 1.2$ pb for $m_a = 30$ GeV. In our benchmark, $\sigma (H_{3,4}) \times \text{Br}(b \bar b \mu^+ \mu^-) \lesssim  0.67$ fb, three orders of magnitude smaller. At the Tevatron, the CDF Collaboration performed a search for pair production of new particles $Y$, each decaying into two jets, $p \bar p \to YY \to jjjj$. The mass range explored $M_{Y} \geq 50$ GeV does not cover $M_{H_2} = 30$ GeV, but for illustration we can take the limit for $M_Y = 50$ GeV, namely $\sigma(YY \to jjjj) \leq 200$ pb. In our benchmarks, the prediction is $\sigma(H_{3,4}) \times \text{Br}(jjjj) \lesssim 0.1$ fb, four orders of magnitude below that limit.

\subsection{Scenario 2}

This scenario is similar to scenario 1 but with heavier masses, $M_{Z'} = 3.3$ TeV, $M_{H_3} \simeq M_{H_4} \simeq 400$ GeV, and $M_{H_2} \simeq 80$ GeV. For the signal coupling we choose $g_{Z'} z = 0.2$, yielding a total $Z'$ production cross section of 20.1 fb. The total $Z'$ width is $\Gamma = 70.2$ GeV (model 1) and $\Gamma = 127.7$ GeV (model 2). We set the mixing factor $R_{34}=1$ in eq.~(\ref{ec:Rij}), leading to the branching ratios
\begin{align}
& \text{Br}(Z' \to H_3 H_4) = 0.10 \quad \quad \text{(model 1)} \,, \notag \\
& \text{Br}(Z' \to H_3 H_4) = 0.51 \quad \quad \text{(model 2)} \,.
\end{align}
Again, $\text{Br}(H_{3,4} \to H_2 H_2)\simeq 1$, and for the decay of $H_2$ into quark pairs
\begin{align}
& \text{Br}(H_2 \to b \bar b) = 0.89 \,, \notag \\
& \text{Br}(H_2 \to c \bar c) = 0.065 \,.
\end{align}
The combined branching ratio factor for the four $H_2$ decays into quark pairs is $\text{Br}(H_2 \to b \bar b,c\bar c)^4 = 0.843$. The event selection for the generic analysis is the same as for the dijet search, but requiring groomed jet masses $m_J \geq 250$ GeV. We apply the tagger performance efficiency factors obtained in ref.~\cite{Aguilar-Saavedra:2017rzt} of $0.01$ for QCD jets and 0.33 for the signal jets ($M_{H_{3,4}} = 400$ GeV, $M_{H_2} = 80$ GeV).

\begin{figure}[p]
\begin{center}
\begin{tabular}{cc}
\includegraphics[height=6cm,clip=]{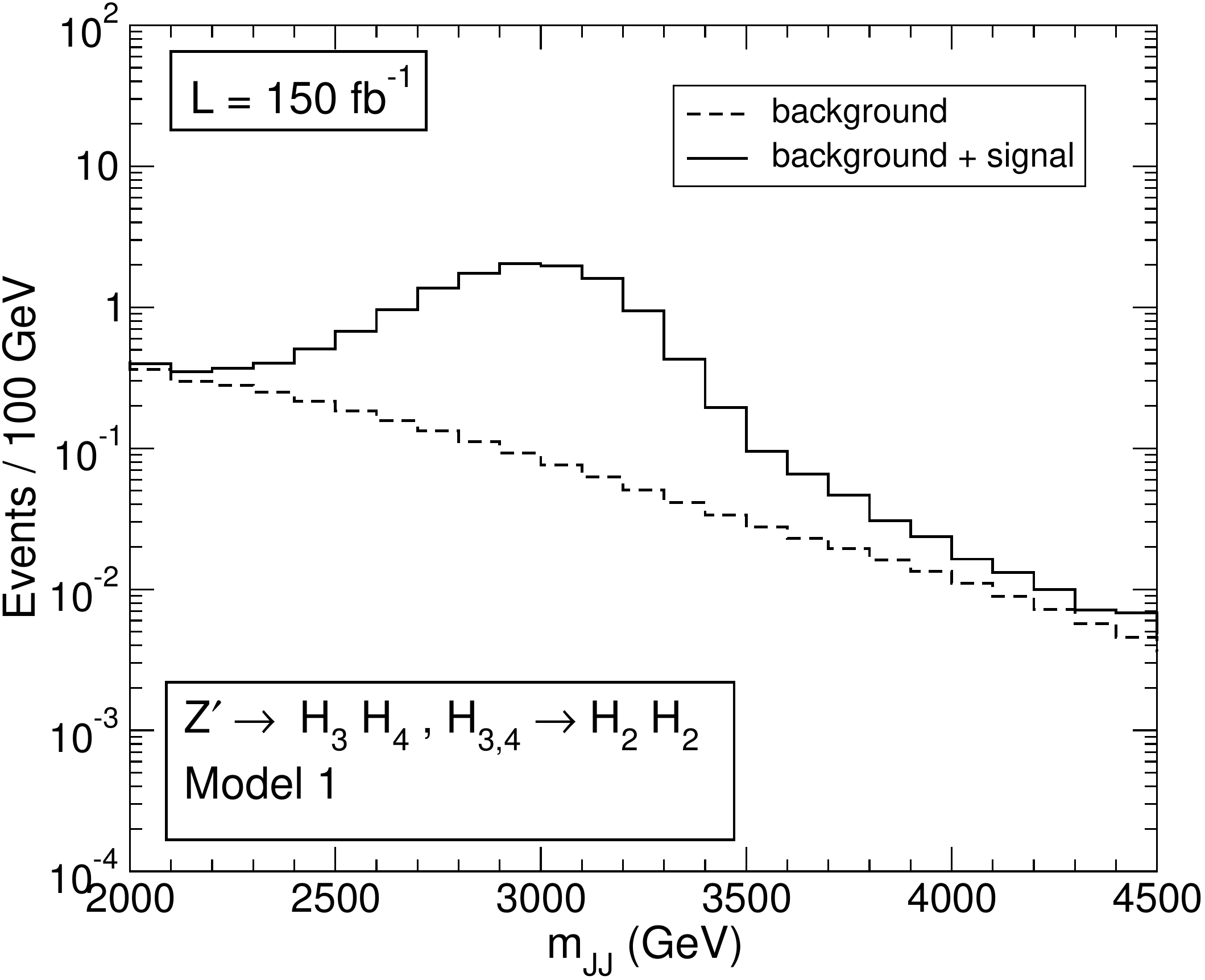} & 
\includegraphics[height=6cm,clip=]{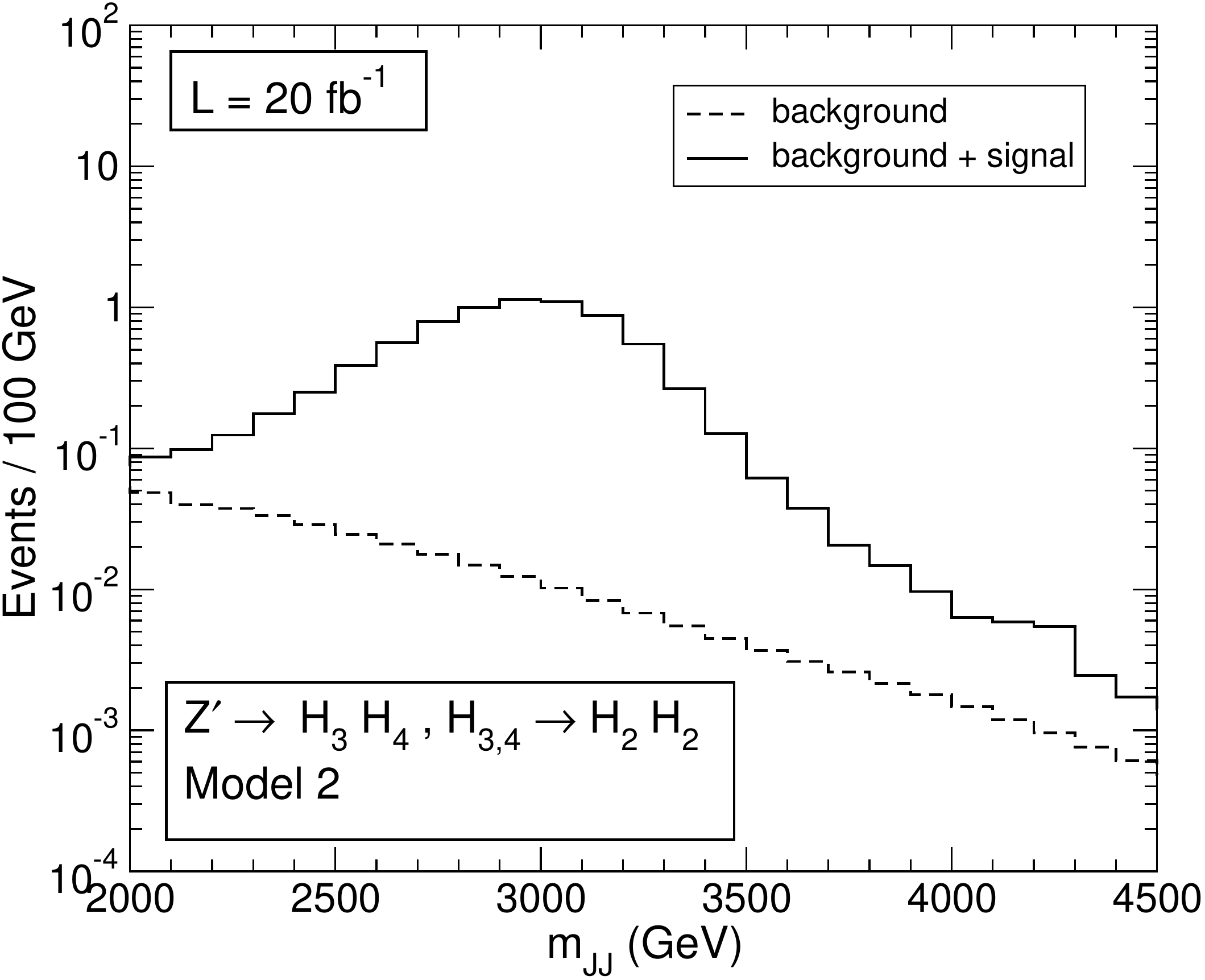} \\
\includegraphics[height=6cm,clip=]{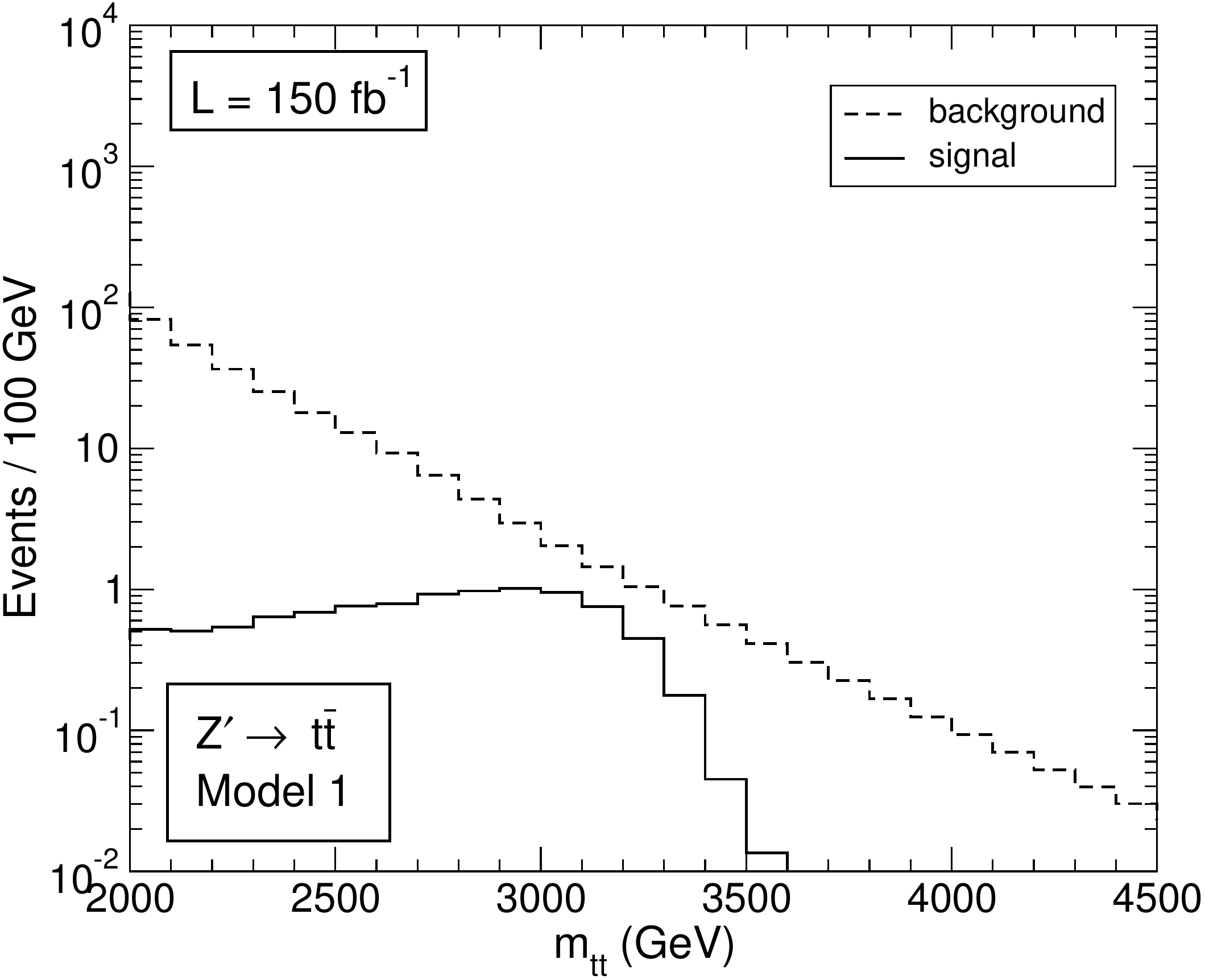}  &
\includegraphics[height=6cm,clip=]{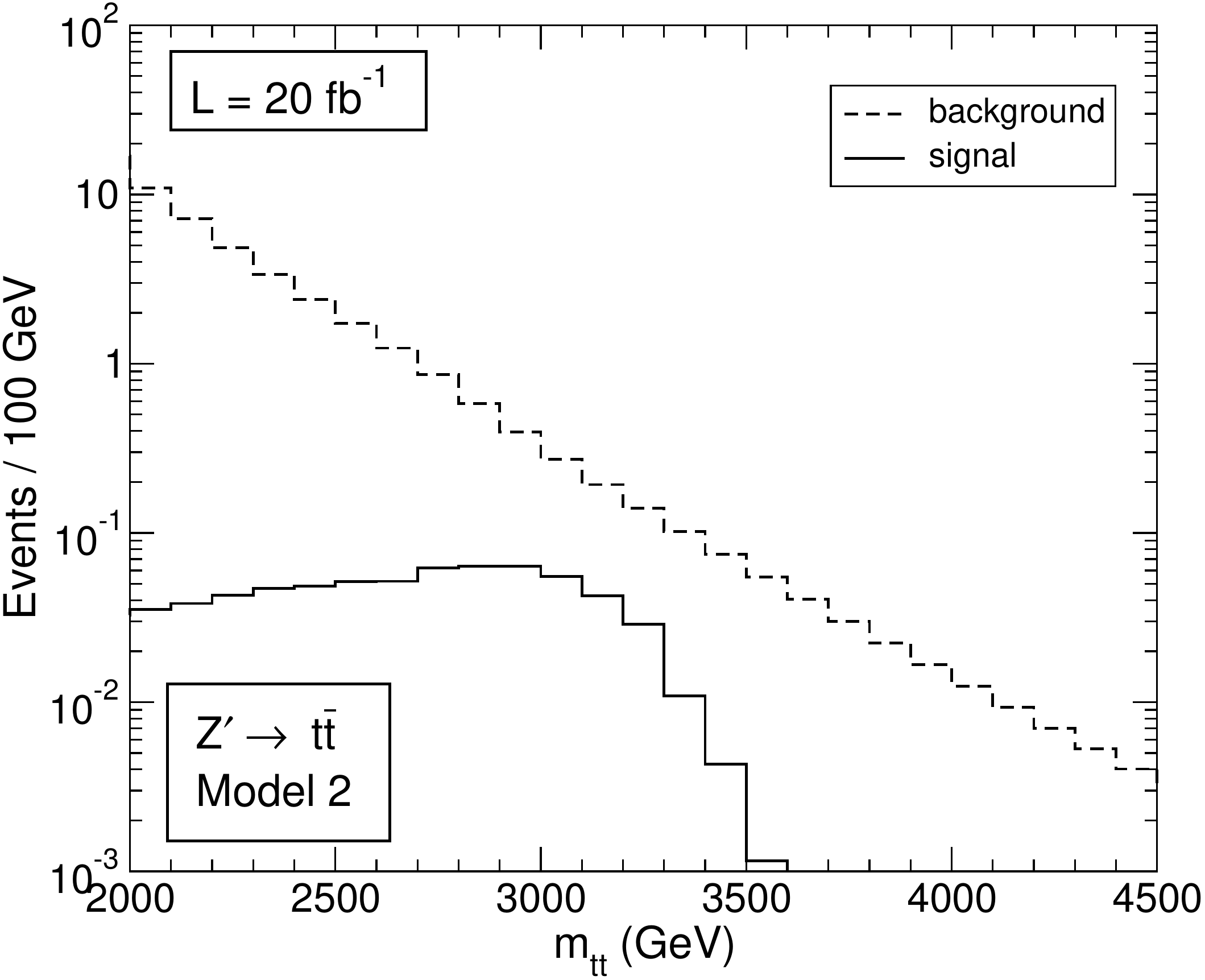} \\
\includegraphics[height=6cm,clip=]{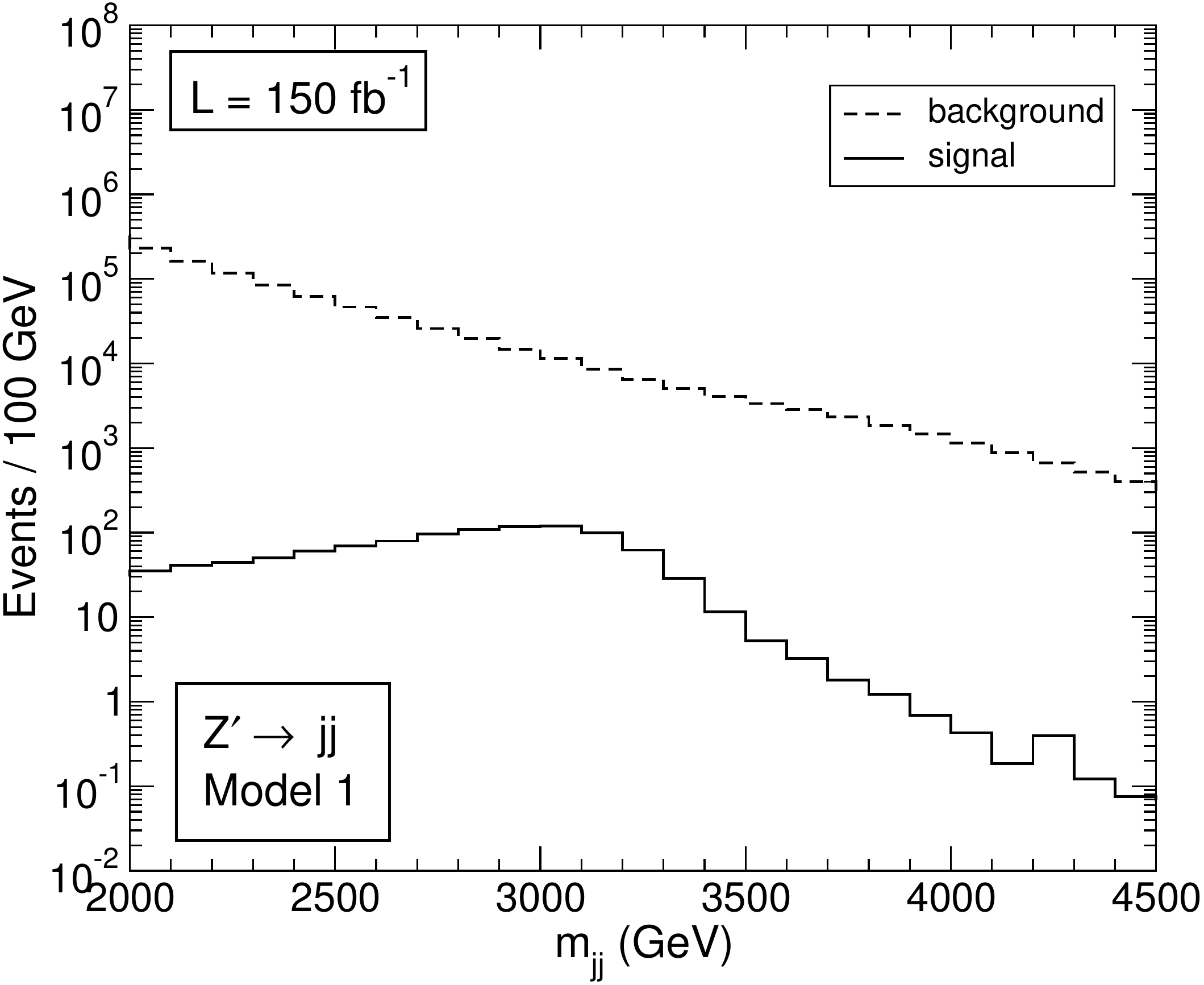}  &
\includegraphics[height=6cm,clip=]{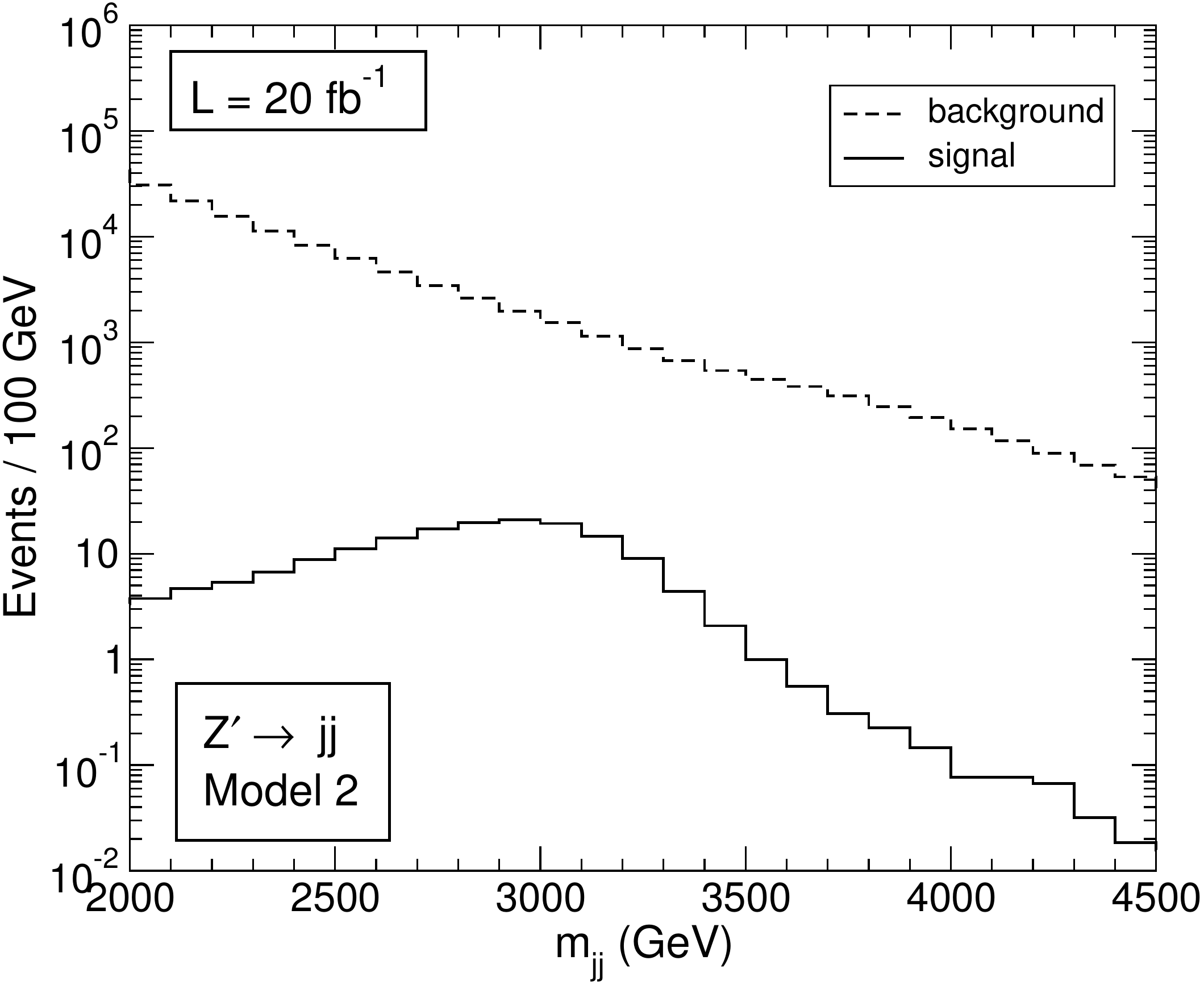}
\end{tabular}
\caption{Invariant mass distribution for the $Z'$ signals in scenario 2 and their backgrounds, in the generic (top), $t \bar t$ (middle) and dijet (bottom) analyses, for model 1 (left) and model 2 (right).}
\label{fig:inv2}
\end{center}
\end{figure}

The dijet / $t \bar t$ invariant mass distributions are presented in figure~\ref{fig:inv2} for the generic (top), $t \bar t$ (middle) and dijet (bottom) analyses. Given that the cross sections for $Z'$ production are smaller to those of scenario 1, we present our results for integrated luminosities $L = 150$ fb$^{-1}$ for model 1, and $L = 20$ fb$^{-1}$ for model 2. The signal and background cross sections for the different event selections are collected in table~\ref{tab:sig2}.

\begin{table}[t]
\begin{center}
\begin{tabular}{lccc}
                        & Generic  & dijet& $t \bar t$  \\
$Z'$ (model 1) & 0.079 fb & 8.1 fb & 0.080 fb  \\
$Z'$ (model 2) & 0.37 fb   & 9.3 fb & 0.043 fb  \\
$jj$                   & 0.032 fb & 280 pb & 21.5 fb  \\
$b \bar b$        & 0.03 ab     & 0.89 pb & 1.9 fb  \\
$t \bar t$          & ---        & ---          & 78 fb  
\end{tabular}
\caption{Signal and background cross sections for the $Z'$ signal in scenario 2 and main SM backgrounds (in rows) under the three different event selections for generic, dijet, and $t \bar t$ resonance searches. The event selection for dijet and $t \bar t$ is the same as in scenario 1, and the quoted background numbers are the same as in table~\ref{tab:sig1}. }
\label{tab:sig2} 
\end{center}
\end{table}

\begin{figure}[htb]
\begin{center}
\begin{tabular}{cc}
\includegraphics[height=5cm,clip=]{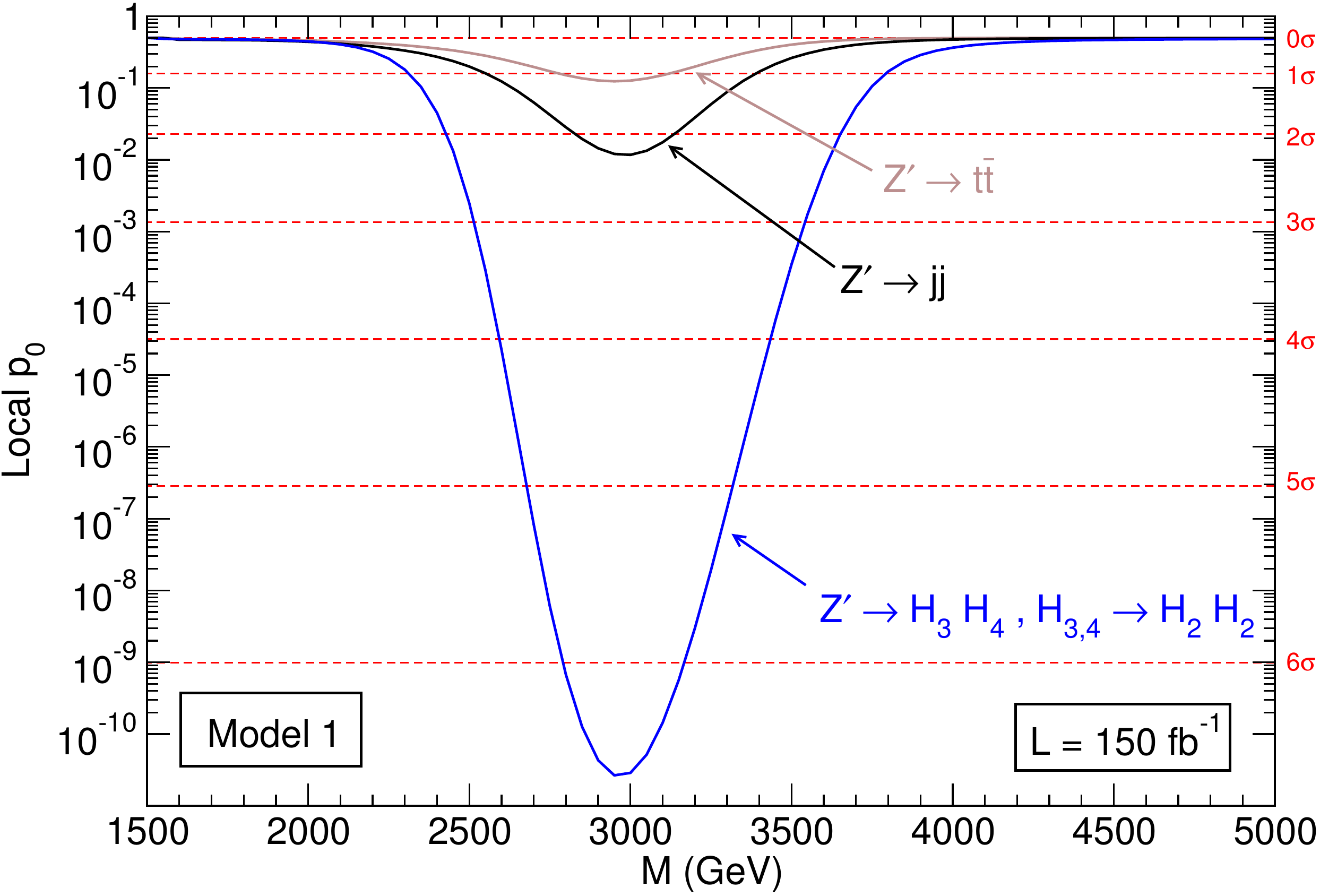} & 
\includegraphics[height=5cm,clip=]{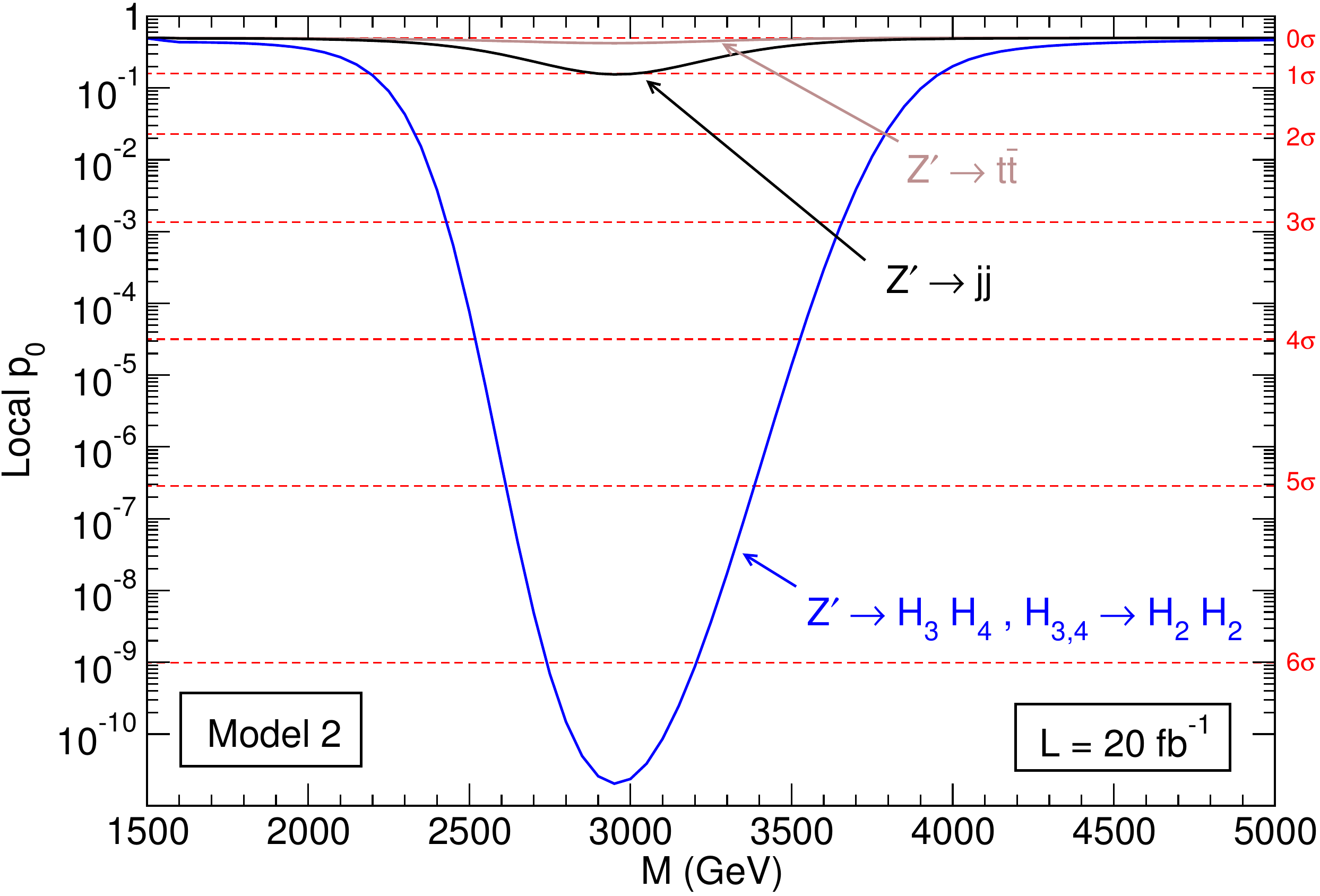}
\end{tabular}
\caption{Expected local $p$-value for the $Z'$ signal in the various searches, for scenario 2 of model 1 (left) and model 2 (right).}
\label{fig:Pval2}
\end{center}
\end{figure}
The expected significance of the $Z'$ signal in the different searches is presented in figure~\ref{fig:Pval2}, assuming luminosities of 150 fb$^{-1}$ in model 1, and 20 fb$^{-1}$ in model 2. The difference between stealth boson modes and standard $t \bar t$ and dijet decays is quite pronounced. The significance of the signals in the generic search (expressed in terms of standard deviations) is 3 and 6 times larger than in dijets, for model 1 and model 2, respectively. Still, we remind the reader that we have not taken advantage of $b$ tagging, which would improve the signal significance by a factor of $2-5$ in the generic search.

Let us also comment on signals of the direct production of the new scalars within this scenario.
Direct production of the lightest scalar $H_2$ with decay into SM particles (with its decay branching ratios corresponding to a SM Higgs with a mass of 80 GeV) can be constrained from Higgs boson searches. At LEP, the non-observation of a signal constrains $O_{1i}^2 \leq 0.04$ for $M_{H_2} = 80$ GeV~\cite{Barate:2003sz}, two orders of magnitude above the bound $O_{1i}^2 \lesssim 10^{-4}$ in our benchmark. At the Tevatron, searches by the D0 and CDF Collaborations~\cite{Abazov:2013gmz,Aaltonen:2013ipa} only cover masses above 90 GeV, and also are less sensitive.

For the heaviest scalars the production and decay chain is $p p \to H_{3,4} \to H_2 H_2$, with subsequent decay of $H_2$. A search for pair produced resonances decaying into quark pairs by the CMS Collaboration~\cite{Sirunyan:2018rlj} is sensitive to (but no optimised for) this process. The limits corresponding to stop masses $m_{\tilde t} = 80$ GeV are $\sigma(\tilde t \tilde t^* ) \leq 400$ pb, assuming 100\% branching ratio for the $R$-parity violating decay $\tilde t \to b \bar q$. In our benchmark, the cross section times branching ratio for final states with $b$ quarks is $\sigma (H_{3,4}) \times \text{Br}(b \bar b b \bar b) \lesssim  0.20$ fb for $M_{H_{3,4}} = 400$ GeV. This is six orders of magnitude below the above experimental limit.

\subsection{Scenario 3}

\begin{figure}[t]
\begin{center}
\begin{tabular}{cc}
\includegraphics[height=5.5cm,clip=]{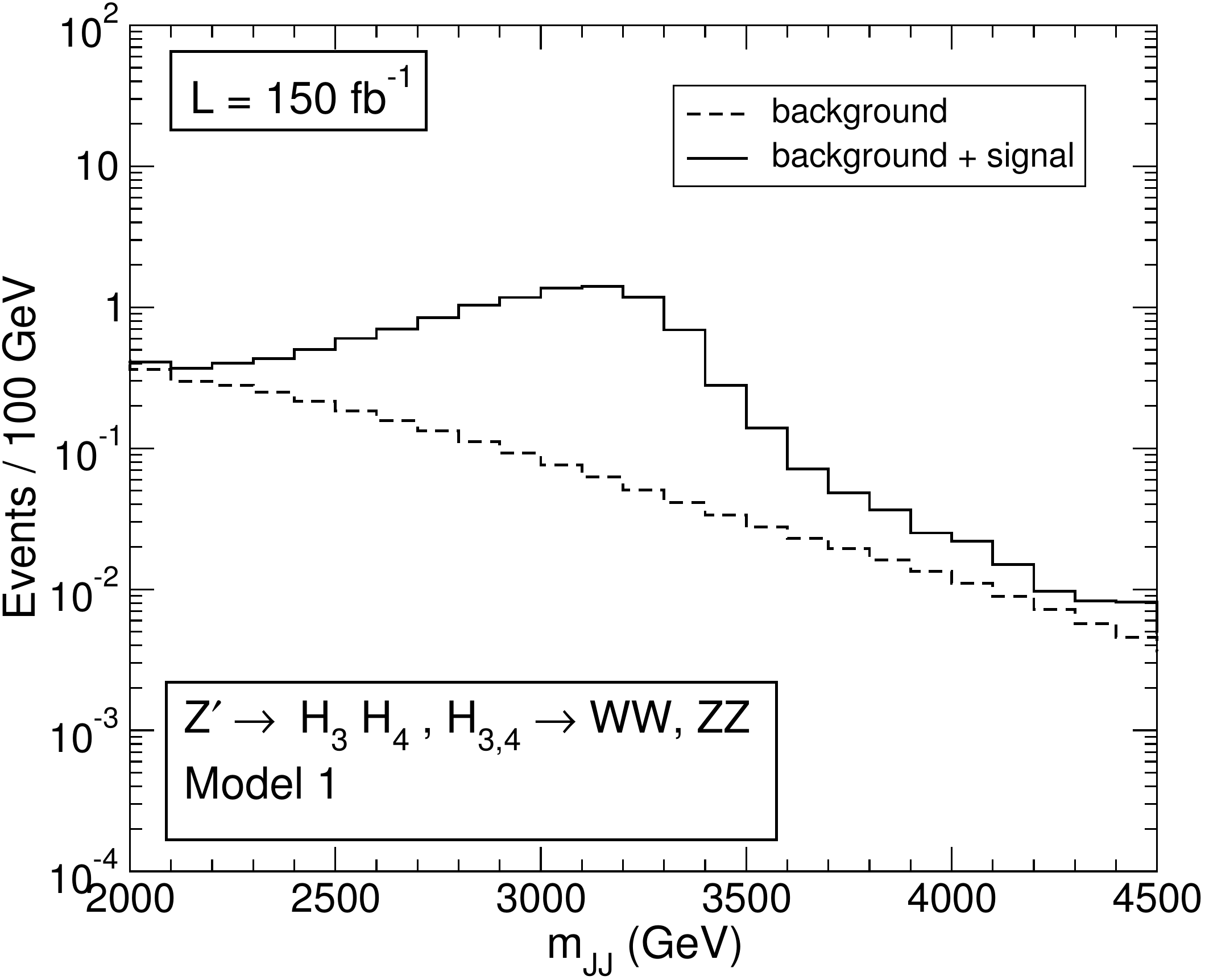} & 
\includegraphics[height=5.5cm,clip=]{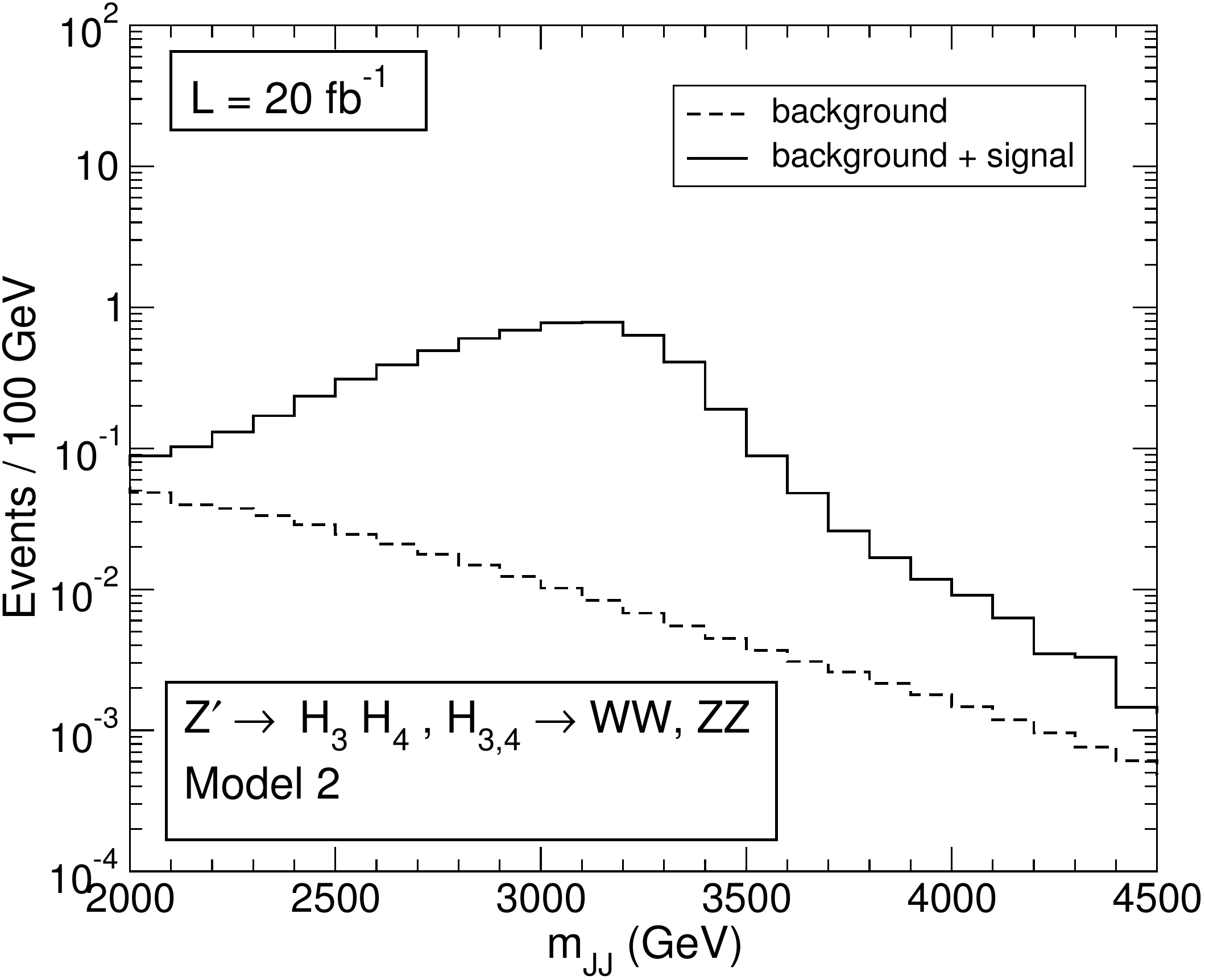} \\
\includegraphics[height=5.5cm,clip=]{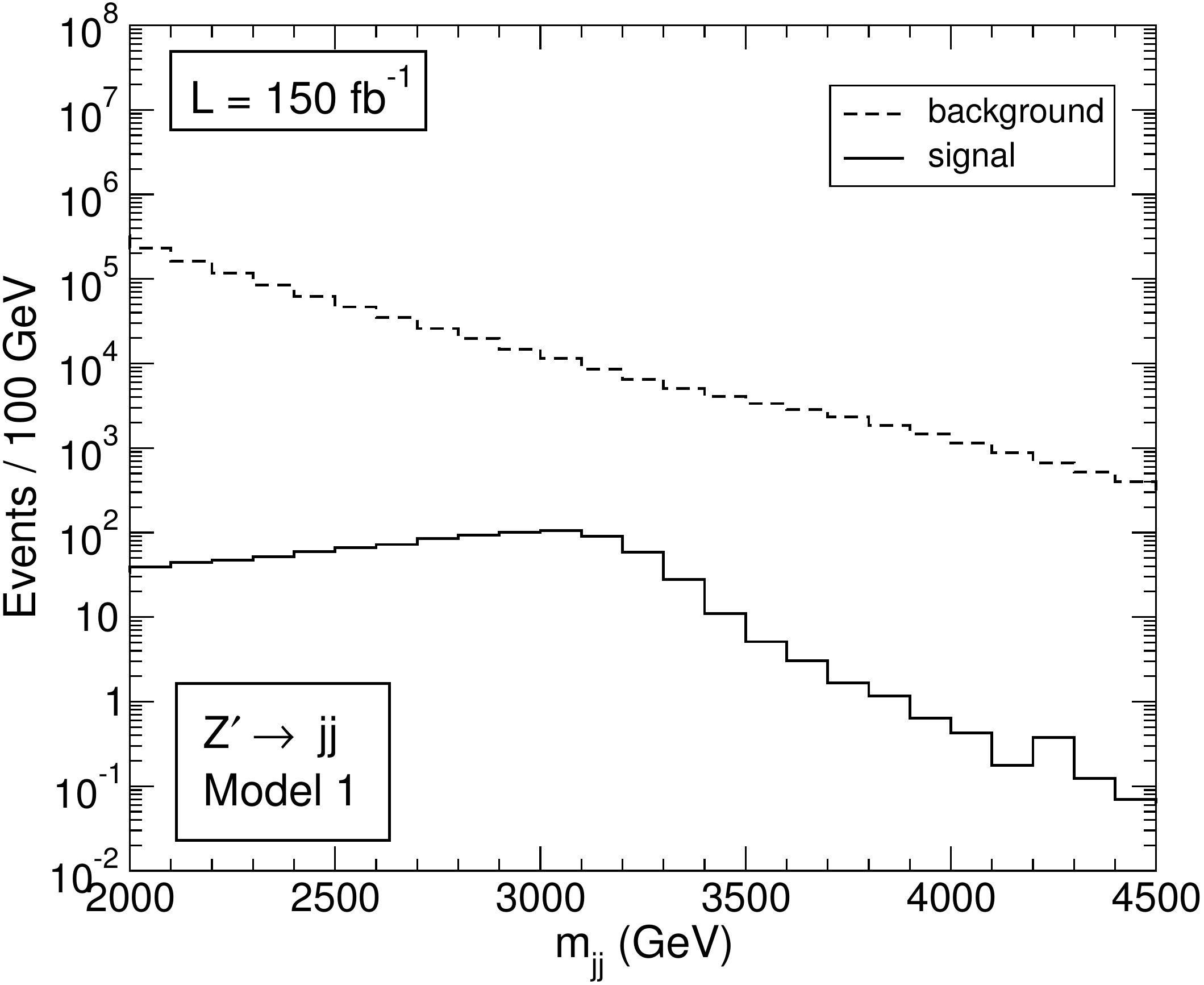}  &
\includegraphics[height=5.5cm,clip=]{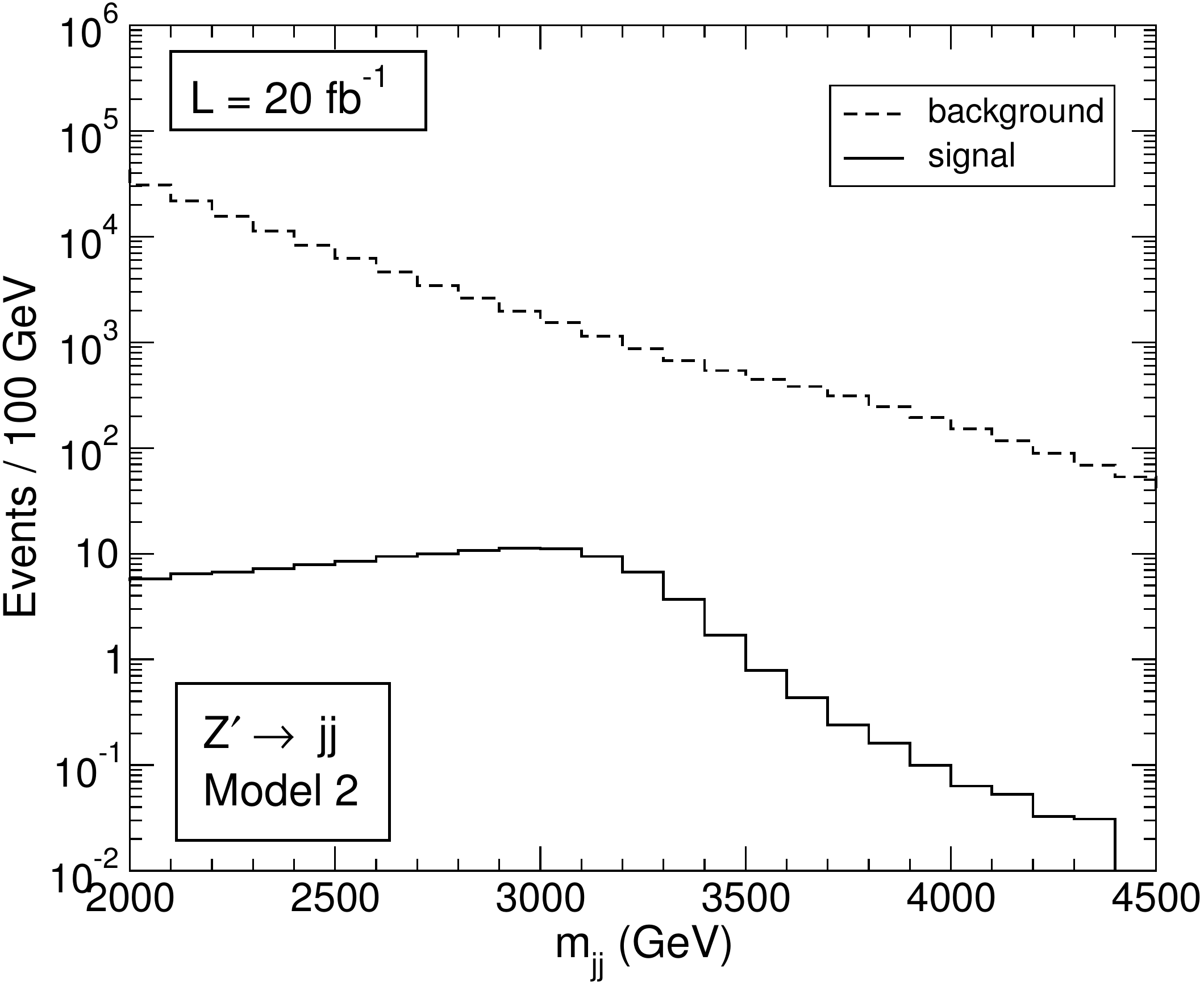}
\end{tabular}
\caption{Invariant mass distribution for the $Z'$ signals in scenario 3 and their backgrounds, in the generic (top) and dijet (bottom) analyses, for model 1 (left) and model 2 (right). }
\label{fig:inv3}
\end{center}
\end{figure}

Here we take $M_{Z'} = 3.3$ TeV and $M_{H_3} \simeq M_{H_4} \simeq 400$ GeV, as in scenario 2, and we keep the same signal coupling $g_{Z'} z = 0.2$. Therefore, the $Z'$ production cross section and width are the same, $\sigma(Z') = 20.1$ fb, and $\Gamma = 70.2$ GeV in model 1, $\Gamma = 127.7$ GeV in model 2. 
We set $R_{34}$ to unity, hence the branching ratios are the same as in scenario 2,
\begin{align}
& \text{Br}(Z' \to H_3 H_4) = 0.10 \quad \quad \text{(model 1)} \,, \notag \\
& \text{Br}(Z' \to H_3 H_4) = 0.51 \quad \quad \text{(model 2)} \,.
\end{align}
The differences with respect to scenario 2 stem from the fact that $H_2$ is now heavier, namely $M_{H_2} \simeq 300$ GeV. This forbids the decays $H_{3,4} \to H_2 H_2$. We therefore focus on $H_2$ decays into gauge boson pairs, taking (see figure~\ref{fig:scan2H})
\begin{align}
& \text{Br}(H_{3,4} \to WW) = 0.57  \,, \notag \\
& \text{Br}(H_{3,4} \to ZZ) = 0.26  \,.
\end{align}
The event selection for the three analyses is the same as in scenario 2 (for the dijet and $t \bar t$ searches it is also the same as in scenario 1). The dijet invariant mass distributions are presented in figure~\ref{fig:inv3} for the generic (top panel) and dijet (bottom panel) analyses, for integrated luminosities $L = 150$ fb$^{-1}$ (model 1) and $L = 20$ fb$^{-1}$ (model 2). The results for the $t \bar t$ analysis are the same as in scenario 2, and were already shown in figure~\ref{fig:inv2}. The signal and background cross sections for the different event selections are collected in table~\ref{tab:sig3}.

\begin{table}[t]
\begin{center}
\begin{tabular}{lccc}
                        & Generic  & dijet& $t \bar t$  \\
$Z'$ (model 1) & 0.062 fb & 7.8 fb & 0.080 fb  \\
$Z'$ (model 2) & 0.30 fb   & 7.6 fb & 0.043 fb  \\
$jj$                   & 0.032 fb & 280 pb & 21.5 fb  \\
$b \bar b$        & 0.03 ab     & 0.89 pb & 1.9 fb  \\
$t \bar t$          & ---        & ---          & 78 fb  
\end{tabular}
\caption{Signal and background cross sections for the $Z'$ signal in scenario 3 and main SM backgrounds (in rows) under the three different event selections for generic, dijet, and $t \bar t$ resonance searches. The event selection is the same as in scenario 2, and the quoted backgrounds are the same as in table~\ref{tab:sig2}. The signal in the $t \bar t$ selection is also the same.}
\label{tab:sig3} 
\end{center}
\end{table}
\begin{figure}[t]
\begin{center}
\begin{tabular}{cc}
\includegraphics[height=5cm,clip=]{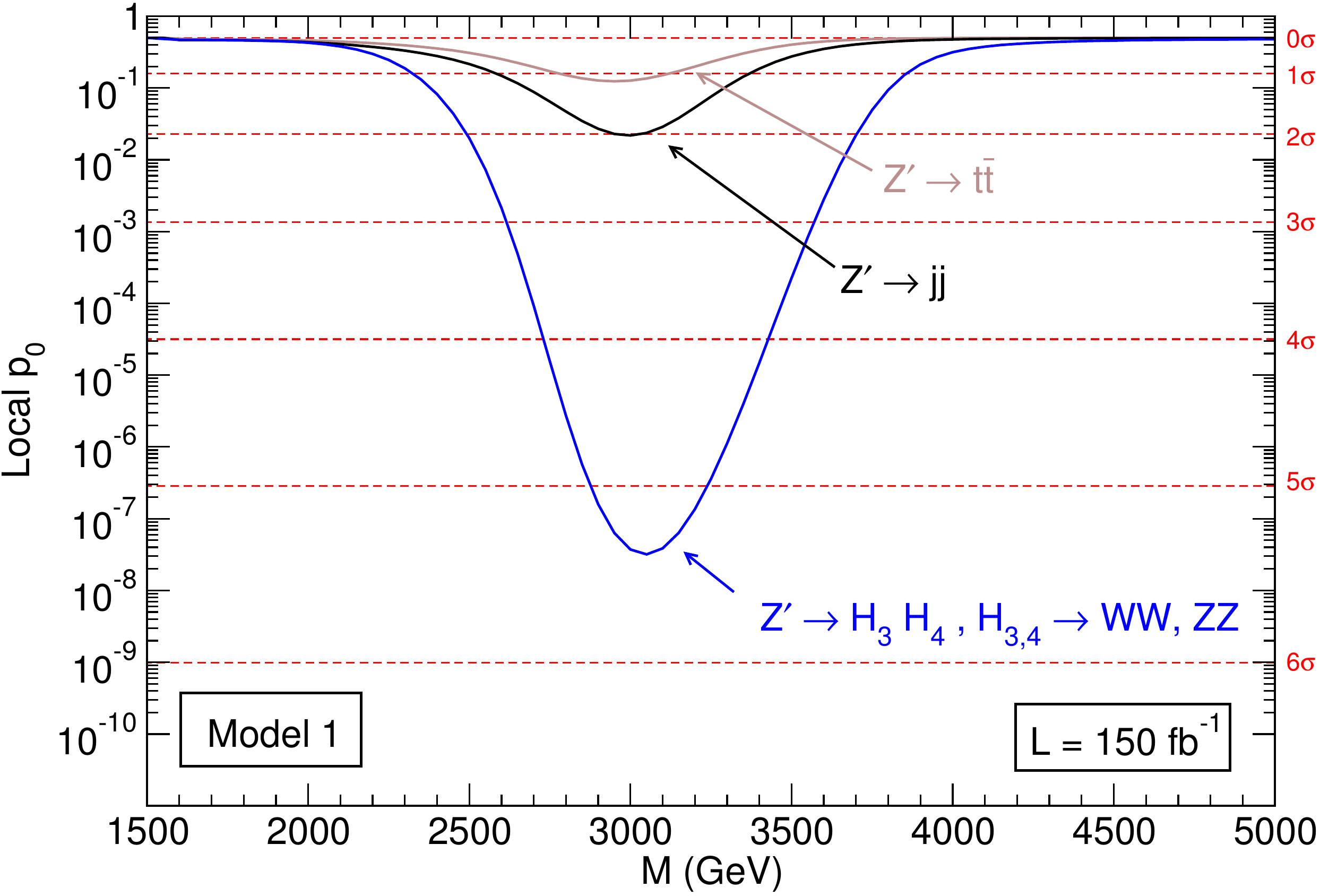} & 
\includegraphics[height=5cm,clip=]{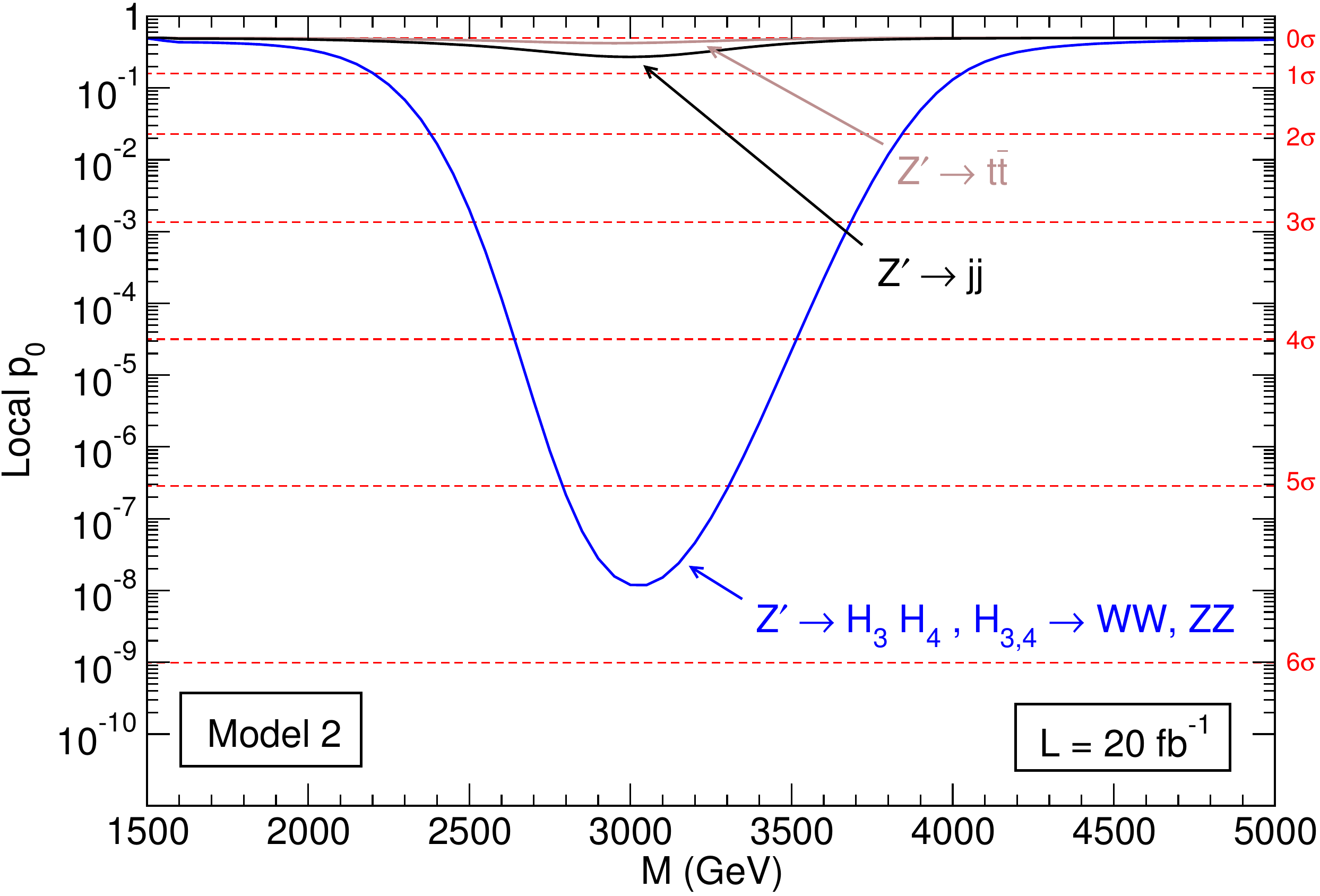}
\end{tabular}
\caption{Expected local $p$-value for the $Z'$ signal in the various searches, for scenario 3 of model 1 (left) and model 2 (right).}
\label{fig:Pval3b}
\end{center}
\end{figure}
The expected significance of the $Z'$ signal in the different searches is presented in figure~\ref{fig:Pval3b}, assuming luminosities of 150 fb$^{-1}$ in model 1 and 20 fb$^{-1}$ in model 2. 
The significance of the signals in the generic search (expressed in terms of standard deviations) is 2.5 and 9 times larger than in dijets, for model 1 and model 2, respectively.
Regarding the results for the generic analysis, it is worth remarking a few points.
\begin{itemize}
\item The requirement of jet masses $m_J \geq 250$ GeV in the generic analysis filters the hadronic decays of both gauge bosons, which have branching ratio of 0.45 for $WW$ and 0.49 for $ZZ$, reducing the signal with respect to scenario 2. This explains why the expected sensitivity in the generic analysis is slightly worse, despite the larger efficiency of the tagger for jets corresponding to $H_{3,4} \to WW$ (0.49 for the signal for a background rejection of $10^2$) than for jets with $H_{3,4} \to H_2 H_2$ in scenario 2.
\item The decays $H_{3,4} \to HH$ have a branching ratio of 0.13, and the generic search would also be sensitive to jets containing two SM Higgs bosons. Because there is no benchmark working point of the tagger available, we have not included these signal contributions in our analysis.
\item The tagger has an efficiency of 0.21 for jets containing two top quarks, resulting from $H_{3,4} \to t \bar t$. For simplicity we have not included this signal contribution to the generic search, given the small branching ratio $\text{Br}(H_{3,4} \to t \bar t ) = 0.04$.
\end{itemize}
Therefore, although the significance of the signals simulated for this scenario is smaller than in scenario 2, it is expected that the significances become similar when all possible heavy scalar decay channels are included. Let us also comment on the direct production of the scalars. Searches for new scalars produced in gluon gluon fusion and decaying into $VV$ pairs set a limit of approximately $\sigma \times \text{Br}(VV) \leq 350$ fb for a scalar mass of 400 GeV. In this benchmark we have $\sigma(H_{3,4}) \times \text{Br}(VV) \leq 0.26$ fb, more than two orders of magnitude below the limit. The lightest new scalar $H_2$ with $M_{H_2} = 300$ GeV is not covered by this analysis, which only considers masses above 400 GeV. Still, $\sigma(H_{2}) \times \text{Br}(VV) \leq 0.66$ fb is well below potential constraints at this mass.

\section{Discussion}
\label{sec:5}

The search for elusive new physics signals yielding various types of multi-pronged jets requires a model-independent approach, with the use of novel tools like the anti-QCD tagger~\cite{Aguilar-Saavedra:2017rzt} or non-supervised learning methods~\cite{Collins:2018epr,Heimel:2018mkt,Farina:2018fyg}. In order to contextualise the relevance of these signals as discovery channel for new (leptophobic) resonances, it is crucial to provide examples of consistent models that may produce them. We have done so for stealth bosons, boosted particles with a cascade decay giving a four-pronged fat jet. We have worked out the minimal implementation, adding to the SM a leptophobic $Z'$ boson, two complex scalar singlets and extra matter, either new vector-like quarks (model 1) or new vector-like leptons (model 2), to cancel anomalies. In these models one can compare the potential significance of stealth boson signals, still unexplored at the LHC, with the standard signals (dijets, top pairs and dibosons) already searched for. Depending on the model and benchmark scenario considered, the significance of the former may be up to 9 times larger than the most sensitive of the latter. Therefore, it is clear that stealth boson signals might well be hidden in LHC data, yet invisible to current searches. Besides, direct production of the new light scalars is suppressed by the square of the small mixing, and signals are too small to be observed. 

In the two models considered in this work the branching ratios of $Z'$ decays into scalars are sizeable (around 10\% in model 1 and 50\% in model 2). Moreover, cascade decays of the new scalars are likely to happen, provided one of the following conditions are fulfilled:
\begin{itemize}
	\item There is a hierarchy among the masses of the new scalars, so that the decays of one into others are possible. These decays are not suppressed by mixing with the SM scalar doublet, and will therefore be dominant, as in our scenario 1.
	\item The scalars are heavy enough to decay into $W^+ W^-$ (and possibly $ZZ$, $HH$ and $t \bar t$). If the decay into other new scalars is kinematically allowed, it will be the dominant channel, as in our scenario 2. Otherwise, decays into pairs of SM bosons will be dominant, as in our scenario 3.
\end{itemize}
Therefore, as it has been shown with a scan on parameter space, it is natural to have stealth bosons as decay products of the $Z'$. For simplicity, we have restricted our detailed simulations to $Z'$ decays into a pair of stealth bosons giving two four-pronged jets. Still, those processes giving one stealth boson (four-pronged jet) and one scalar that subsequently decays into quarks (two-pronged jet) are also possible and interesting. A generic search would be sensitive to all these possibilities at once, and this is one of the main virtues of the anti-QCD tagger. 

In conclusion, we stress that despite the fact that stealth bosons are rather stealth for current LHC searches, they would be quite conspicuous in a generic search. Moreover, these signals may well appear in decays of heavy $Z'$ resonances. These facts already provide a strong motivation for model-independent searches.

\section*{Acknowledgements} F.R.J. thanks the AHEP group at IFIC/CSIC (Valencia) and the Theoretical Physics Department of the University of Valencia for the warm hospitality and for financial support during the final stage of this work. J.A.A.S. thanks A. Casas and C. Mu\~noz for useful discussions. The work of F.R.J. is supported by Funda\c{c}{\~a}o para a Ci{\^e}ncia e a Tecnologia (FCT, Portugal) through the projects CFTP-FCT Unit 777 (UID/FIS/00777/2013), CERN/FIS-PAR/0004/2017 and PTDC/FIS-PAR/29436/2017, which are partly funded through POCTI (FEDER), COMPETE, QREN and EU. The work of J.A.A.S. is supported by Spanish Agencia Estatal de Investigaci\'on through the grant `IFT Centro de Excelencia Severo Ochoa SEV-2016-0597' and by MINECO project FPA 2013-47836-C3-2-P (including ERDF).

\appendix

\section{Triple scalar couplings}
\label{sec:b}

In the weak interaction basis $H'_i = (\rho_0 \; \rho_1 \; \rho_2 \; A^0)$ the trilinear scalar interactions can be expressed in the condensed form:
\begin{equation}
\mathcal{L}_{3H} = - \sum u \, C_{pqr} H'_p H'_q H'_r \,, 
\end{equation}
with the sum over $p \leq q \leq r$ running from from 1 to 4. The $C_{ijk}$ coefficients are explicitly given by:
\begin{align}
& C_{111} = \frac{v}{2u} \lambda_0 \,, \notag \\
& C_{112} = \frac{1}{4}\, [\,\lambda_5 \cos \beta  + \RE (\lambda'_9)\sin \beta\, ] \,, \notag \\
& C_{113} = \frac{1}{4}\, [\,\RE (\lambda'_9) \cos \beta + \lambda_6\sin \beta  \,] \,, \notag \\
& C_{114} = - \frac{1}{4} \IM( \lambda'_9) \,, \notag \\
& C_{122} =  \frac{v}{4u} \lambda_5 \,, \notag \\
& C_{123} = \frac{v}{2u} \RE (\lambda'_9) \,, \notag \\ 
& C_{124} = - \frac{v}{2u} \IM( \lambda'_9)\cos \beta  \,, \notag \\
& C_{133} = \frac{v}{4u} \lambda_6 \,, \notag \\
& C_{134} = - \frac{v}{2u} \IM( \lambda'_9)\sin \beta  \,, \notag \\
& C_{144} = \frac{v}{4u}\, [\, \lambda_6\cos^2 \beta  - \RE (\lambda'_9)\sin (2 \beta)  + \lambda_5\sin^2 \beta\, ]  \,, \notag \\ 
& C_{222} =  \frac{1}{4}\, [\, 2 \lambda_1\cos \beta  + \RE (\lambda'_7)\sin \beta \,] \,, \notag \\
\displaybreak
& C_{223} = \frac{1}{4}\, \left\{\, 3  \RE (\lambda'_7)\cos \beta + 2\,  [\lambda_3 + \RE (\lambda'_4)\,]\sin \beta\right\} \,, \notag \\ 
& C_{224} =  - \frac{1}{4}\, \left\{\,\IM (\lambda'_4)\sin( 2 \beta)  + \IM( \lambda'_7)[2+\cos (2\beta)\,] \, \right\} \,, \notag \\
& C_{233} = \frac{1}{4} \left\{3 \RE( \lambda'_8)\sin \beta +2\,  [\,\lambda_3 + \RE (\lambda'_4)\,]\cos \beta  \right\} \,, \notag \\
& C_{234} = -\frac{1}{4} \left\{4 \IM (\lambda'_4) + \IM( \lambda'_7 + \lambda'_8)\sin 2\beta  \right\}  \,, \notag \\
& C_{244} = \frac{1}{4}\, \left\{ 2\,[\lambda_3 - \RE (\lambda'_4) \cos^3 \beta\, ]
+ \RE( \lambda'_8 - 2 \lambda'_7)\cos^2 \beta  \sin \beta  \notag\right. \\
& \quad \quad \left.+ 2\, [\,\lambda_1 - 2 \RE (\lambda'_4)\,] \sin^2 \beta \cos \beta 
+ \RE (\lambda'_7)\sin^3 \beta  \right\} \,, \notag \\
& C_{333} =  \frac{1}{4} [\,2 \lambda_2\sin \beta  + \RE( \lambda'_8)\cos \beta\, ] \,, \notag \\
& C_{334} =  \frac{1}{4} [- \IM (\lambda'_4)\sin (2 \beta)  +\IM (\lambda'_8) (-2 + \cos 2\beta)  \,] \,, \notag \\
& C_{344} = \frac{1}{4}\, \{\, \RE( \lambda'_8)\cos^3 \beta 
+ 2 [\lambda_2 - 2 \RE (\lambda'_4)]\cos^2 \beta \sin \beta 
+  \RE (\lambda'_7- 2  \lambda'_8)\cos \beta \sin^2 \beta  \notag \\
& \quad \quad + 2 [\lambda_3 - \RE (\lambda'_4)]\sin^3 \beta  \} \,, \notag \\
& C_{444} =  \frac{1}{4} [\, \IM (\lambda'_4)\sin (2\beta)  - \IM( \lambda'_7)\sin^2 \beta  - \IM (\lambda'_8)\cos^2 \beta \, ] \,.
\end{align}
In the $H_a$ mass eigenstate basis, with $H'_i = O_{ia} H_a$, 
\begin{equation}
\mathcal{L}_{3H} = - \sum u \, C_{pqr} O_{pa} O_{qb} O_{rc} H_a H_b H_c \,,
\label{ec:L3Hm}
\end{equation}
where the sums over $a,b,c$ and $p \leq q \leq r$ run from 1 to 4. From this, one can read the interactions $H_i H_j H_k$ which, when the indices $i,j,k$ are different, can be written as $- u \lambda_{ijk} H_i H_j H_k$, with
\begin{equation}
\lambda_{ijk} =  \sum_{p\leq q \leq r,(s)} \, C_{pqr} O_{p s_1} O_{q s_2} O_{r s_3} \,.
\label{ec:lambda}
\end{equation}
The sum above runs over all permutations
\begin{equation}
(s_1 , s_2 , s_3) = \{ (i , j , k) \,, (i , k , j) \,, (j , k , i) \,, (j , i , k) \,, (k , i , j) \,, (k , j , i) \} \,.
\label{ec:perm}
\end{equation}
When two of the indices $i,j,k$ are equal, the sum (\ref{ec:perm}) contains each of the three independent permutations twice, thus introducing a double counting. When the three indices are equal, $i=j=k$, this sum counts six times the single term $H_i H_i H_i$ present in the sum (\ref{ec:L3Hm}). One can take this fact into account by introducing a symmetry factor $S_{ijk}$, which is one if the three indices are different, two if two of the indices are equal, and six if $i=j=k$. With this convention, the interaction is (no sum over indices)
\begin{equation}
 - u \, \frac{\lambda_{ijk}}{S_{ijk}} H_i H_j H_k \,,
\end{equation}
keeping the definition (\ref{ec:lambda}) for $\lambda_{ijk}$ and all permutations (\ref{ec:perm}), even repeated ones. When deriving the Feynman rule for the three-scalar interaction, one has to multiply by a symmetry factor for the presence of identical particles, which is precisely $S_{ijk}$. Therefore, the Feynman rule for the vertex is simply $-i u \lambda_{ijk}$.

\section{Model with two scalar doublets and one singlet}
\label{sec:a}

An attractive SM extension which apparently could lead to stealth boson decays would be that with and extra scalar doublet and a scalar singlet. However, this model does not produce any of the desired processes
\begin{align}
& Z' \to H_i Z \,, \notag \\
& Z' \to H_i H_j \,.
\label{ec:Zdec}
\end{align}
For illustration and completeness we summarise why. We label the two existent scalar doublets as $\Phi_1 = (\phi_1^+ \; \phi_1^0)^T$ and $\Phi_2 = (\phi_2^+ \; \phi_2^0)^T$, and the singlet as $\chi$. The scalar potential compatible with the $\text{SU}(2)_L \times \text{U}(1)_Y$ symmetry is
\begin{eqnarray}
V & = & m_{11} \Phi_1^\dagger \Phi_1 + m_{22} \Phi_2^\dagger \Phi_2 + \frac{m_0^2}{2} \chi^\dagger \chi 
- \left[ m_{12}^2  \Phi_1^\dagger \Phi_2 + \text{h.c.} \right] \notag \\
& &+ \frac{\lambda_1}{2} (\Phi_1^\dagger \Phi_1)^2  + \frac{\lambda_2}{2} (\Phi_2^\dagger \Phi_2)^2
+ \lambda_3 (\Phi_1^\dagger \Phi_1) (\Phi_2^\dagger \Phi_2)
+ \lambda_4 (\Phi_1^\dagger \Phi_2) (\Phi_2^\dagger \Phi_1) \notag \\
& & + \frac{1}{2} \left[  \lambda_5 (\Phi_1^\dagger \Phi_2) (\Phi_1^\dagger \Phi_2)
+ \lambda_6 (\Phi_1^\dagger \Phi_2) (\Phi_1^\dagger \Phi_1)
+ \lambda_7 (\Phi_1^\dagger \Phi_2) (\Phi_1^\dagger \Phi_1) + \text{h.c.} \right] \notag \\
& & + \frac{\lambda_8}{8} ( \chi^\dagger \chi)^2
+ \frac{\lambda_9}{2}  (\Phi_1^\dagger \Phi_1) ( \chi^\dagger \chi)
+ \frac{\lambda_{10}}{2}  (\Phi_2^\dagger \Phi_2) ( \chi^\dagger \chi) \notag \\
& & + \frac{1}{2} \left[ \lambda_{11} (\Phi_1^\dagger \Phi_2) + \text{h.c.} \right] ( \chi^\dagger \chi) \,.
\end{eqnarray}
Among the parameters above, $m_{11}$, $m_{22}$, $m_0$, $\lambda_{1-4}$ and $\lambda_{8-10}$ are real, while $m_{12}$, $\lambda_{5-7}$ and $\lambda_{11}$ can be complex. Writing the neutral scalar fields in the usual way:
\begin{align}
& \phi_1^0 = \frac{1}{\sqrt 2} (\rho_1+v_1 + i \eta_1 ) \,, \quad \phi_2^0 = \frac{1}{\sqrt 2} (\rho_2+v_2 + i \eta_2 ) \,, \quad
\chi = \frac{1}{\sqrt 2} (\rho_3 +u+ i \eta_3 ) \,, 
\end{align}
the would-be Goldstone bosons associated to the breaking of the $\text{U}(1)'$ and electroweak symmetries are $\eta_3$, and a combination of $\eta_1$ and $\eta_2$ (as in the usual two-Higgs doublet model), respectively. Therefore, we have three scalars and a pseudo-scalar, which can in principle mix.

At least one of the two doublets must have vanishing hypercharge $Y'$, as required by the existence of Yukawa terms (\ref{ec:Y}). We choose it to be $\Phi_1$. If the other doublet $\Phi_2$ has hypercharge $Y'_{\Phi_2}  \neq 0$, then invariance under $\text{U}(1)'$ requires $m_{12} = 0$, $\lambda_{5-7} = 0$, $\lambda_{11} = 0$. After applying the potential minimisation conditions it is found that the physical pseudoscalar is massless, which is unacceptable. Besides this obvious drawback, we note that the vacuum expectation value $\langle \phi_2^0 \rangle = v_2/\sqrt{2}$ contributes to $Z-Z'$ mixing~\cite{Langacker:2008yv}, which is constrained to be very small. Because the $Z-Z'$ coupling to the scalars in $\Phi_2$ is proportional to $v_2$, the width for $Z' \to H_i Z$, which would be characteristic for this model, is also very small. If both doublets have $Y' = 0$, neither of the decays in (\ref{ec:Zdec}) is present, the former because of the vanishing doublet hypercharges, and the latter because the only $Z'$ coupling to scalars is $Y'_\chi Z^{\prime \mu} \eta_3 \overleftrightarrow{\partial_\mu}  \rho_3 $ and $\eta_3$ is not physical.

\section{Alternative scenario 1}
\label{sec:c}

We present here results for an alternative scenario 1 for model 2, with $M_{H_2} = 15$ GeV, in which the substructure of the four-pronged fat jets resulting from $H_{3,4} \to H_2 H_2 \to q \bar q q \bar q$ resembles more a two-pronged structure because of the lighter $H_2$. For this mass, we have
\begin{align}
& \text{Br}(H_2 \to b \bar b) = 0.81 \,, \notag \\
& \text{Br}(H_2 \to c \bar c) = 0.12 \,.
\end{align}
\begin{figure}[t]
	\begin{center}
		\begin{tabular}{cc}
			\includegraphics[height=5.5cm,clip=]{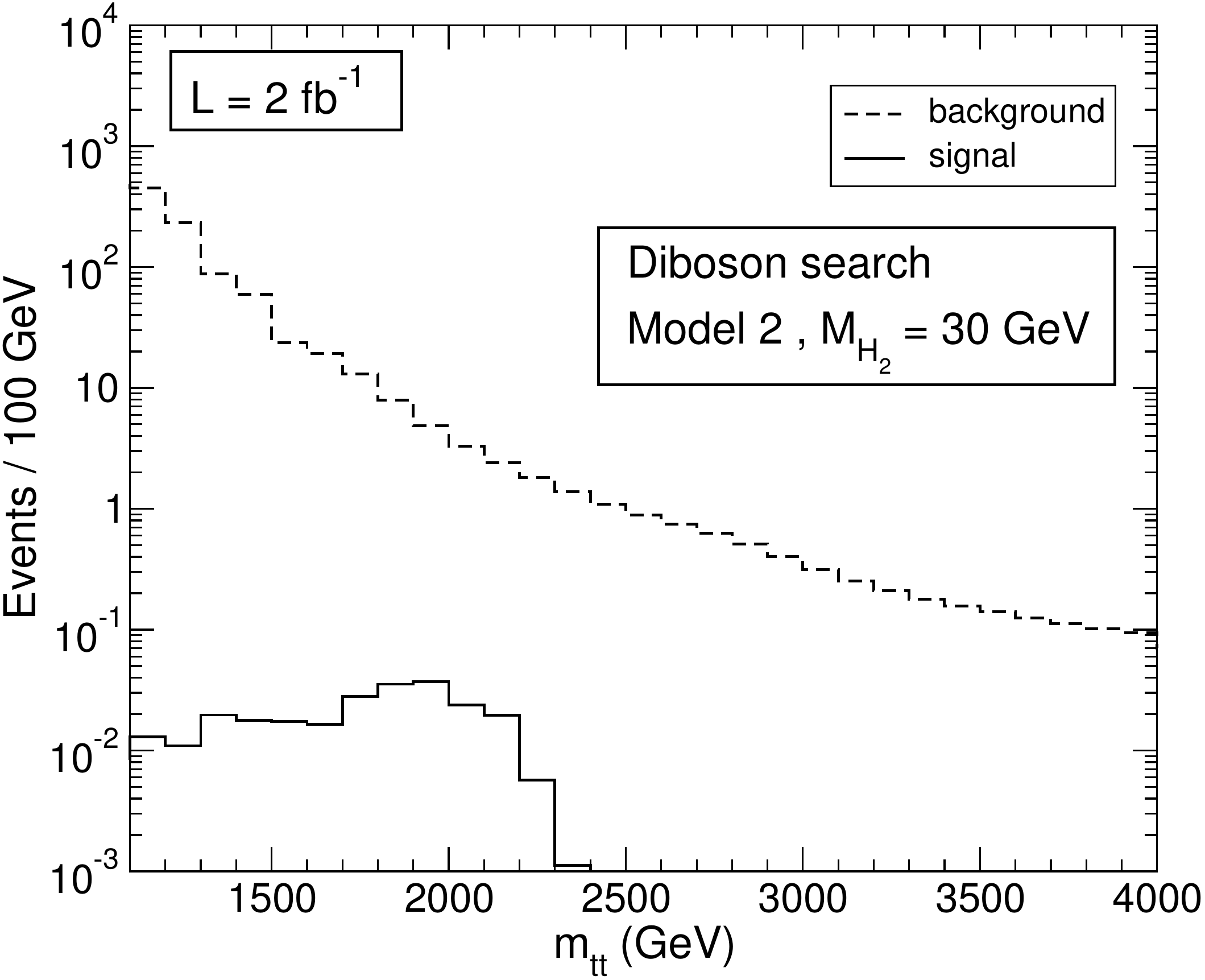} & 
			\includegraphics[height=5.5cm,clip=]{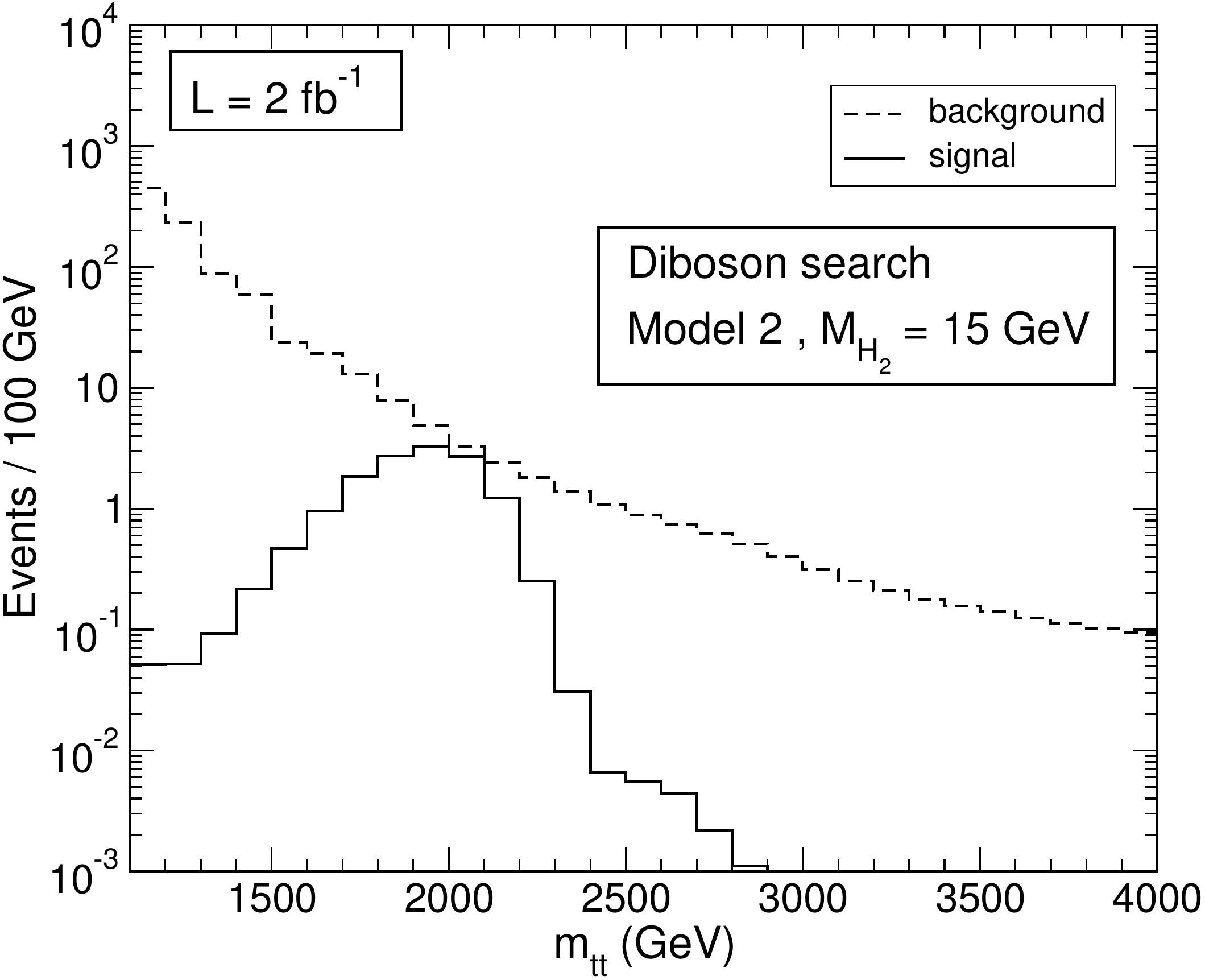} 
		\end{tabular}
		\caption{Invariant mass distribution for the $Z'$ signals and their backgrounds in the diboson analysis, for and alternative scenario 1 of model 2 with $M_{H_2} = 15$ GeV. }
		\label{fig:inv1b}
	\end{center}
\end{figure}
\begin{figure}[t]
	\begin{center}
		\includegraphics[height=5cm,clip=]{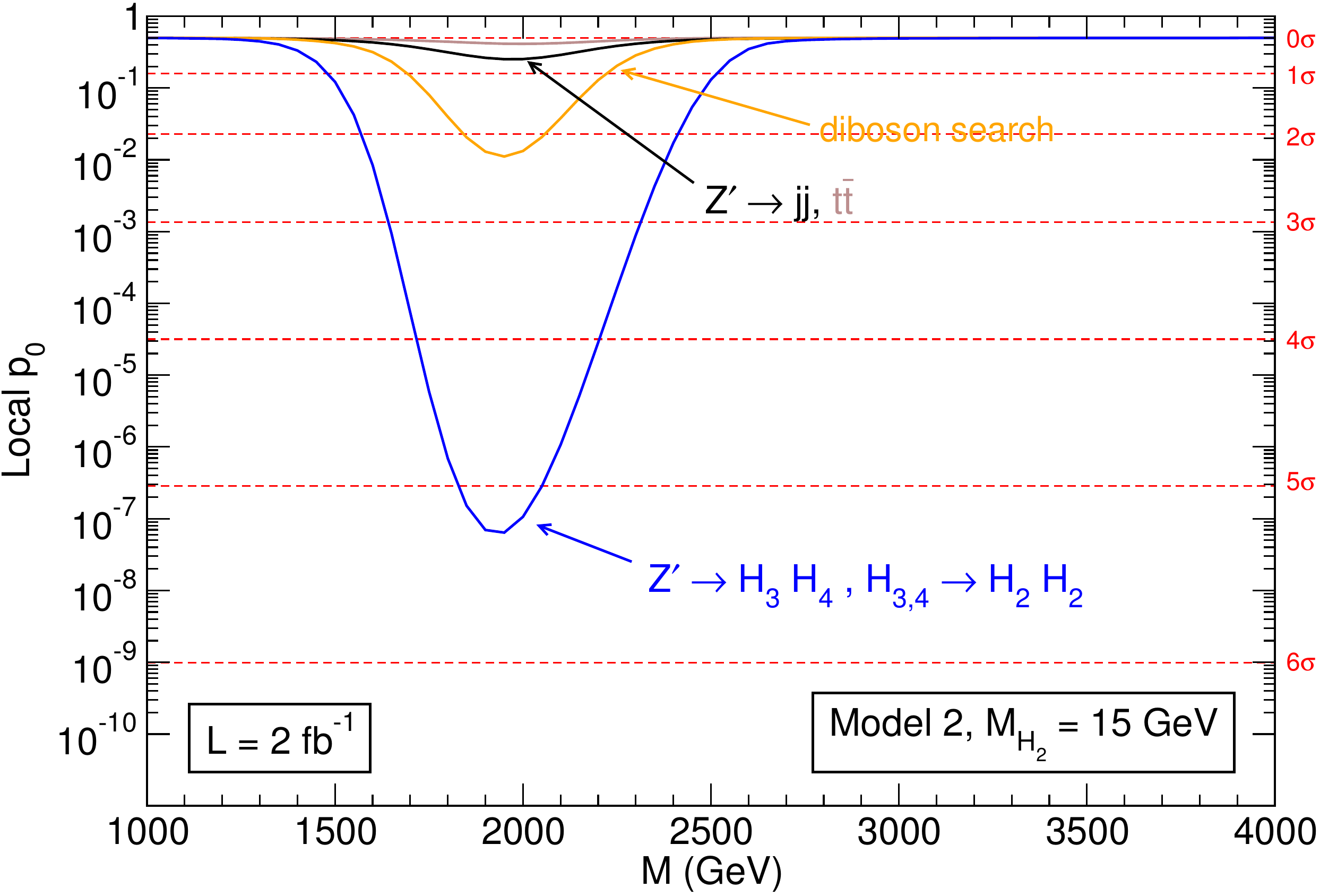}
		\caption{Local $p$-value for the $Z'$ signal in the different searches for and alternative scenario 1 of model 2 with $M_{H_2} = 15$ GeV.}
		\label{fig:Pval1b}
	\end{center}
\end{figure}
so that $\text{Br}(H_2 \to b \bar b,c\bar c)^4 = 0.732$, being the signals slightly smaller. The efficiency of the tagger for this signal is practically the same as in scenario 1. The dijet mass distribution for the diboson analysis is shown in figure~\ref{fig:inv1b} for $M_{H_2} = 30$ GeV (left panel ) and $M_{H_2} = 15$ GeV (right panel). Besides the large differences in the cross section, due to the larger acceptance for $Z' \to H_3 H_4$, in the latter case we observe a resonant signal structure that is not present in the former. The $p$-value for the different $Z'$ signals is given in figure~\ref{fig:Pval1b}. Notice that, despite the fact that a possible signal would be more visible in the diboson resonance searches than in $t \bar t$ and dijet final states, it is by far surpassed by the signal that would be visible in a generic search using the anti-QCD tagger.

\end{document}